\documentclass[twocolumn]{aastex62}
\pdfoutput=1 %for arXiv submission
\usepackage{epstopdf} %for arXiv submission
\usepackage{comment}
\usepackage[caption=false]{subfig}  
\usepackage{amsmath}
\usepackage{tabularx}
\usepackage{url}

\graphicspath{{./}{Astars/}{Bstars/}{Fstars/}{Gstars/}{Kstars/}{LBVs/}{Mstars/}{Ostars/}{WRstars/}{bad\_nifty/}{short\_tch/}}

\shorttitle{Variability of massive stars in M31 from PTF}
\shortauthors{Soraisam et al.}

\begin{document}

\title{Variability of massive stars in M31 from the Palomar Transient Factory}

\correspondingauthor{Monika Soraisam}
\email{soraisam@illinois.edu}

\author{Monika D.~Soraisam}
\affiliation{National Center for Supercomputing Applications, University of Illinois at Urbana-Champaign, Urbana, IL 61801, USA}
\affiliation{Department of Astronomy, University of Illinois at Urbana-Champaign, Urbana, IL 61801, USA}
\author{Lars Bildsten}
\affiliation{Department of Physics, University of California, Santa Barbara, CA 93106, USA}
\affiliation{Kavli Institute for Theoretical Physics, University of California, Santa Barbara, CA 93106, USA}
\author{Maria R.~Drout}
\affiliation{Department of Astronomy and Astrophysics, University of Toronto, 50 St. George Street, Toronto, Ontario, M5S 3H4 Canada}
\affiliation{The Observatories of the Carnegie Institution for Science, 813 Santa Barbara St., Pasadena, CA 91101,USA}
\author{Thomas A.~Prince}
\affiliation{Division of Physics, Mathematics, and Astronomy, California Institute of Technology, Pasadena, CA 91125, USA}
\author{Thomas Kupfer}
\affiliation{Kavli Institute for Theoretical Physics, University of California, Santa Barbara, CA 93106, USA}
\author{Frank Masci}
\affiliation{Infrared Processing and Analysis Center, California Institute of Technology, Pasadena, CA 91125, USA}
\author{Russ R.~Laher}
\affiliation{Infrared Processing and Analysis Center, California Institute of Technology, Pasadena, CA 91125, USA}
\author{Shrinivas~R.~Kulkarni}
\affiliation{Division of Physics, Mathematics, and Astronomy, California Institute of Technology, Pasadena, CA 91125, USA}

\begin{abstract}
Using data from the (intermediate) Palomar Transient Factory (iPTF), we characterize the time variability of $\approx500$ massive stars in M31. Our sample is those stars which are spectrally typed by Massey and collaborators, including Luminous Blue Variables, Wolf-Rayets, and warm and cool supergiants. We use the high-cadence, long-baseline ($\approx5$~years) data from the iPTF survey, coupled with data-processing tools that model complex features in the light curves. We find widespread photometric ($R$-band) variability in the upper Hertzsprung Russell diagram (or CMD) with an increasing prevalence of variability with later spectral types. Red stars ($V-I>1.5$) exhibit larger amplitude fluctuations than their bluer counterparts. We extract a characteristic variability timescale, $t_{\rm ch}$, via wavelet transformations that are sensitive to both continuous and localized fluctuations. Cool supergiants are characterized by longer timescales ($>100$~days) than the hotter stars. The latter have typical timescales of tens of days but cover a wider range, from our resolution limit of a few days to longer than 100~days timescales.  Using a 60-night block of data straddling two nights with a cadence of around 2~minutes, we extracted $t_{\rm ch}$ in the range 0.1--10~days with amplitudes of a few percent for 13 stars. Though there is broad agreement between the observed variability characteristics in the different parts of the upper CMD with theoretical predictions, detailed comparison requires models with a more comprehensive treatment of the various physical processes operating in these stars such as pulsation, subsurface convection, and the effect of binary companions.

\end{abstract}


\section{Introduction} \label{sec:intro}
Massive stars, characterized by initial main-sequence masses greater than around $10~M_{\odot}$ and thus being relatively rare and short-lived \citep{Massey-2003, Massey-2013}, are 
important players in varied astrophysical phenomena, shaping the Universe at scales both small and large. For example, their strong stellar winds and radiation fields, and their eventual supernova explosions, cause the chemical enrichment, ionization, and turbulent motion of their local interstellar medium (ISM; \citealt{Abbott-1982,Freyer-2003,Freyer-2006,Matzner-2002,Nomoto-2013}); the collective radiation energy of their population acts as a key source of ionization in galaxies \citep{Haffner-2009,Wolfire-2003}, their populations in high-redshift galaxies are considered to significantly contribute to the cosmic re-ionization budget \citep[e.g.,][]{Robertson-2015}, and the collective energy from their final explosions provides vital feedback regulating galaxy formation and evolution \citep{Veilleux-2005,Governato-2010}.

These stars are expected to undergo variability at various stages in their evolution---instability in the radiation-dominated envelopes for the hot stars driven by the iron opacity bump \citep{Paxton-2015,YanFei-2015}, pulsational instability driven by the opacity of the partial ionization zone of hydrogen for the cool supergiants \citep{Li-1994,Yoon-2010} as well as the warm supergiants (the classical instability strip; \citealt{Bono-1999}), and convectively driven low-amplitude stochastic variability both in hot and cool stars \citep{Cantiello-2009, Stothers-2010, Aerts-2015, Diaz-2018}. These different instabilities in a star manifest as observable photometric and/or spectroscopic variability. In particular, pulsations play an important role in the evolution of massive stars through their effect on mass loss \citep[e.g.,][]{Neilson-2008,Yoon-2010,YanFei-2018, Ouchi-2019}. Observations of variability from massive stars, particularly photometric variability, easily afforded by the multitude of current/upcoming time-domain surveys, will provide powerful constraints on their theoretical models. Since it will be impossible to observationally pursue the evolution of a single massive star, statistical studies of large samples of massive stars representing many spectral types are needed to gain important insights into their populations.

A wide range of timescales is expected to characterize the photometric variability of massive stars \citep[e.g.,][]{Lovy-1984, Heger-1997, Yoon-2010, YanFei-2015, YanFei-2018}. Accordingly, wide-field, high-cadence time-domain surveys from the ground with a long baseline extending to a decade, e.g., ASAS \citep{Pojmanski-2002}, OGLE \citep{Udalski-2003}, and the more recent Palomar Transient Factory (PTF) survey \citep{Law-2009, Rau-2009}, are well suited for observational studies of the variability of these stars. 
In fact, statistical studies of variability have been performed for massive stars in the Large Magellanic Cloud (LMC; \citealt{LMC-2010}) and in the Small Magellanic Cloud (SMC; \citealt{Kourniotis-2014}) based on ASAS and OGLE data, respectively. Space-based surveys (e.g., {\it HST, Kepler, TESS}), with their exquisite photometry, are desirable, but their narrow field of view and baseline generally limit such analyses to probing short timescales for a handful of stars, with the exception of a few fields \citep{Wallenstein-2019, Bowman-2019}.

Recently \citet{Conroy-2018} presented their work based on {\it HST} data on the variability of massive stars in M51, which has a high star formation rate. \citet{Spetsieri-2018} also used $\approx 3$~months of archival {\it HST} data to catalog massive stars in the Virgo Cluster galaxy NGC 4535 and analyze their variability; \citet{Gaia-2018} performed a similar study based on 22~months of {\it Gaia} data for the Milky Way's stellar populations extending from luminous stars down to dwarf stars. \citet{Conroy-2018} extended the baseline to a decade using archival {\it HST} data of M51. All these studies examined the light curves of the stars to determine the fraction showing observable variability, and extracted variability features such as amplitudes and periods; the latter were also used to label the periodic variables. These quantities were mapped out in the color-magnitude diagram (CMD), and in agreement with our previous study (\citealt{Soraisam-2018}, hereafter Paper~I), \citet{Conroy-2018} found almost all cool supergiants to be variable. The latter authors also made a direct comparison of the observations with theoretical predictions for instabilities (based on, for example, \citealt{Paxton-2013, Paxton-2015, Yoon-2010}) and found agreement for most parts of the CMD, except for some regions occupied by the fainter blue and red variables that require further investigation.

M31 is one of the most imaged galaxies, where we are able to obtain a wealth of information for its resident stellar populations in the different parts of the galaxy. It has, thus, also been frequently targeted by time-domain surveys, such as Pan-STARRS \citep{Kaiser-2010, clee-2012}, PTF.  Given the relatively shallow depth ($\approx20$~mag) for these surveys, variability of luminous stars constitutes the main detection from these surveys. 
In this paper, we use the wealth of data---both spectroscopic and time-resolved photometric, obtained from \citet{Massey-2016}, hereafter MNS16, and the PTF survey, respectively---available for M31 to study variability of its massive-star population. Our study thus provides a detailed summary of the observed variability characteristics of massive stars in one of the important galaxies in the Local Universe. For instance, M31 oftentimes stands as a representative of super-solar-metallicity environment when studying the effect of metallicity on the various physical processes of these stars, such as stellar winds \citep{Massey-2017, Neugent-2019}, pulsation \citep{Ren-2019}, convection \citep{Chun-2018}, as well as binarity \citep{Neugent-2014}. Yet, a comprehensive and consistent characterization of the photometric time-series observations of these stars, which will be influenced by these processes, did not exist (barring for Red Supergiants presented in Paper~I) prior to this work.

Given the importance of massive stars in the evolution of galaxies, performing such studies for a large sample of diverse host-galaxy environments (e.g., with varying metallicities and star-formation rates) will inform us both on the physics of the variability phenomena and on their relation to galactic evolution. This work substantially extends the sample size of galaxies with the variability of massive stars characterized in the CMD. Furthermore, timescales of variability were only estimated for periodic light curves in the preceding similar studies. Here, by using a wavelet-based analysis, we also determine the characteristic  timescale $t_{\rm ch}$ for variability that may be localized (non-periodic) or unlocalized (periodic) in time.

This paper is organized as follows. In Sect.~\ref{sec:data}, we describe the sample of massive stars in M31 used for this study and the time-domain data from PTF used for constructing their light curves. We present the method for analyzing the light curves for variabilities and their characteristic timescales in Sect.~\ref{sec:lc_analysis}, and show the results of our analysis in Sect.~\ref{sec:result}. We discuss our results in Sect.~\ref{sec:discuss} and conclude with Sect.~\ref{sec:conclude}.

\section{Data} \label{sec:data}
We perform the variability analysis of massive stars in M31 using two complementary data sets: (1) a catalog of spectroscopically typed luminous stars in M31 (MNS16) and (2) data from the long-baseline optical time-domain survey of M31 by PTF. These data sets are described below. 

\subsection{Spectroscopic sample of massive stars in M31} \label{sec:spectra}
The Local Group Galaxy Survey (LGGS), by providing an extensive catalog of stars in M31 with multiband photometry \citep{Massey-2006}, has enabled drawing up lists of bright and isolated targets for spectroscopic observations and their subsequent characterization \citep[e.g.,][]{Massey-2007, Cordiner-2011, Neugent-2012}. MNS16 doubled the number of massive stars with spectroscopic labels, particularly those extending from O- to G-type. MNS16 further updated the LGGS catalog of M31, including the spectral classifications from their study as well as those available from earlier efforts; this updated catalog also includes the majority of the spectral classifications for the stars studied by \citet{Humphreys-2017}. \citet{Massey-2016b} also performed a study exclusively focused on the Red Supergiant (RSG) content of M31, contemporaneous with MNS16, obtaining the spectral types for $\approx 200$ RSGs. This RSG catalog was the source for our study of RSG variability in Paper~I. The spectral type information for these RSGs is also included in the updated LGGS catalog.

In total, 1050 stars in the LGGS catalog of MNS16 have spectral classifications with membership flag (Mm) greater than 0, where a value of 0 indicates a foreground star, 1 a confirmed member of M31, 2 a probable member, and 3 unknown membership. These 1050 stars form our reference sample for study of massive star variability in M31, around 70\% of which have an Mm flag value of 1. We expect the foreground contamination in the remaining 30\% to be minimal; as may be expected, we find many of them to match objects in 
{\it Gaia}-DR2 \citep{Gaia-2018b,Gaia-2016} but the signal-to-noise ratio of the measured proper motions for almost all of the matches is low, indicating that the stars are likely extragalactic. A detailed analysis of foreground contamination is beyond the scope of this paper.

Our reference sample includes (candidate) luminous blue variables (LBVs), Wolf-Rayet stars, OB stars with luminosity classes I, III, and V (for O-type), and supergiants with spectral types ranging from A to M (see MNS16 for more details).

\subsection{M31 PTF time-domain data} \label{sec:iptf}
The (intermediate) PTF survey \citep{Masci-2017} provides a rich optical time-domain data set for M31 \citep{Soraisam-2017}, with almost daily sampling for about five years. We use the same iPTF/PTF\footnote{We use `PTF' and `iPTF' interchangeably in this paper.} data set in the $R$-band as in Paper~I, covering around $1.8\times2.4~{\rm deg}^2$ at depth reaching $m_{R}\approx 21$, to construct the forced-photometry-based light curves of our 1050 reference sample stars. More details about the survey data as well as the photometry process are given in Paper~I and references therein. 
Once the Zwicky Transient Facility \citep{Bellm-2019}, the successor to iPTF, completes operation, the combined baseline will extend to over a decade.

Of the 1050 stars in the reference sample, 1015 are in the PTF footprint that we have analyzed. We construct their light curves via forced photometry on the difference images (see \citealt{Masci-2017} for details of the iPTF difference-image pipeline). To convert these measured flux differences to absolute fluxes, we add the corresponding subtracted fluxes of the stars obtained by cross-matching (using a search radius of $2''$--the typical PTF FWHM) our reference sample with the PSF-fit photometry catalog of the template image used in the subtraction \citep{Laher-2014}. For 63 of the stars in our reference sample, we do not find counterparts in the template image catalog within the search radius threshold of $2''$. Around 40 of these are tagged as Wolf-Rayets in MNS16. The LGGS $R$-band magnitudes for almost half of these 63 stars are in the range 18--20 while the remaining half have values greater than 20~mag, which is the typical depth of the iPTF survey. For these 63 stars, the distance to the nearest neighbor in the template image extends up to $8.5''$, where the chance of random association increases. It is likely that a combination of faint magnitudes and/or offsets of more than the standard $2''$ between the position of the detection on the difference images and the tabulated source position in the template image catalog prevents identification of their counterparts. We drop these 63 sources (just around 5\% of the reference sample) from further consideration.

The iPTF light curves for the stars in our reference sample are available at DataLab hosted by the National Optical Astronomy Observatory.{\footnote{The light curve data can be accessed programmatically on DataLab (\url{https://datalab.noao.edu}) at \url{soraisam://public/m31PTFmassivestars/}}}

\begin{figure}
\includegraphics[width=88mm]{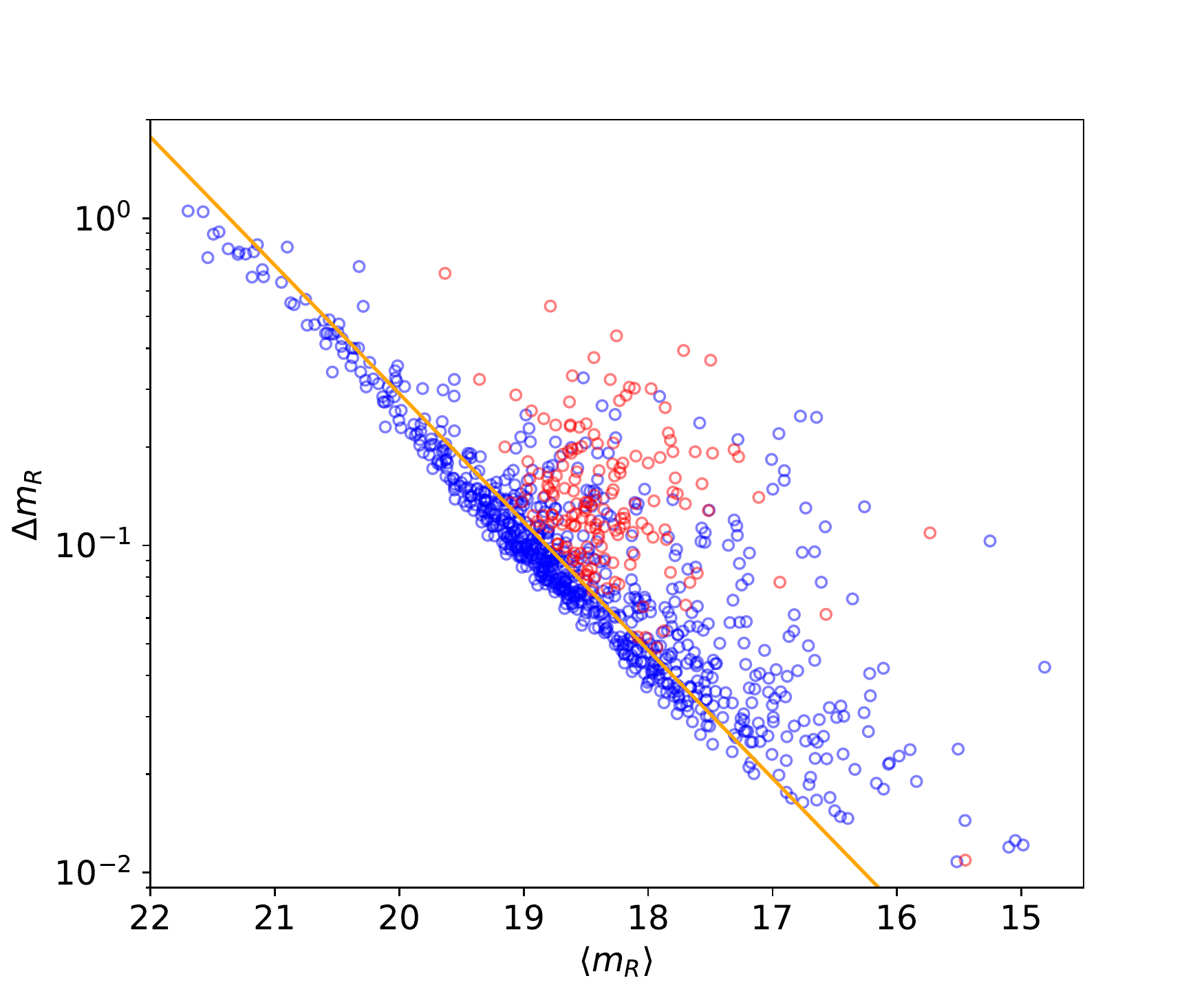}
\caption{Distribution of standard deviation $\Delta m_{R}$ against mean magnitude $\left<m_{R}\right>$ for the massive stars in M31 based on iPTF light curves. The red points mark the RSGs with observed variability from Paper~I. Points above the orange line are stars with their computed RMS variability exceeding that from noise.}
\label{fig:filter}
\end{figure}

\section{Analysis of light curves} \label{sec:lc_analysis}
\subsection{Photometric variability} \label{sec:var}
We use the iPTF $R$-band light curves of the massive stars constructed above in order to asses their photometric variability. Similar to Paper~I, we compute the mean magnitude $\left<m_{R}\right>$ and standard deviation about the mean ($\Delta m_{R}$) for each star. Figure~\ref{fig:filter} shows the variation of $\Delta m_{R}$ with the mean magnitude of the star. The figure shows absolute variability in magnitude units, equivalent to relative variability in flux units. The orange line shows the noise level of PTF from Paper~I, which was computed considering only stars without PTF-detectable variability and with $\left<m_{R}\right><20$ and extrapolated to fainter stars. As may be expected, the sensitivity of the PTF survey to relative photometric variability increases with the mean brightness of the stars. In absolute flux units, the noise floor is approximately constant and does not depend on the brightness of the star. It is to be noted that the majority ($\sim 90\%$) of the stars in our reference sample are isolated/uncrowded (MNS16) and we find no significant difference between the distributions of the low-level $\Delta m_{R}$ values for the crowded and uncrowded stars in our sample.

The $\Delta m_{R}$ values of the few sources at $\left<m_{R}\right>>20$ above the orange line (Fig.~\ref{fig:filter}) are suspect since the typical depth of the iPTF single-exposure observations is around 20~mag. Indeed, visually examining the light curves of these sources reveals that they are dominated by large error bars. We therefore ignore sources whose mean magnitudes are fainter than 20~mag. 
Given the sensitivity of the iPTF survey, we find 502 stars with $\left<m_{R}\right><20$ above the orange line, i.e., with observable variability. 
Under the spectral type column of MNS16, 8 of these 502 stars are labelled as HII and 3 listed as belonging to clusters without a proper spectral specification. One of the stars belonging to HII regions is likely a W~UMa-type contact binary (see Appendix~\ref{append:wuma}). 
We drop these 11 stars, thus bringing our final sample of massive stars with observed variability to 491. For 16 of the stars in our sample, we find the angular separation amongst pairs of them to be less than $5''$, which is comparable to the typical iPTF FWHM of $2''$, and therefore effects of blending are present for them. Nevertheless, we still include them in our analysis, since in this paper we are particularly interested in the population statistics. The total number of stars in our reference sample with $\left<m_{R}\right><20$ not belonging to clusters or HII regions and with no discernible variability from iPTF is 384.

\begin{figure*}
\includegraphics[width=60mm]{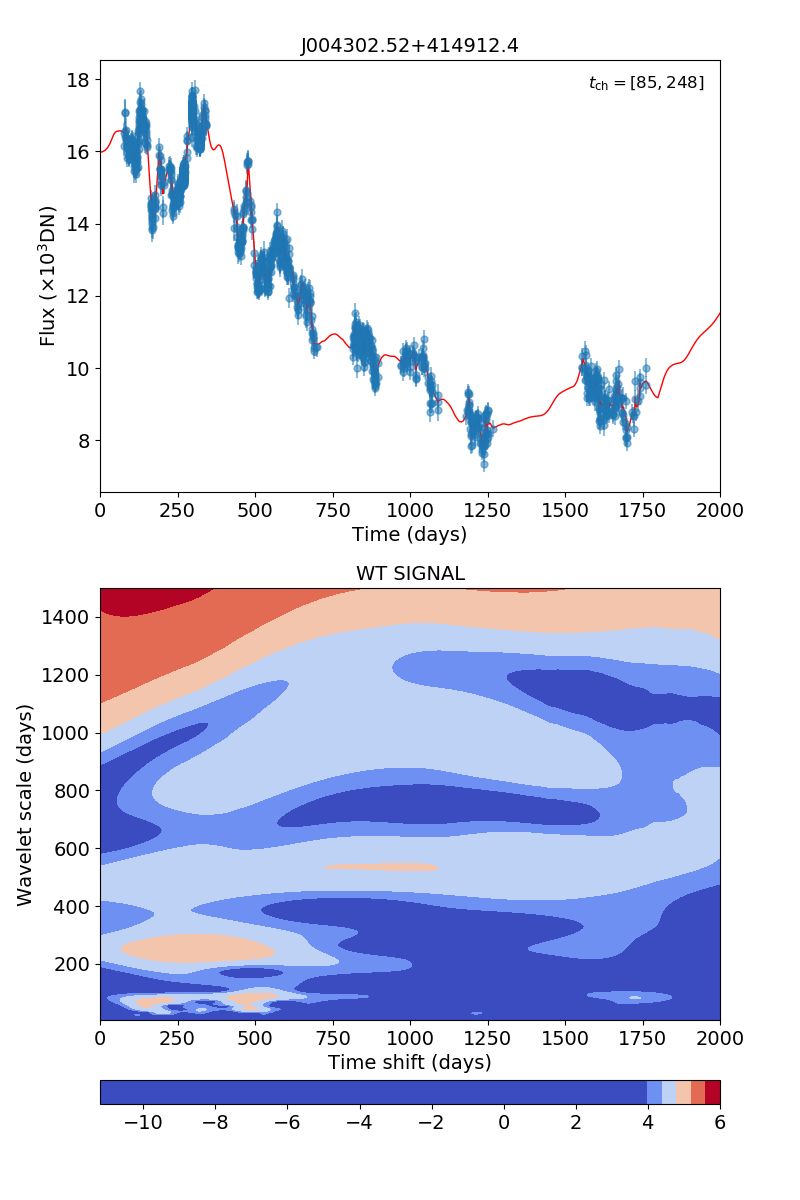}\hfill\includegraphics[width=60mm]{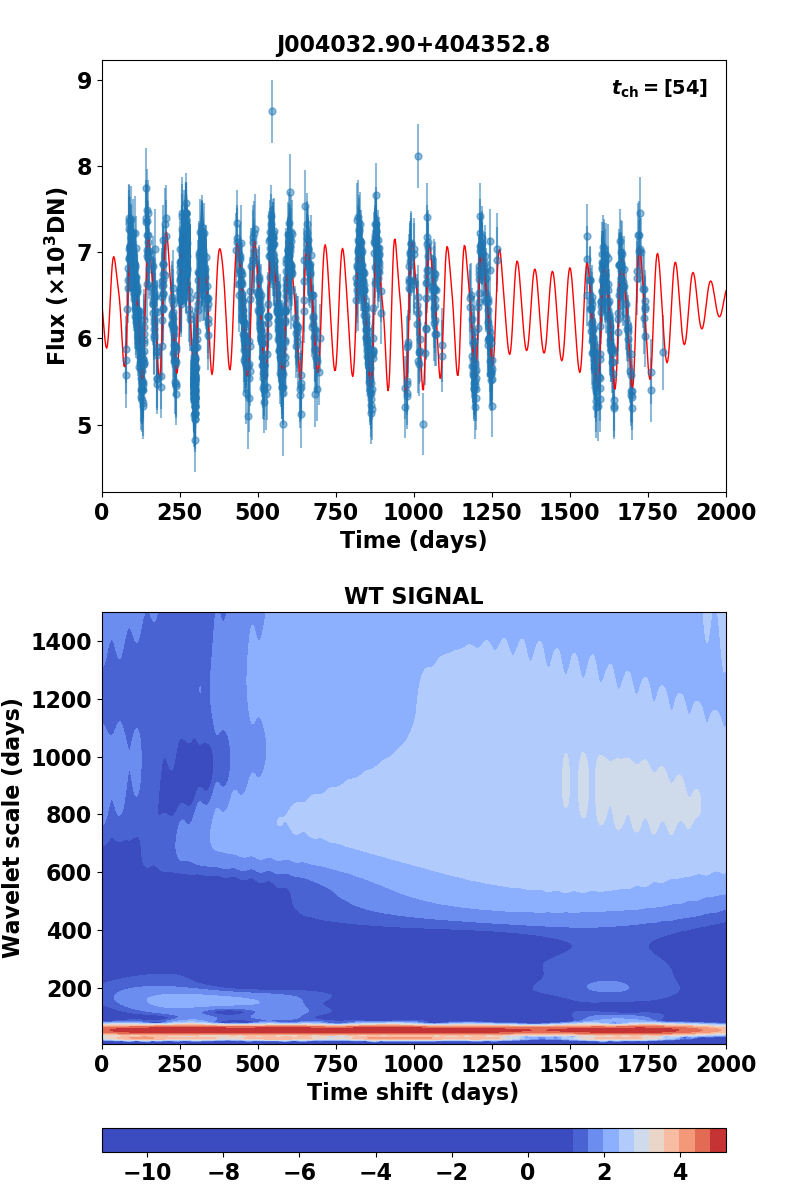}\hfill\includegraphics[width=60mm]{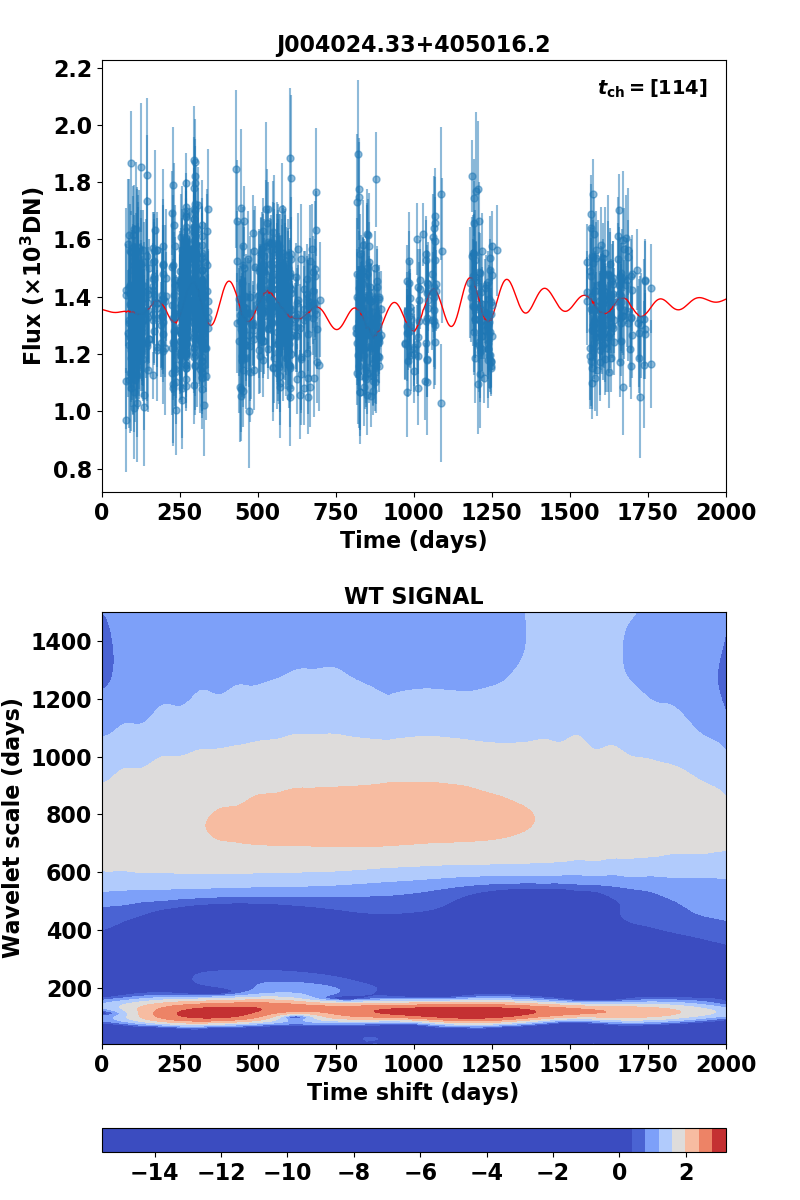}
\includegraphics[width=60mm]{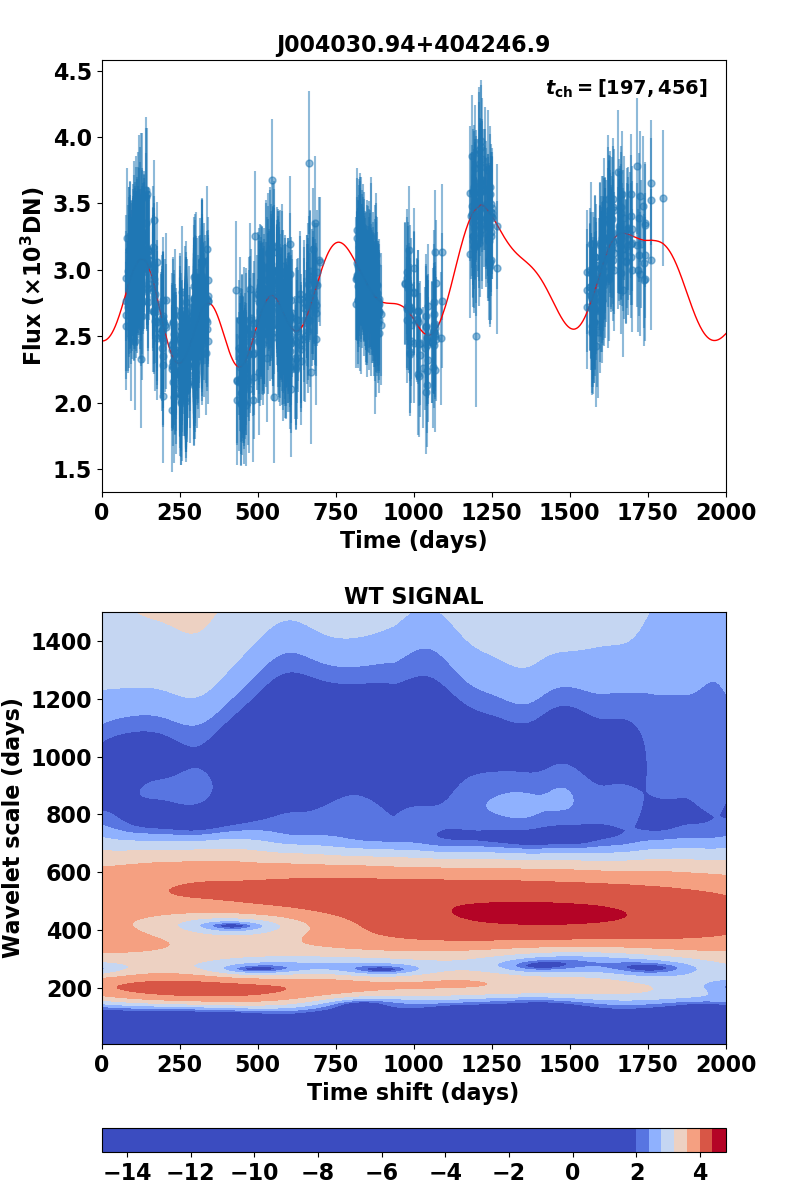}\hfill\includegraphics[width=60mm]{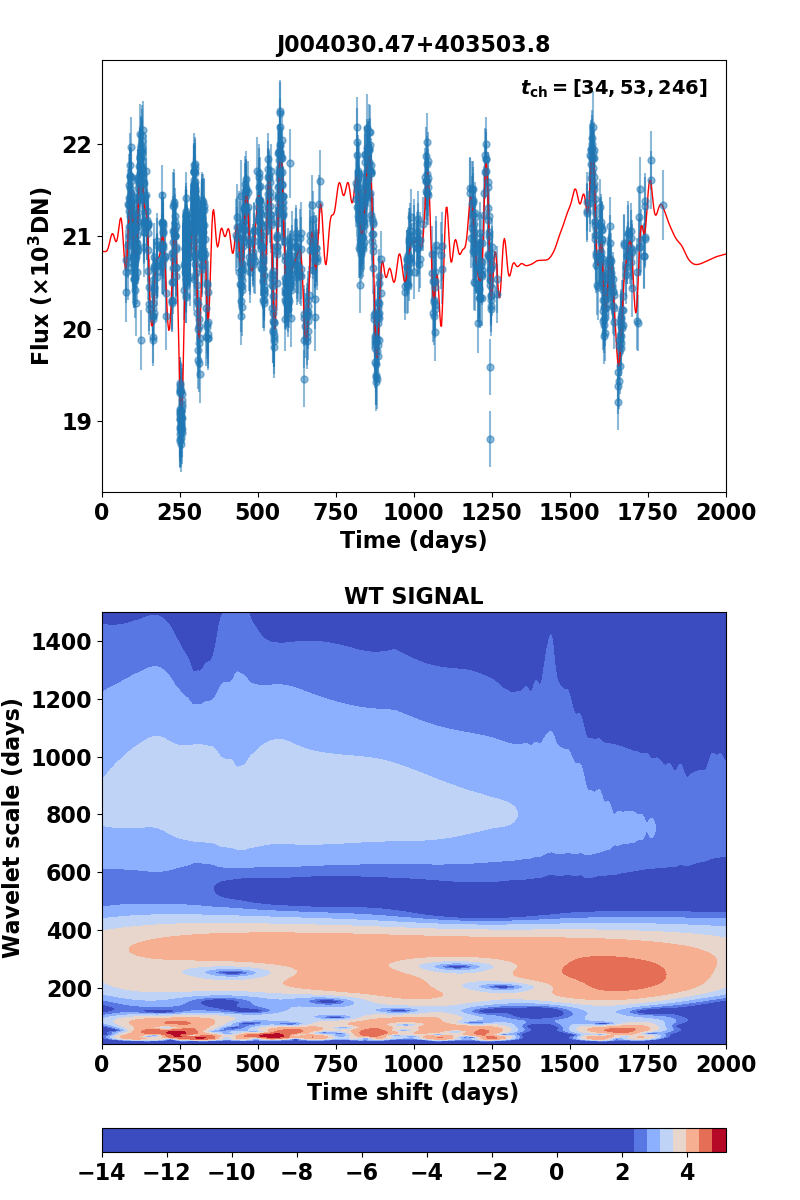}\hfill\includegraphics[width=60mm]{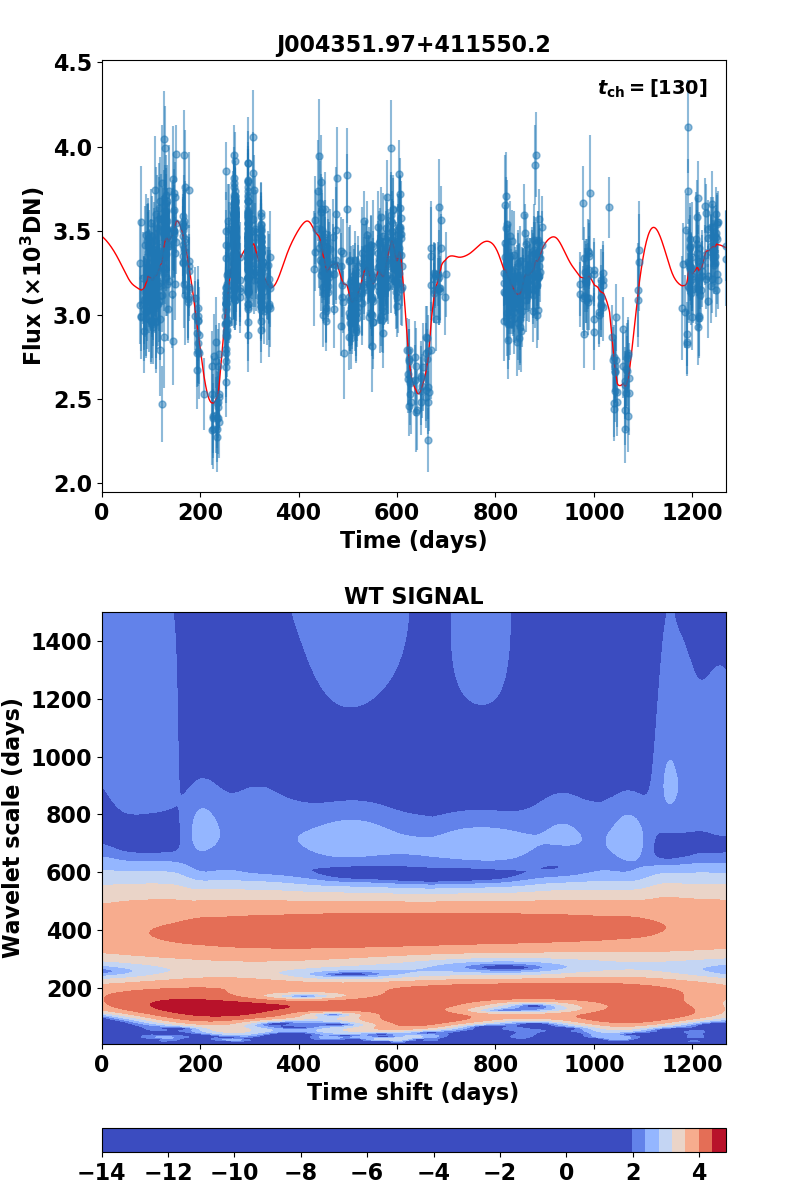}
\caption{Examples of iPTF light curves along with the corresponding wavelet transform maps for an LBV (cross-id AE And), YSG, and WR star ({\it top}, left to right), and for supergiants of type O, B, A ({\it bottom}, left to right). The red curve in each panel with the observed light curve shows the reconstruction (see text). The ID of the star from MNS16 is shown on top of each plot. The time axis in the light curve plots is with respect to a reference value of MJD 56000. The wavelet transform values are shown in log scale.} 
\label{fig:time_eg}
\end{figure*}

\begin{figure}
\includegraphics[width=88mm]{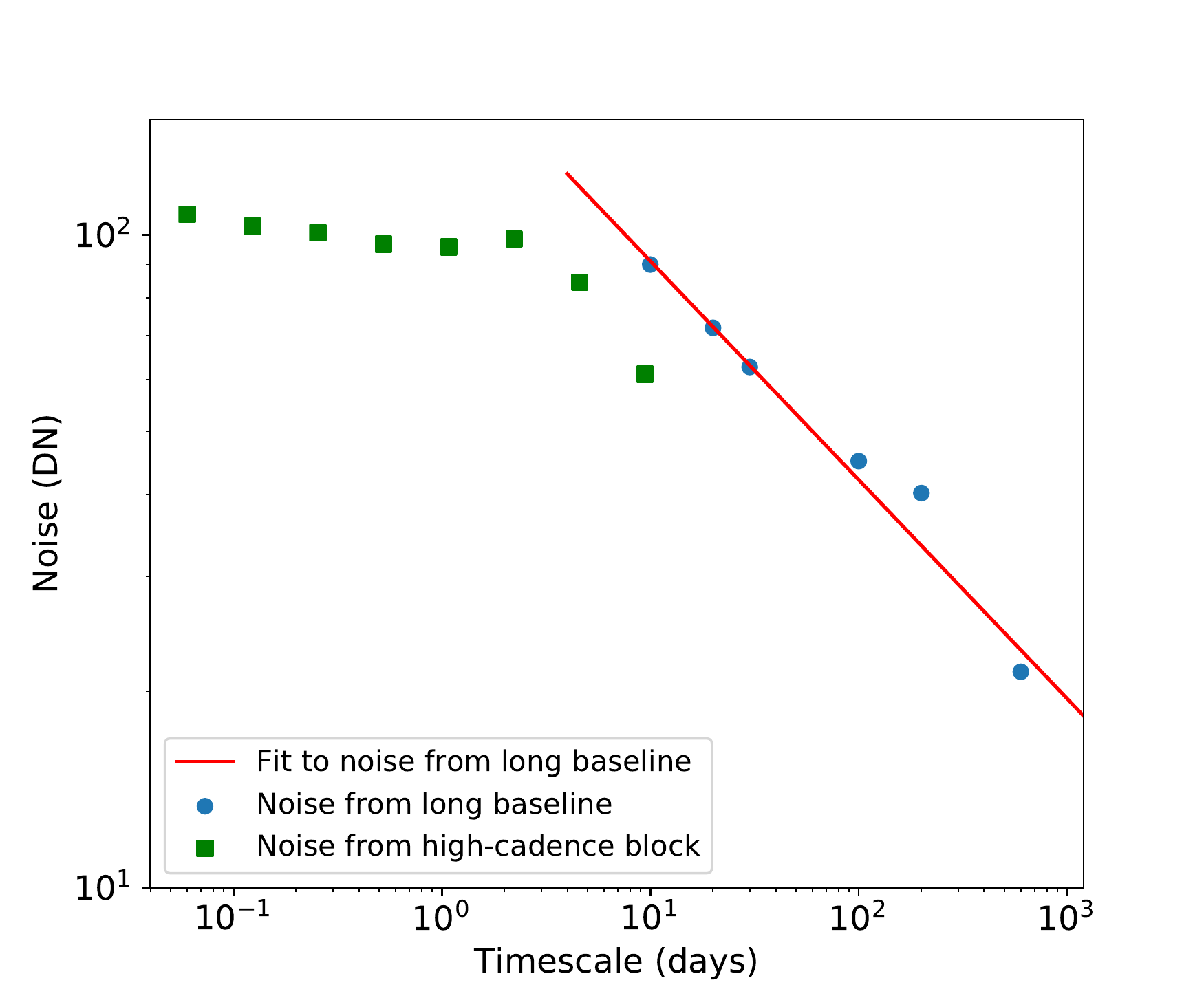}
\caption{Approximate values of noise (in flux units) for PTF data against characteristic timescales over which they are applicable, obtained by smoothing out the contributions from smaller timescales. The blue circles denote noise estimates extracted using the long baseline PTF light curves of static stars from Paper~I  and the red line is the fit to them. The green square symbols denote noise estimates obtained using the high-cadence block of the light curves (see Sect.~\ref{sec:short_tch}). 
}
\label{fig:timescale_noise}
\end{figure}

\subsection{Characteristic timescales for variable stars} \label{sec:time}
The nature of the photometric variability exhibited by these massive stars is quite diverse: it can be completely stochastic, perfectly periodic, or anything in between. 
Some examples highlighting the different flavors of variation in the light curves are shown in Appendix~\ref{append}. As mentioned before, a large range in the characteristic timescales of variability of massive stars is expected.

We ascribe the timescale(s) found from a time-frequency or, more precisely, a time-scale length analysis of the light curve using a wavelet transform, as the characteristic timescale(s) of variability for the star, $t_{\rm ch}$. The conventional form of timescale determination based on Fourier analysis (as used in previous studies, see Sect.~\ref{sec:intro}), is limited in its application---it is meant for stationary signals without a time-varying frequency. For periodic signals, Fourier-based analysis is well-justified, but not for a more generic analysis of timescales covering different forms of variation in the light curves. On the other hand, a wavelet transform using oscillatory functions with finite duration, i.e., wavelets, offers a reasonably convenient tool that allows investigating not just a regular mode of variation (i.e., periodicity) but also fluctuations that may be evolving or even transient, along with their timescales via the scale parameter values used to stretch or dilate the wavelet (see Eq.~\ref{eg:WT}).

Despite the efficacy of the wavelet transform formalism, it finds limited application in the analysis of time-series that are unevenly sampled, which is almost always the case for astronomical time-domain data, including iPTF. Some examples of tackling the gaps in data for the purposes of employing a wavelet transform are provided by \citet{WWZ-1996} and \citet{Frick-1997}. We take an alternative approach here, by first reconstructing the light curve to fill the gaps and then simply performing a continuous wavelet transform using the Morlet wavelet. We make use of a Gaussian Process model to reconstruct the signal via the critical filter tool of \citet{Oppermann-2013} in the NIFTy package of \citet{Selig-2013}. We successfully implemented this approach in Paper~I for analyzing the RSG light curves characterized by varied forms of fluctuations (see Paper~I for more details). Some examples of such reconstructions are shown in Fig.~\ref{fig:time_eg} and also in Appendix~\ref{append}, which also show the diverse shapes of the light curves of massive stars.

We convolve the reconstructed light curve with the Morlet wavelet, which is a harmonic function with a Gaussian envelope suited for analysis of variable-star light curves, taking scales ranging from 1 day to 1500 days, and extracting the transformation coefficient at every convolution step. The continuous wavelet transform of a light curve $f(t)$ is 
\begin{align}
w(a,b)&=a^{-1}\int_{-\infty}^{\infty}f(t')\,\psi^{*}\left(\frac{t'-b}{a}\right)dt',\\ 
\psi(t)&=e^{\frac{-t^2}{2}}e^{i\omega_{o}t},
\label{eg:WT}
\end{align}
where $a$ and $b$ represent the wavelet scale and translation (time shift) parameter, respectively, $^*$ denotes complex conjugation, and $\psi$ is the Morlet wavelet where we take the dimensionless constant $\omega_{o}=6$ such that the wavelet decays significantly in a single cycle (see, for example, \citealt{WWZ-1996,Frick-1997}). Note that the transformation is normalized here with $1/a$, instead of the conventional $1/\sqrt{a}$, following \citet{Lilly-2017} to ensure that the transform coefficient scales proportionally to the RMS deviation about the mean of the signal, hereafter referred to as amplitude.

The resulting map of the correlation power (transformation coefficient-squared, i.e., $\left|w(a,b)\right|^2$) is shown in the bottom panel of the corresponding light curve in Fig.~\ref{fig:time_eg}, where a large correlation value (indicated in red in the figure) at a given wavelet scale $a$ indicates the presence of such a timescale in the light curve. Because of the uncertainty principle, we cannot both localize the frequency and time. Therefore, significant power at small wavelet scales (with better time resolution) appears as narrow streaks or blobs; in the alternative case, power excess at large wavelet scales corresponding to poor time resolution, result in big blobs in the maps. These features are evident in the bottom-middle panel of Fig.~\ref{fig:time_eg}.

\begin{figure*}[t]
    \centering
    \includegraphics[width=88mm]{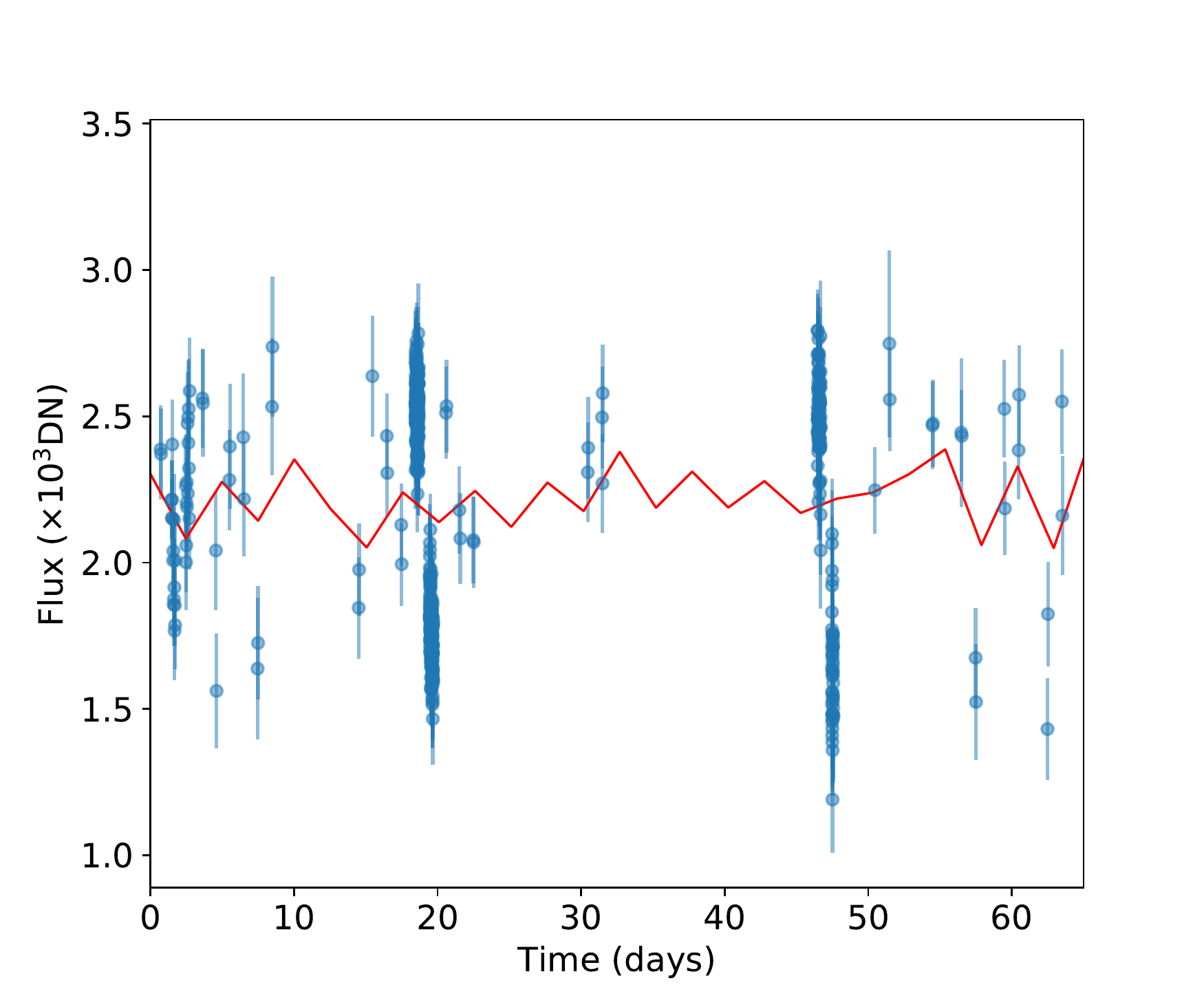}\hfill\includegraphics[width=88mm]{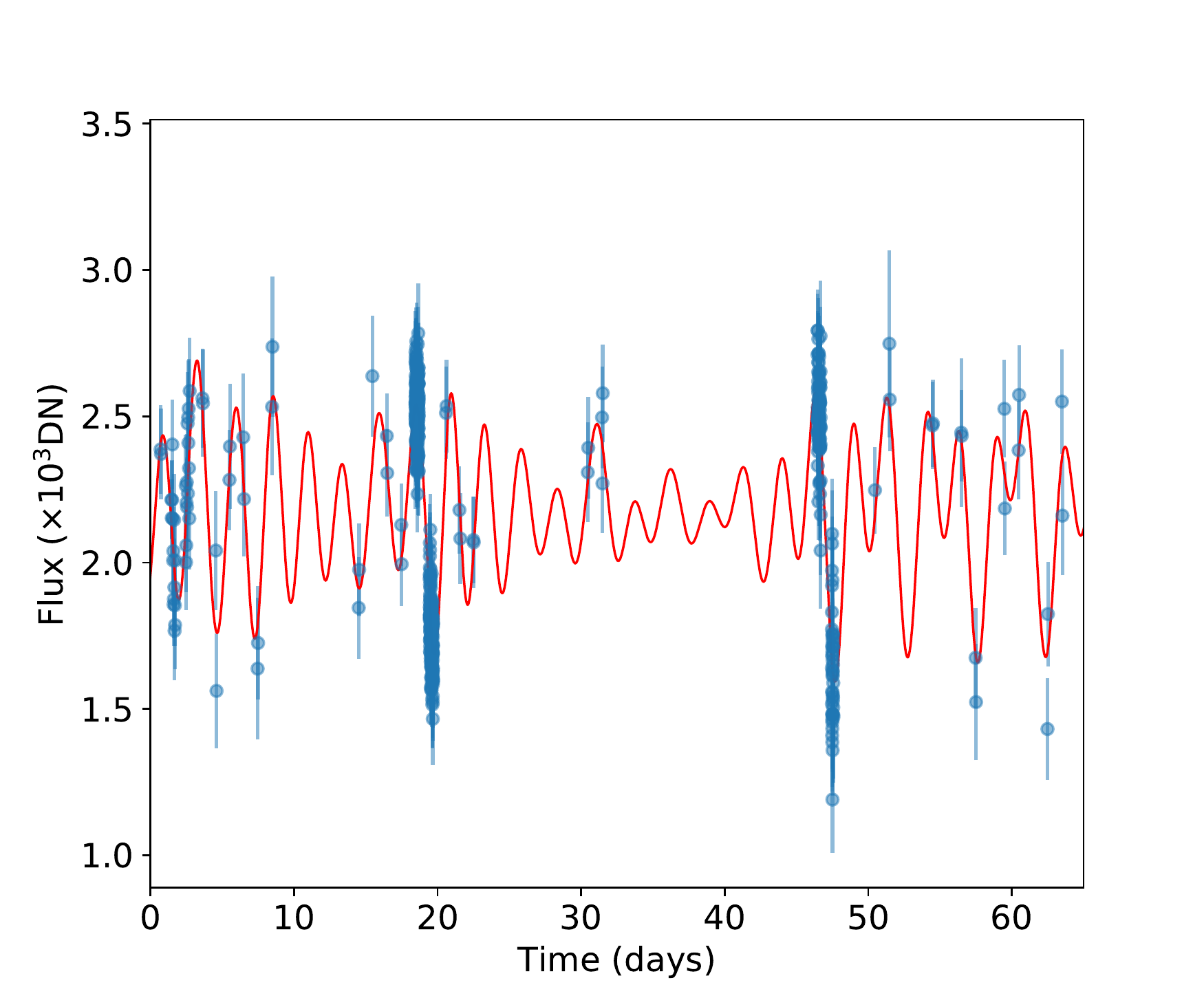}
    \includegraphics[width=88mm]{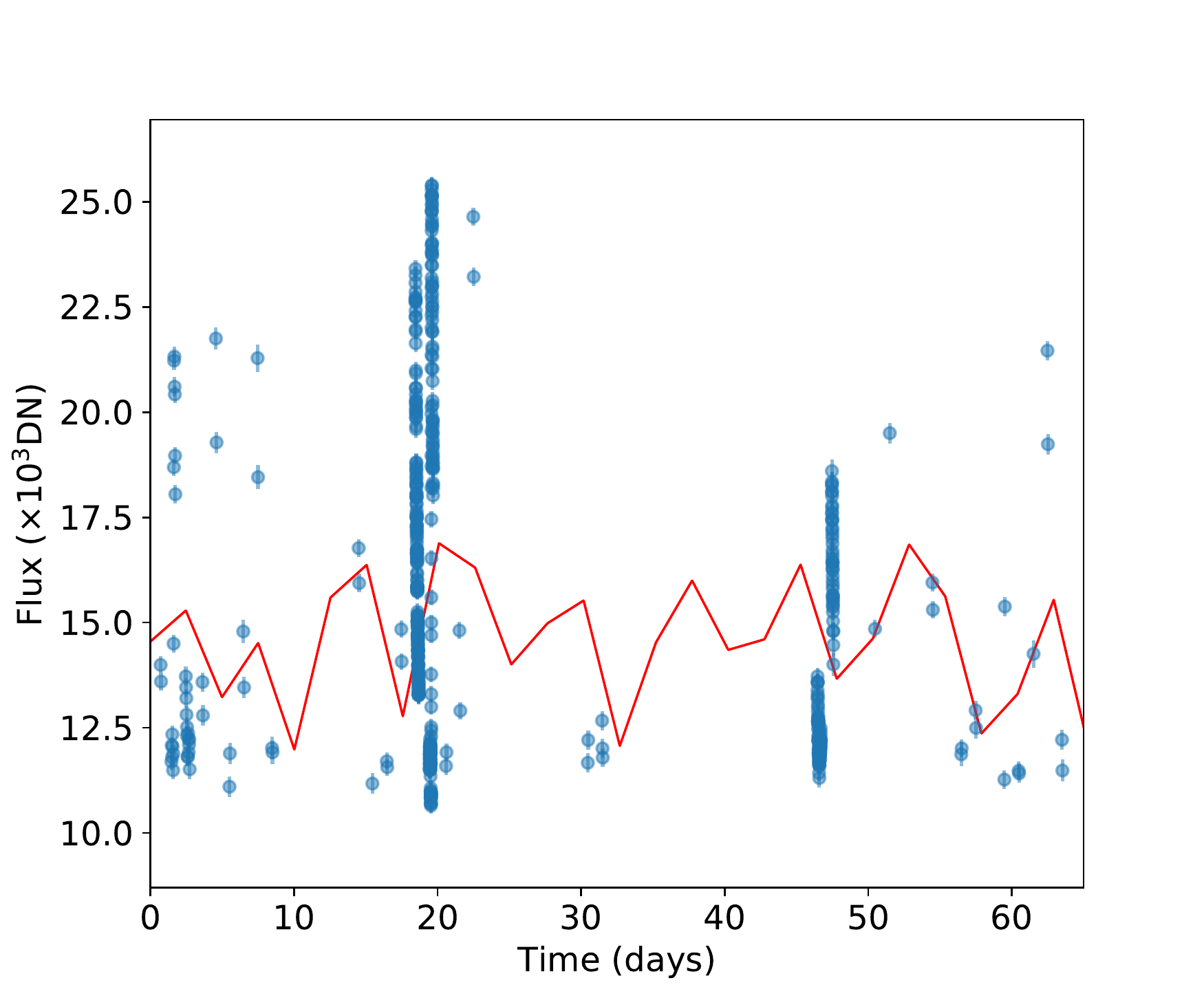}\hfill\includegraphics[width=88mm]{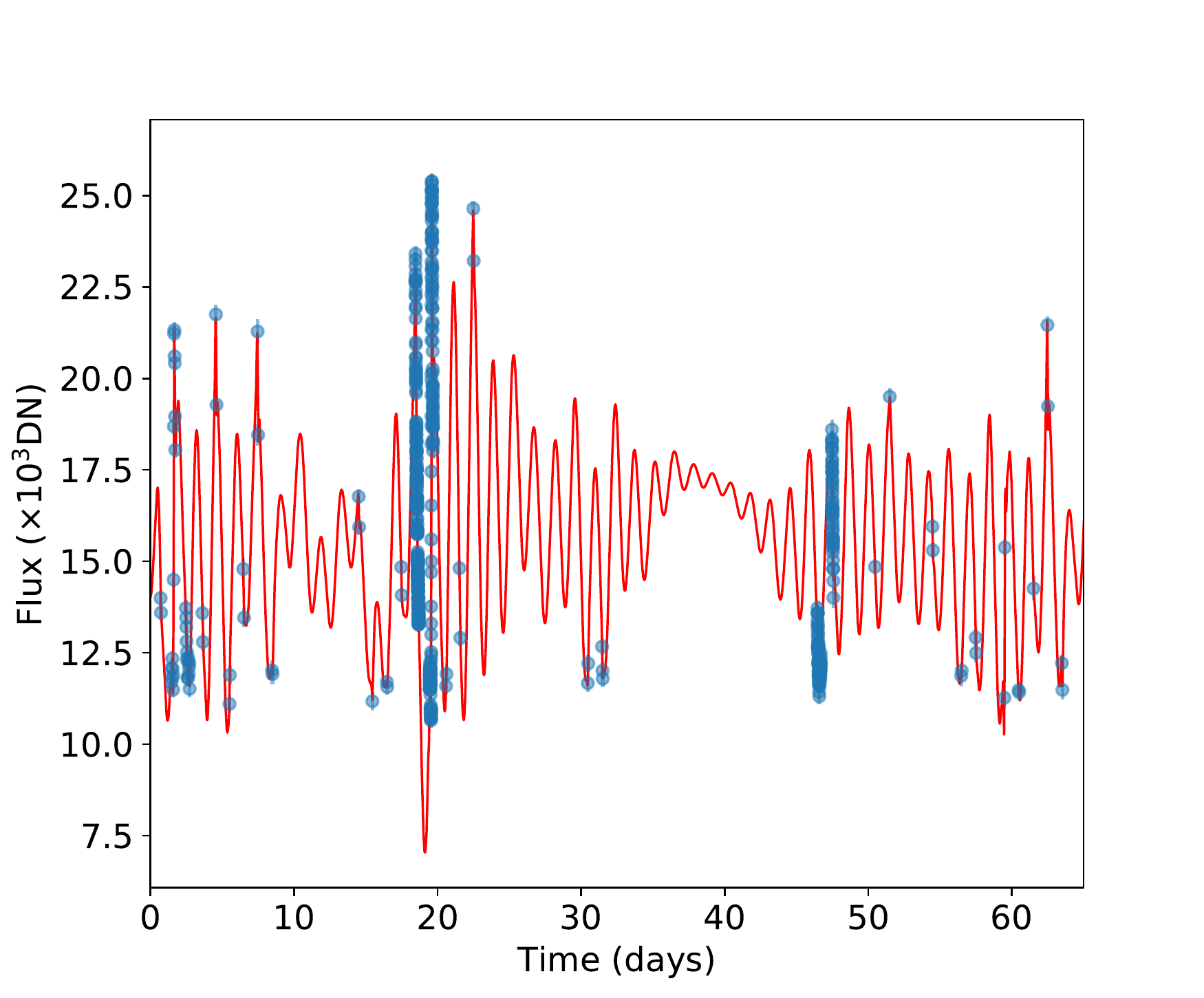}
    \caption{iPTF light curves for the two stars---J004026.84+403504.6 ({\it{top}}) and J004509.86+413031.5 ({\it{bottom}})---vetted out as discussed in Sect.~\ref{sec:time}. Their wavelet transforms are shown in Fig.~\ref{fig:small_tch} of Appendix~\ref{append}. The reconstructed signals, shown in red, in their respective left panels are generated with a maximum resolution of 3~days, while the ones in the right panels are for a resolution of around 30 minutes constructed using the high-cadence block (see text). The time axes in all panels are shown with respect to a reference MJD of 56250.621783.}
    \label{fig:high_cad}
\end{figure*}

\begin{figure}
    \centering
    \includegraphics[width=88mm]{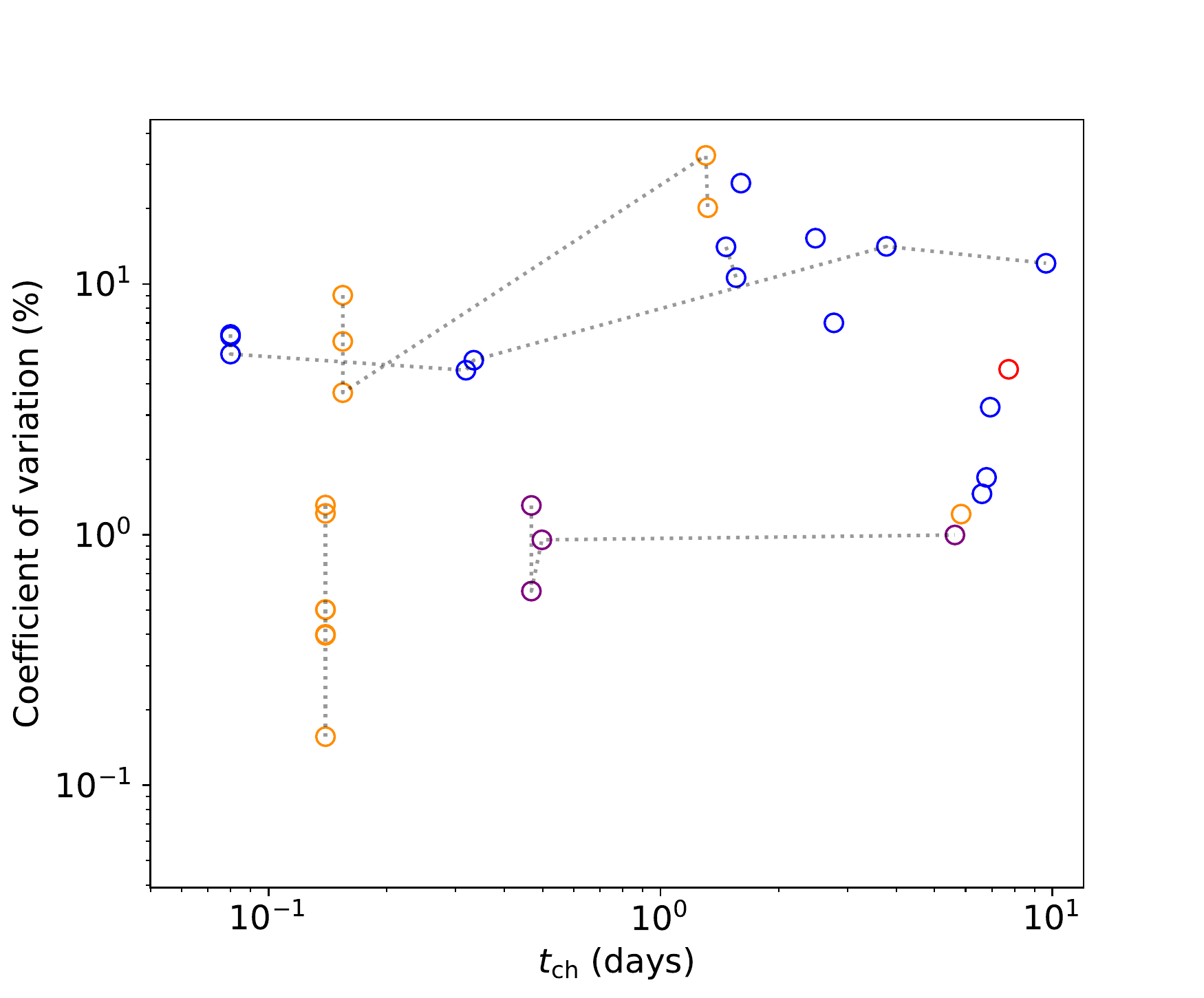}
    \caption{Coefficient of variation, i.e., the ratio of the localized RMS amplitude to the average flux value of the light curve, as a function of the extracted $t_{\rm ch}$ for the 13 stars having significant variations in the high-cadence blocks. The data points from the same star are connected by a dotted line. The O- and B-type stars are shown in blue, A-type supergiants in purple, yellow supergiants in orange, and M-type supergiant in red.}
    \label{fig:small_tch_cv}
\end{figure}

\begin{figure*}
\includegraphics[width=88mm]{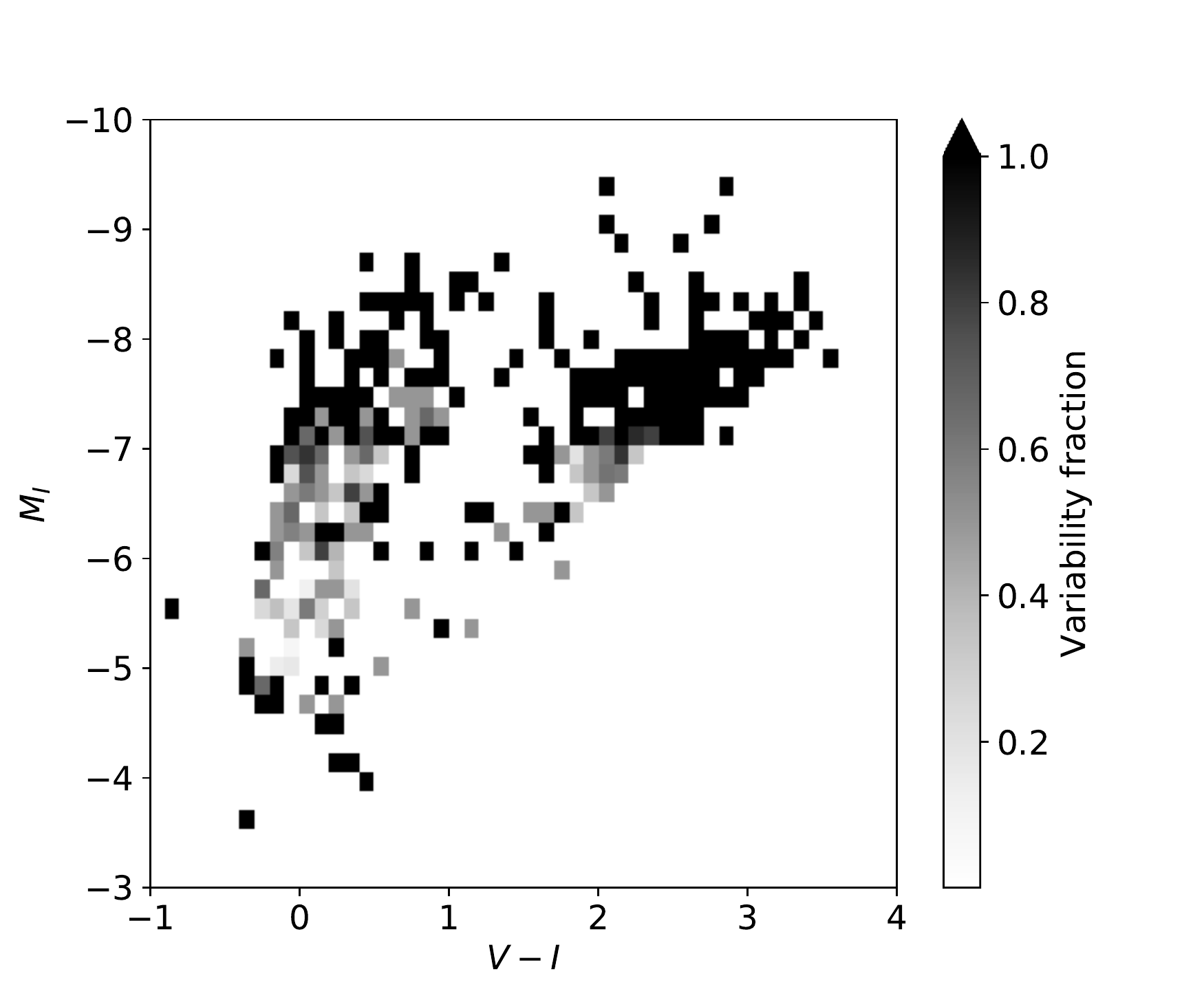}\hfill\includegraphics[width=88mm]{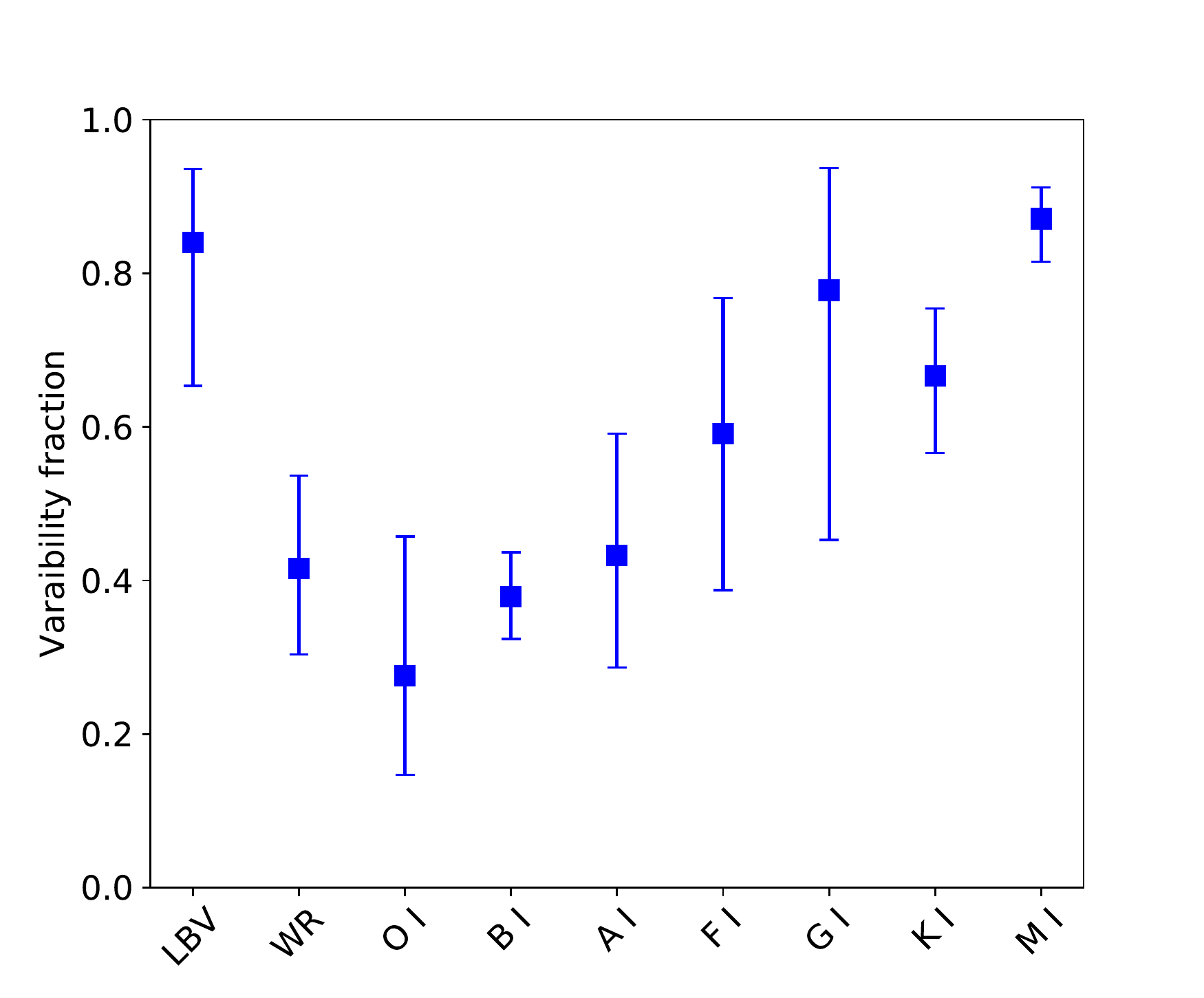}
\caption{CMD ({\it left}) showing the fraction of massive stars in M31 identified by MNS16 with observed variability on timescales $\geq10$~days from iPTF, and as a function of spectral types for the supergiants ({\it right}). The vertical error bars in the right panel indicate the 95\% binomial confidence interval.
}
\label{fig:conroy}
\end{figure*}

We extract the characteristic timescales $t_{\rm ch}$ of the stars from the correlation-power maps, by identifying connected regions of power above a threshold, which we define as $5\sigma$ above the background fluctuations (the background is obtained by taking the mean of the pixel values lower than the 75th percentile of the distribution and $\sigma$ is their standard deviation). For each of the resulting connected regions, we identify the wavelet scale outside the cone of influence{\footnote{Cone of influence is the region in the wavelet transform map affected by edge effects given that we are dealing with finite time series.}} corresponding to the pixel with the maximum power. We term this scale $a$ as $t_{\rm ch}$. If a star's wavelet transform map has no region above its corresponding threshold, then no timescales will be reported for it, whereas if there are multiple islands of excess power above the background, then we will obtain multiple timescales. Because of this automated detection of $t_{\rm ch}$, in some cases, an apparently contiguous region of excess wavelet transform power could be broken up into multiple regions.

We further filter the extracted values of $t_{\rm ch}$ on the basis of their corresponding {\it amplitudes}. We define the amplitude for a characteristic timescale $t_{\rm ch}$ of the given light curve, hereafter termed as $t_{\rm ch}$-specific amplitude, as $\alpha\,P^{1/2}$, where $P$ is the wavelet-transform correlation power evaluated at its local maximum. The constant $\alpha$ is chosen such that if the signal is itself a wavelet with timescale $t_{\rm ch}$, this amplitude matches the RMS of the signal within $\pm t_{\rm ch}$ of its center (see Appendix~\ref{append:wt_sim}). The $t_{\rm ch}$-specific amplitudes are then compared with the expected noise at similar timescales. The timescale-specific noise values are estimated by using the light curves of the static/non-varying stars from Paper~I and smoothing out the noise contributions from smaller timescales. At a given timescale $t$ we smooth the light curves by binning them using bin widths of size $t/2$ and compute the RMS deviations ($\Delta {\rm flux}$) of the smoothed light curves, assuming uncorrelated noise. Finally, we take the median of the $\Delta {\rm flux}$ values of the static stars to define the noise at timescale $t$. These noise estimates are shown as a function of the timescales they are applicable to in Fig.~\ref{fig:timescale_noise}. We compare the amplitude obtained for a $t_{\rm ch}$ to the corresponding noise and drop those $t_{\rm ch}$ values with associated variability consistent with noise (i.e., amplitude below the red line in Fig.~\ref{fig:timescale_noise}).

The maximum resolution of the reconstructed light curve is $\approx 3$~days for the long baseline that we are considering. Increasing the resolution is not possible because of computational limitations. Hence, we only consider timescales recovered from the maps that are larger than 10~days and therefore well sampled at our resolution. 
To ascertain that the reconstructions represent the data reasonably well, we visually examine all reconstructed light curves for which $t_{\rm ch}$ was obtained. In two cases, the star consistently shows large amplitude fluctuations on a timescale less than our maximum resolution of 3~days.     
One of these two stars, J004509.86+413031.5, is labeled as a Yellow Supergiant candidate (YSG:) in the MNS16 catalog and 
the other, J004026.84+403504.6, as a composite spectral type O7+O9f: without a luminosity class (`:' denotes candidate in MNS16). Their light curves are shown in Fig.~\ref{fig:high_cad}, where the data taken on a few nights with an ultra-high cadence of around 2~minutes highlight the insufficient modelling resolution. 
For a uniform analysis of all stars, we drop the two sources J004026.84+403504.6 and J004509.86+413031.5  when evaluating for $t_{\rm ch}\geq10$~days, thereby decreasing our sample size to 489 (however, see the following section for determining $t_{\rm ch}<10$~days). Of these, we determine $t_{\rm ch}\geq10$~days for 356 stars. Typically there are two timescales from a single star for those with any timescales. A tally of the stars is given in Table~\ref{tab:summy}.

\begin{table*}[ht]
\caption{Summary of our reference sample of massive stars in M31}\label{tab:summy}
\renewcommand\arraystretch{1.0}
\renewcommand{\tabcolsep}{12pt}
\centering
\begin{tabularx}{0.6\textwidth}{lr}
\hline
Total spectroscopic sample 	&1050\\
In the PTF footprint 			&1015\\
Brighter than $m_{R}=20$ and with PTF-detectable variability 		&502\\
Long-timescale wavelet analysis performed			&489\\
Significant $t_{\rm ch}\geq10$~days detected			&356\\
Short-timescale wavelet analysis performed			&182\\
Significant $t_{\rm ch}<10$~days detected 			&13\\
\hline
\end{tabularx}
\end{table*}

\subsection{High-cadence data} \label{sec:short_tch}
In light of the few high-cadence nights included in the PTF data set, we repeated the $t_{\rm ch}$ analysis considering only around 60~nights as the baseline, straddling all the high-cadence ones, and increasing the maximum resolution in the light curve reconstruction to $\approx 30$~minutes. We term the data from this shortened baseline as the high-cadence block.

For the majority of the stars (around 63\% of the 491 sources), the fluctuations on this short baseline are hidden under the noise (orange line in Fig.~\ref{fig:filter}). For those above the noise floor (182 stars), we perform wavelet transformation of their reconstructed light curves from the high-cadence block. This set of 182 stars includes J004026.84+403504.6 and J004509.86+413031.5 discussed above, and we obtain reasonable reconstructions using their high-cadence blocks, as shown in the right panels of Fig.~\ref{fig:high_cad}.

We extract $t_{\rm ch}$ from the wavelet-correlation-power maps obtained using the high-cadence block signal, in a similar manner as in Sect.~\ref{sec:time}. As can be seen from  Fig.~\ref{fig:timescale_noise}, the noise estimates for the short $t_{\rm ch}$ values from the high-cadence block are approximately constant ($\approx 100$~DN); the noise estimate is biased for the long smoothing timescale similar to the baseline considered (60~days), and hence there is a sharp drop around those timescales ($t_{\rm ch}$ around 10~days). We use a constant threshold of 100~DN to filter short $t_{\rm ch}$ values with corresponding amplitudes consistent with noise. The results are described in the next section.

\section{Results} \label{sec:result}

\subsection{Short characteristic timescales from the high-cadence block} \label{sec:short_tch}

We find significant variations with $t_{\rm ch}<10$~days for 13 stars, two of which are J004026.84+403504.6 and J004509.86+4130\-31.5. The reconstructed light curves along with the wavelet-correlation-power maps for the 13 stars are shown in Fig.~\ref{fig:small_tch} of Appendix~\ref{append}.

For J004026.84+403504.6, we obtain $t_{\rm ch}$ from the high-cadence block of around 2.5~days, while for the YSG candidate J004509.86+413031.5, $t_{\rm ch}$ around 0.15~days and 1.3~days (see Fig.~\ref{fig:small_tch}). As can be seen, the light curve of this candidate YSG is quite regular and its $t_{\rm ch}$ values are consistent with periods of Cepheids, hence it is likely a Cepheid in M31 and its variability can be attributed to pulsation.

\citet{Wallenstein-2019} recently studied the variability of seven massive stars, comprising YSGs and LBVs, in the LMC based on {\it TESS} data, which continuously monitored the stars with a 2-minute cadence for $\approx 27$~days. They find short-timescale ($<10$~days) variability, from analysis of periodograms (hence periodic timescales), for five of the stars with amplitudes $<1\%$ albeit without constraining the nature of variability. In agreement with their result, we also find existence of short-timescale variability in evolved stars in M31.  
For the larger sample size of stars (182) and their data with two times longer baseline we have analyzed, we find significant short $t_{\rm ch}$ values (either periodic and/or non-periodic) for $\approx 7\%$ of the stars in our sample. The spectral types of the remaining 11 stars consist of supergiants of O (including a main-sequence), A, B, M-types and yellow supergiants (Fig.~\ref{fig:small_tch}).

In Fig.~\ref{fig:small_tch_cv}, we show the amplitudes as a function of $t_{\rm ch}$ for the 13 stars. 
For these stars, the variability amplitude at $t_{\rm ch}$ values $\lesssim 1$~day is $\lesssim 10\%$, while for $t_{\rm ch}$ between 1 and 10~days, the amplitude ranges from 1\% to a ${\rm few}\times10\%$. Given that the ground-based PTF data have much larger noise than the {\it TESS} data, we can observe low levels of variability ($<1\%$) only for the brightest stars. On the other hand, we find high levels of short-timescale variability ($>1\%\mbox{--}10\%$) only for $<7\%$ of the stars in our much larger sample. For the small number of stars \citet{Wallenstein-2019} analyzed, it is thus not surprising that they did not find such variability.

Higher-cadence data (and thus with reduced noise) over an extended baseline for a large sample of stars will be ideal for probing short(er) timescales in these stars, even for the fainter ones.

\begin{figure*}[t]
\includegraphics[width=88mm]{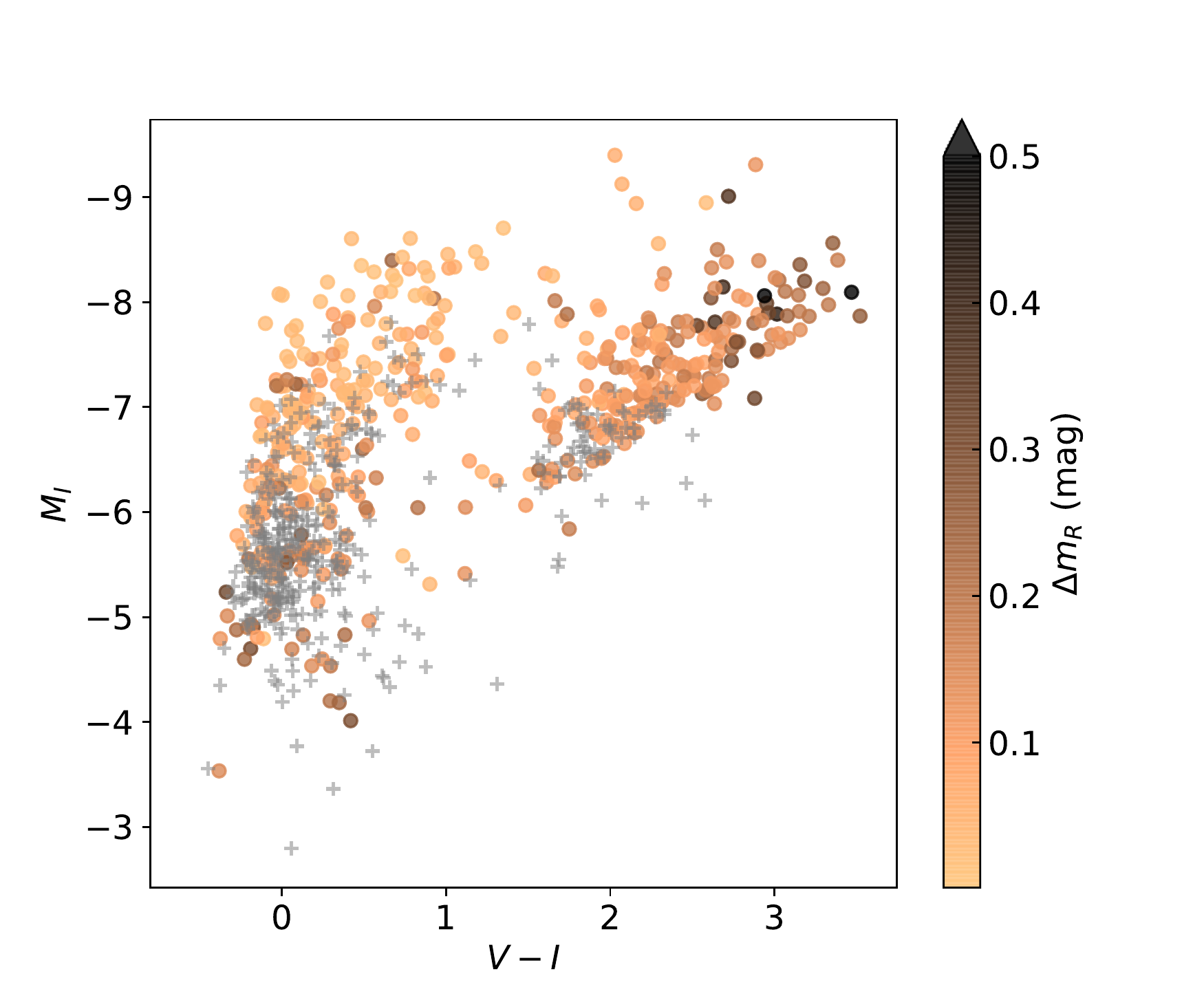}\hfill\includegraphics[width=88mm]{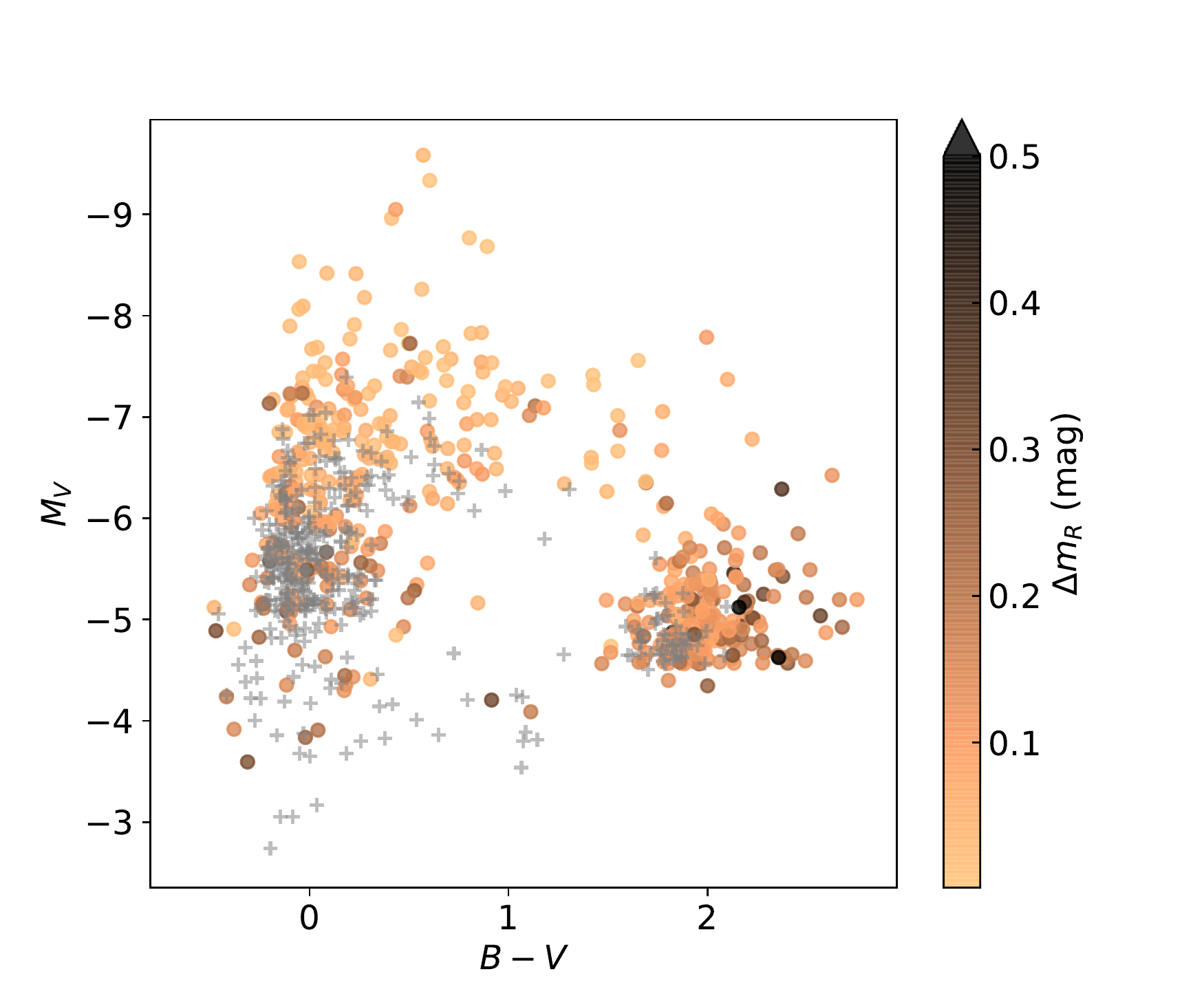}
\caption{CMDs based on the LGGS photometry of the massive stars in M31 from MNS16, color-coded by the observed variability amplitude from iPTF, expressed as the RMS deviation $\Delta{m_{R}}$ from the mean of the light curve. Grey plus symbols signify objects that did not have detectable variability in our data. 
}
\label{fig:vars}
\end{figure*}

\subsection{Results based on the long baseline light curve data}
The variability characteristics we have determined for the massive stars in M31 are listed in Table~\ref{tab:all}. We also list in this table all $t_{\rm ch}>10$~days extracted automatically for a given star, with the caveat mentioned in Sect.~\ref{sec:time} that some of these $t_{\rm ch}$ values reported for a star, in particular, when the values are close to one another, may result from segmentation of a single connected region in our automated analysis.

Using the available multiband photometry from LGGS for the massive stars in M31 studied here, coupled with the variability information we have extracted from their time-domain iPTF data (Table~\ref{tab:all}), we now map out their variability characteristics in the CMD. In particular, we consider the $B$, $V$, and $I$ photometric measurements from LGGS. However, it is to be noted that the LGGS photometries do not represent long-baseline averaged values. Nevertheless, these LGGS-CMD-based variability maps still provide information for the population as a whole because any variable photometry will average out when looking at the whole population.

\subsubsection{Map of variability fraction in the upper CMD of M31}\label{sec:map_varfrac}

We show the variability fraction in the $V-I$ vs.\ $M_{I}$ CMD in the left panel of Fig.~\ref{fig:conroy}, which can be directly compared with Fig.~13 of \citet{Conroy-2018}, and also as a function of the spectral types of supergiants extending from O- to M-type, and including LBVs and Wolf-Rayets, in the right panel of the same figure. In computing both the colors and absolute magnitudes, we have corrected for the foreground extinction using the reddening along the line of sight to M31, $E(B-V)=0.062$ from \citet{Schlegel-1998}; we do not account for the interstellar extinction intrinsic to M31. We use a distance modulus of 24.36 for M31 \citep{Vilardell-2010}. 
Note that the selection effects of our parent data sets propagate into our study---the spectroscopic catalog of MNS16 is likely complete only for the Wolf-Rayets, and the sensitivity of the time-domain PTF data to variability drops for the fainter stars to $>0.1$~mag (Fig.~\ref{fig:filter}). This caveat applies to all results presented in the following.

\citet{Conroy-2018} covered a much larger parameter space in the CMD, particularly the evolved states of late-type main-sequence stars including AGBs. These are not represented in our study since the MNS16 work (and the companion work on RSGs; \citealt{Massey-2016b}) was primarily directed toward massive stars in regions where contamination from the foreground is minimal. This is also the reason for the gap, where there are no stars in our reference sample, between the two groups in the left panel of Fig.~\ref{fig:conroy}. 
Further, the sample of stars used by \citet{Conroy-2018} is complete for the luminous stars, but they probed variability at a level $>0.03$~mag comparable to our study. The very high variability fractions we have found for the brightest stars are not reliant on the increased sensitivity of PTF to photometric variability for such bright stars, since these variations tend to have large amplitudes (see Sect.~\ref{sec:map_amp}, Fig.~\ref{fig:vars}).

For the parts of the CMDs analyzed here, there is good agreement between our results and those of \citet{Conroy-2018}, including the unexplained variability observed in faint blue stars (around $0<V-I<0.5$ and $M_{I}>-8$). Those stars are located between the classical instability strip and the region comprising instability in radiation-dominated envelopes of blue massive stars around the location of iron-opacity peaks \citep{YanFei-2015} deduced by \citet{Conroy-2018} based on a simple parameterization of the instability in 1D evolutionary models. The increase in the variability fraction of these stars toward later types can be clearly seen in the right panel of Fig.~\ref{fig:conroy}, reaching almost 100\% for the M supergiants, in agreement with \citet{Conroy-2018}. LBVs by definition are variables (see Sect.~\ref{sec:lbvs}), hence, their fraction is close to 100\%. We caution, however, that the result in Fig.~\ref{fig:conroy} is subject to incompleteness as described above, particularly for the early type O-stars. But it is also to be noted that variability properties were not part of the selection function for the MNS16 spectral catalog.

\begin{figure*}
\includegraphics[width=88mm]{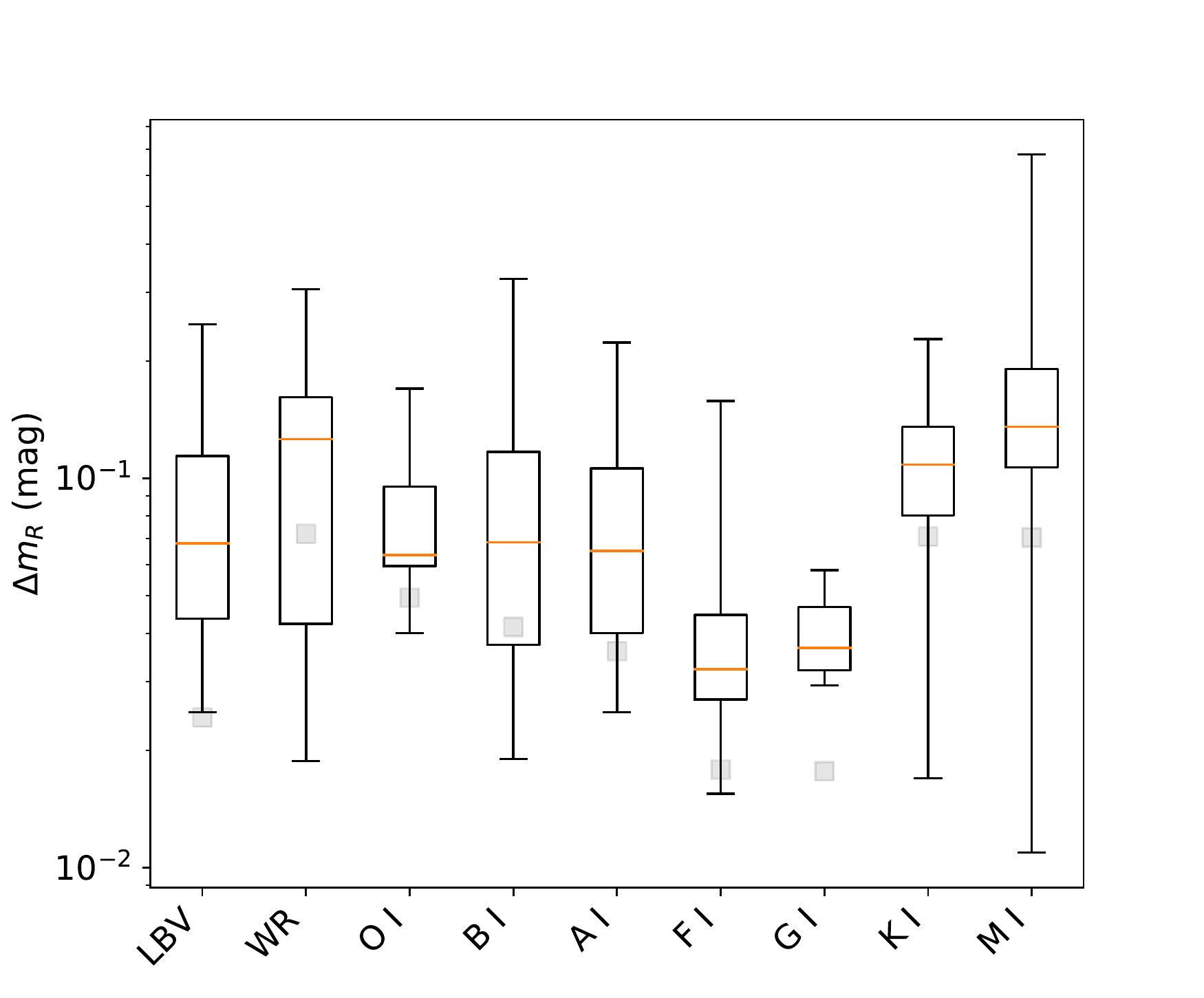}\hfill\includegraphics[width=88mm]{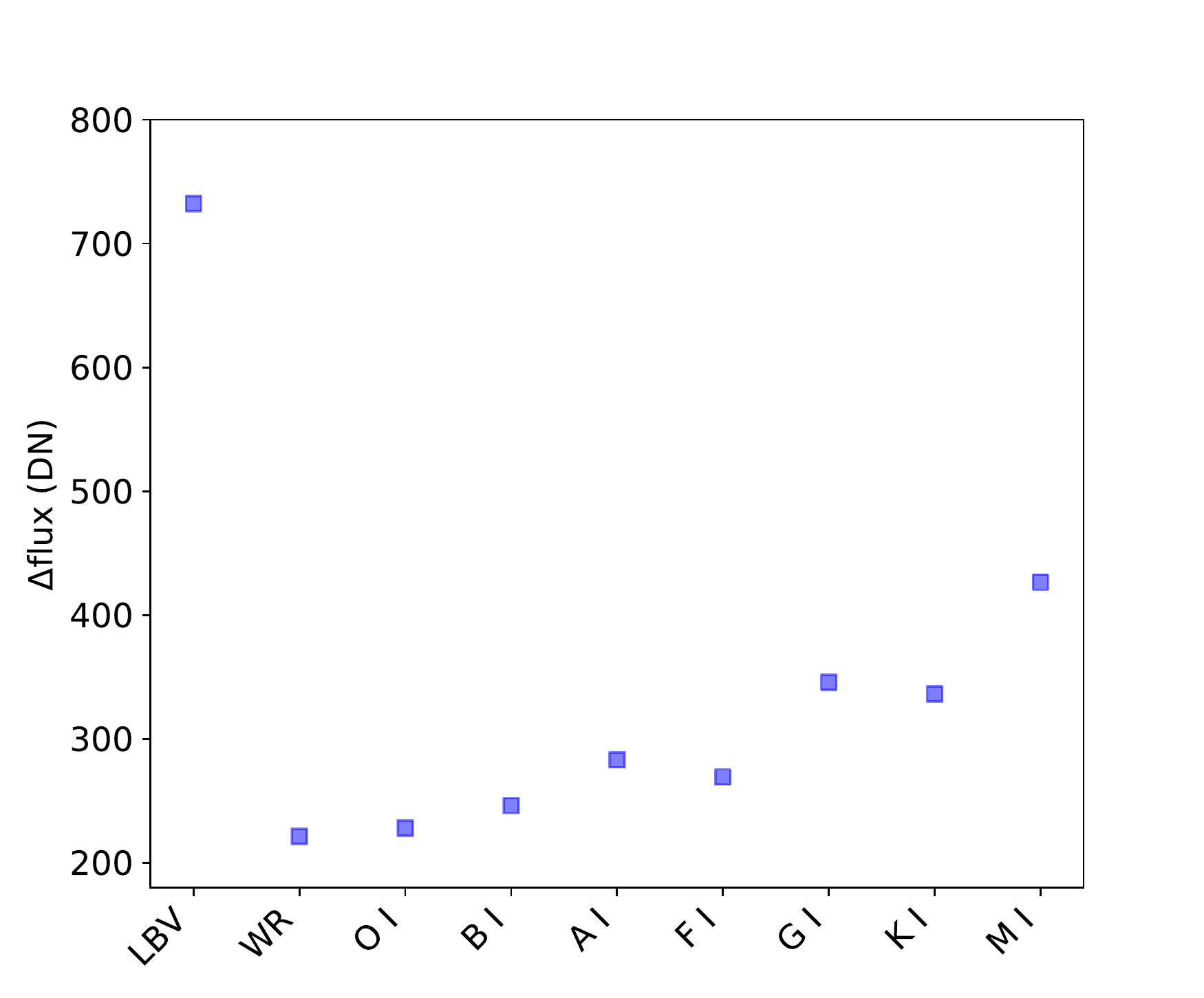}
\caption{{\it Left}: Range of RMS amplitudes for the different spectral types of supergiants in M31. The box extends from the first quartile to the third quartile of the corresponding RMS distribution, and the orange line shows the median while the whiskers extend to the minimum and maximum values of the distribution. The grey squares in the background indicate the typical noise level for the corresponding spectral type, which needs to be considered while comparing the different RMS amplitude distributions. The lower variability tails in the amplitude distributions are formed by the brightest stars in the spectral type. {\it Right}: Average RMS amplitudes in absolute flux units (i.e, the detector unit DN), where the expected noise contribution does not depend on the brightness of the star, as a function of the spectral types of supergiants in M31. We caution that the purpose of this plot is to highlight the relative photometric variability between different spectral classes as opposed to the absolute level of variability of a given class.}
\label{fig:sg_rms}
\end{figure*}

\subsubsection{Map of variability amplitude for massive stars in M31}\label{sec:map_amp}
Figure~\ref{fig:vars} shows the aforementioned variability maps in two CMDs ($M_{I}$ vs.\ $V-I$ and $M_{V}$ vs.\ $B-V$), where the amplitude of fluctuation is expressed as the RMS deviation from the mean of the light curve $\Delta{m_{R}}$. 
As can be seen, the stars without observed variability from iPTF (those below the orange line in Fig.~\ref{fig:filter}; the grey points in the present plot) are largely concentrated at the faint part of the CMD, in particular the blue stars. This is in line with their low-SNR data, and hence we obtain only upper limits on their variability. 

Two groups are clearly evident in this figure--among other things, they reflect the selection cut adopted by MNS16 for their spectroscopic follow-up. The group red-ward of around 1.0 for either color $V-I$ or $B-V$ mainly corresponds to RSGs, while the other group consists of warm (spectral type A, F, G) supergiants and blue stars that include both evolved and non-evolved ones. Our variability maps thus show that the red evolved stars typically exhibit larger variability amplitude than the bluer counterparts on timescales $\geq 10$~days.

Figure~\ref{fig:sg_rms} shows the RMS amplitude distributions as a function of the spectral types of supergiants. Note that all the different timescales of variability of the star contribute to this amplitude. Given that the different supergiant types do not have the same brightness in the optical $R$-band, photometric errors will have an effect on the variability ranges shown (cf.~Fig.~\ref{fig:filter}). The typical noise level for each spectral type is indicated by the grey squares in the background in Fig.~\ref{fig:sg_rms}. Comparison of the RMS amplitude distributions of the different supergiants is difficult without a similar noise level. Nevertheless, as can be seen from the plot, the M supergiants exhibit the largest amplitude photometric variation---the upper tail of the distribution reaches $\Delta m_{R}\approx0.7$. In absolute flux units (right panel of Fig.~\ref{fig:sg_rms}), where the noise floor is constant, the typical variability amplitude for the supergiants can be seen to increase toward later spectral type, barring the LBVs, which have the highest typical amplitude.  

Theoretical predictions of brightness amplitudes for massive stars that can be directly compared with our results are not available in the literature, but our observational results will help to constrain models making such computations in the future.

\subsubsection{Photometric variability of M31 LBVs (and candidates)}\label{sec:lbvs}
LBVs or S~Dor variables are evolved hot stars located in the upper HR diagram with luminosity close to the Eddington limit, generally characterized by the prototype S~Dor-like variability and sometimes giant eruptions like that of $\eta$~Car accompanied by enhanced mass-loss \citep{Hubble-1953, Humphreys-1994}. 
They represent a rare class of variable stars, with less than around 50 of them identified in the Local Group, and their evolution, including the driving mechanism of variability, are not well understood.

Our analysis also allows us to study the photometric variability of the LBVs and candidates in the MNS16 compilation. There are 25 such LBVs in total in their Table 3. Four of them are below the detection threshold of variability for our study (i.e., the orange line in Fig.~\ref{fig:filter}). For the remaining 21 with observed variability, we show all light curves in Fig.~\ref{fig:lbvs} in Appendix~\ref{append}. This includes 17 candidate LBVs. Thus, this study establishes photometric variability for the majority of the LBV candidates in M31 for the first time.

Characterization of observed photometric varaibility in LBVs based on a large sample size has been performed mainly for those in the Galaxy and the Magellanic Clouds, such as the work of \citet{Genderen-2001} using 46 S~Dor variables. The RMS amplitudes of the LBVs in M31 extracted from the nearly five-year-long PTF data are at most a few tenths of a magnitude (cf.~Fig.~\ref{fig:sg_rms}). This is in contrast to the (long) S~Dor- and $\eta$~Car-type variabilities in LBVs that have amplitudes greater than around 1~mag and timescales on the order of decades -- the amplitudes being greater for the latter type of eruptions \citep[e.g.,][]{Genderen-1984, Humphreys-1994, Genderen-1997}. However, the short S~Dor-type variability \citep{Genderen-2001} has timescale on the order of a few years, and the amplitude can be less than 0.5~mag---as exhibited by the so-called weak-active members of the S~Dor variables \citep{Genderen-2001}. Such variability is assumed to be associated with changes in radius and temperature of the star, while the luminosity remains nearly constant. Hence, color information is ideal to ascertain S~Dor-type variability. Various types of instabilities in the stellar envelope have been invoked in different theoretical models as the physical mechanism \citep[e.g.,][]{Stothers-1995, YanFei-2018}, however, detailed long-term predictions that can be directly compared with observations are not yet available.

LBVs also exhibit a third kind of variability, namely microvariations with amplitudes $\lesssim0.1$~mag on timescales of days to even hundreds of days (e.g., \citealt{Genderen-1997, Abolmasov-2011, Mehner-2017}).  Based on a sample of five LBVs in the Galaxy and Magellanic Clouds, \citet{Genderen-1997} grouped the observed microvariations into two types--(1) those with timescales on the order of days to weeks, which show blue colors in the maxima and red in the minima, and (2) longer-timescale ($\sim 100$~days) microvararitions showing the opposite trend in color. Since we do not have color information for the present study, it is not possible to explore the color variation.

\begin{figure*}
\includegraphics[width=88mm]{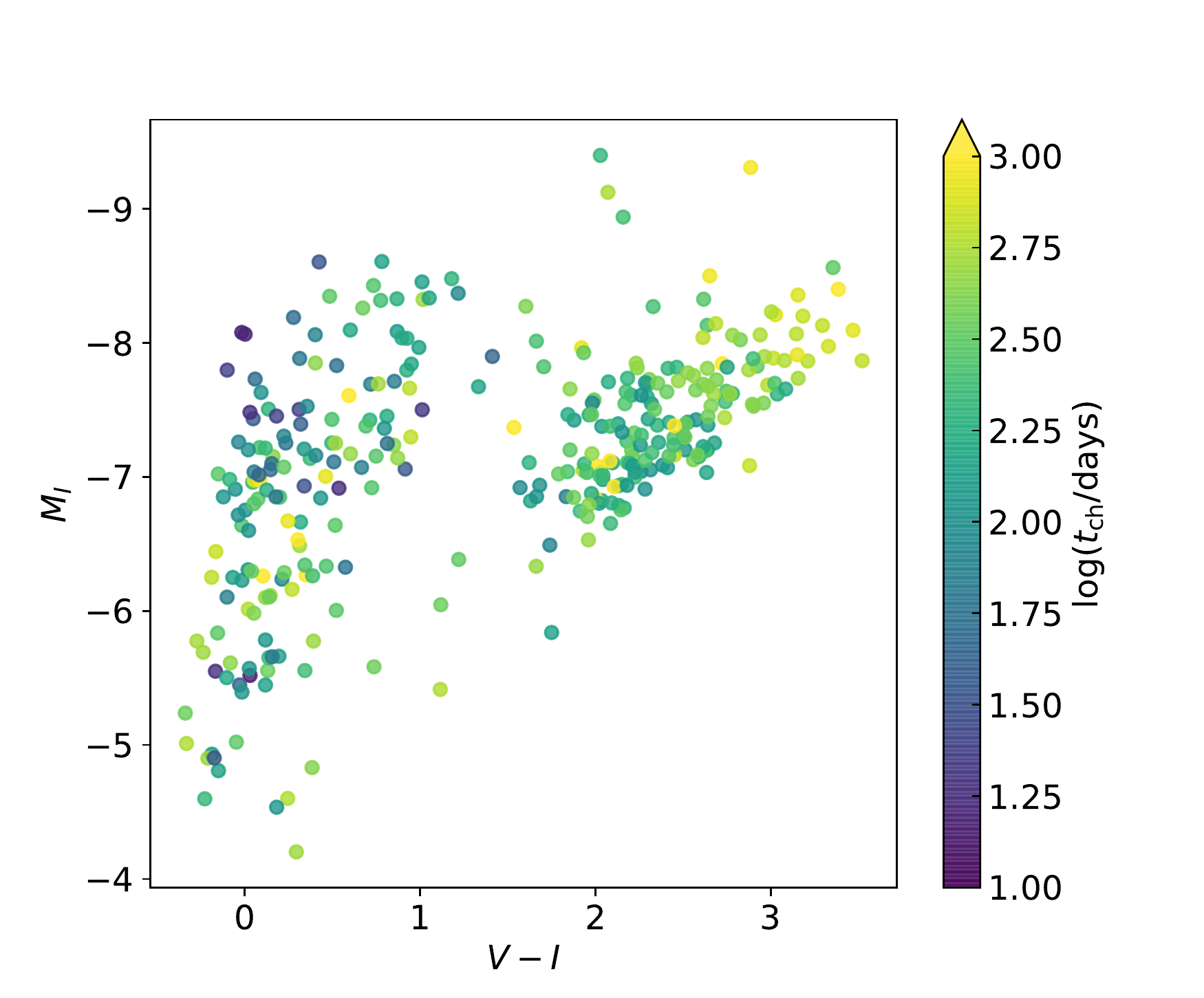}\hfill\includegraphics[width=88mm]{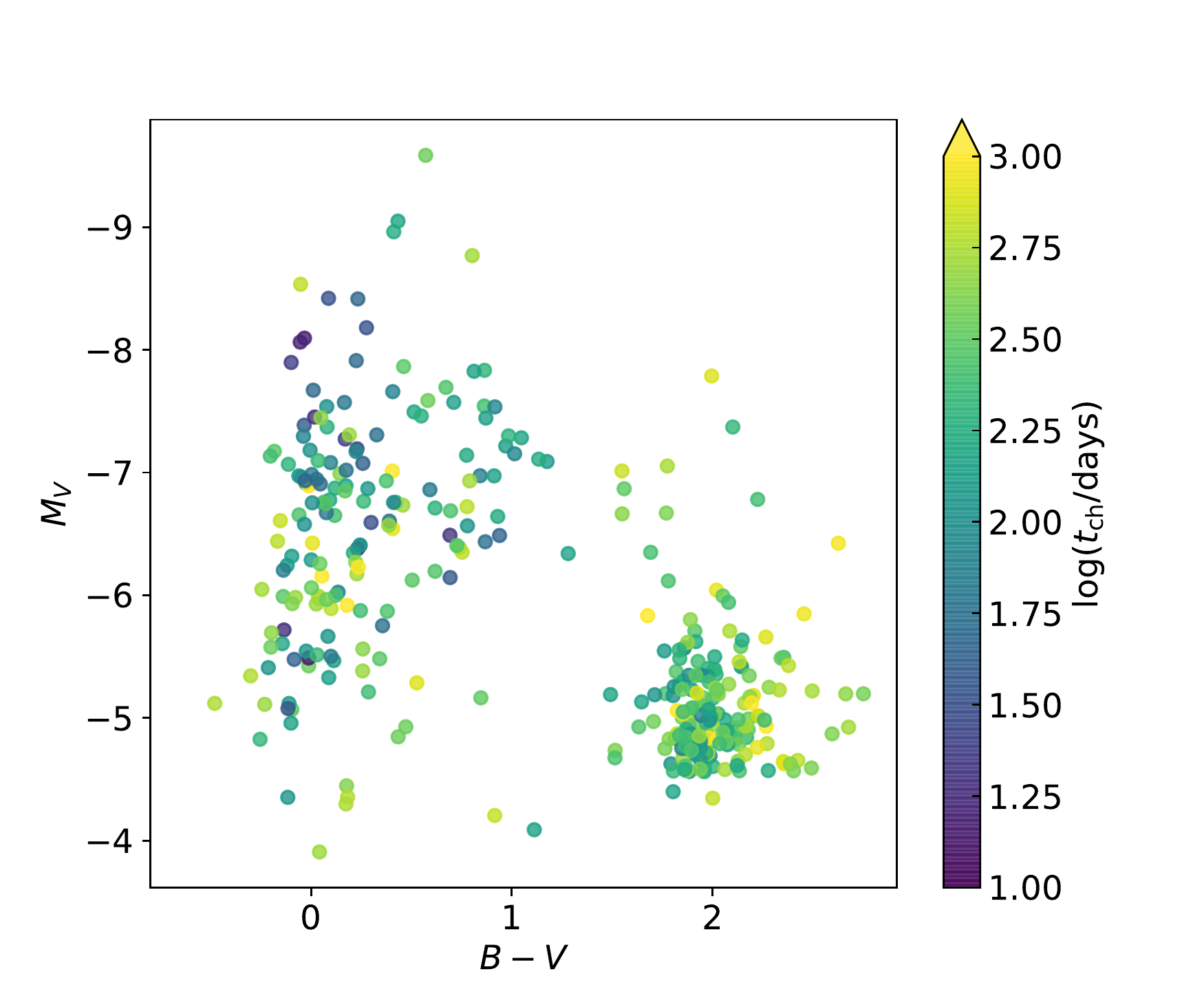}
\caption{CMDs similar to Fig.~\ref{fig:vars}, where the color-coding in this case reflects a single characteristic timescale for each star corresponding to the global maximum power in the star's wavelet transform.}
\label{fig:tch_color}
\end{figure*}

Different authors have proposed different mechanisms to explain these microvariations, including strange mode instabilities \citep{Kiriakidis-1993}, non-radial \textit{g}-mode pulsations \citep{Lamers-1998}, etc. 
Recently, \citet{YanFei-2018} performed 3D hydrodynamical simulations of radiation-dominated envelopes for LBVs and found that convection (owing to the opacity peaks in the envelopes) drives irregular variations in these stars with characteristic timescales of a few days for (10--30)\% variability in luminosity; their simulations covered only less than 1000~hours of the envelope evolution and the variability level includes the contributions from smaller timescales as well. One of the theoretical  light curves computed by Jiang et al. is shown in Fig.~\ref{fig:sims}. Smoothing out the contributions of timescales $\lesssim 10$~days, the variability is at the 5\% level for the longer timescales.

\begin{figure*}
\includegraphics[width=88mm]{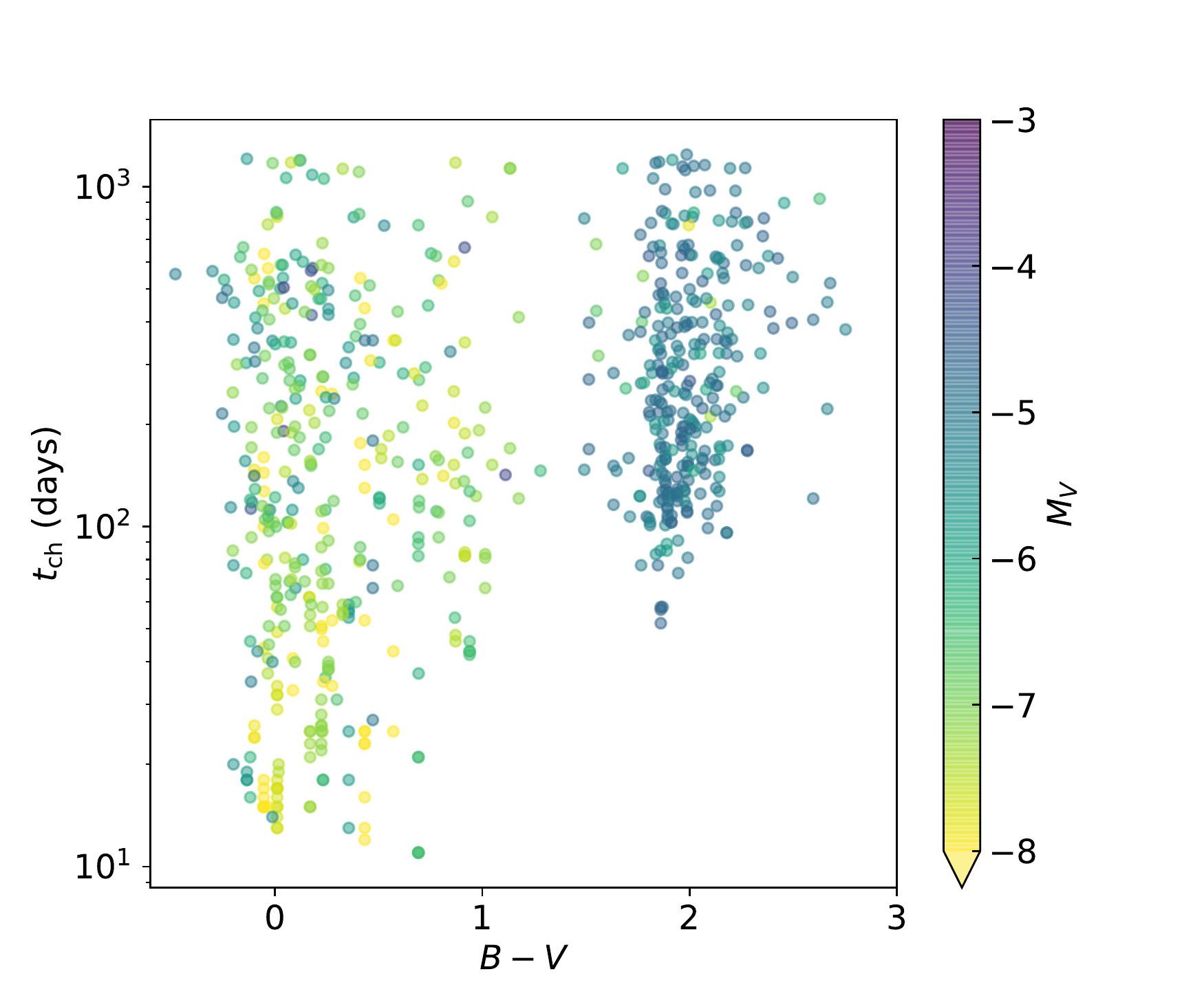}\hfill\includegraphics[width=88mm]{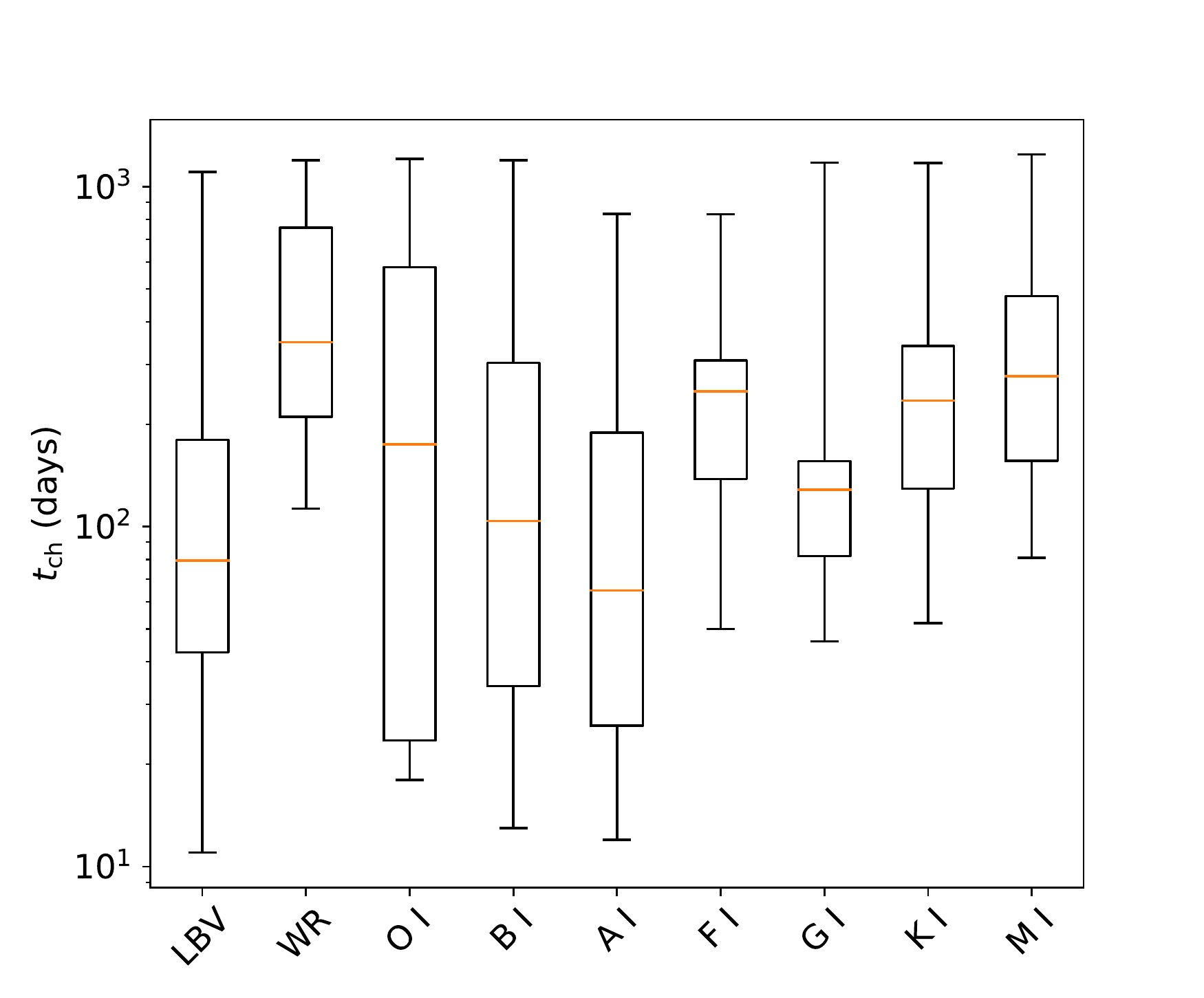}

\caption{{\it Left}: Characteristic timescales $t_{\rm ch}$, obtained from the wavelet transforms of the iPTF light curves (Sect.~\ref{sec:time}), against color obtained from the LGGS photometry of massive stars in M31. It is to be noted that the density of points should not be over-interpreted, as our automated determination of connected regions with high wavelet transform values could in some cases fragment a connected region. {\it Right}: Range of $t_{\rm ch}$ values for different types of supergiants in M31. The box extends from the first quartile to the third quartile values of the corresponding distribution, and the orange line shows the median while the whiskers extend to the minimum and maximum values. 
The upper limit of $t_{\rm ch}$ probed here is $\approx 1200$~days, imposed by the baseline of the iPTF survey in our wavelet transform analysis, while there is a lower limit of $t_{\rm ch}=10$~days dictated by the maximum resolution of our light curve reconstruction.}
\label{fig:tch}
\end{figure*}

The extracted characteristic timescales, $t_{\rm ch}$ for these stars in M31 (Sect.~\ref{sec:map_tch}, Fig.~\ref{fig:tch}) cover a wide range, from tens of days to a thousand days, with the distribution concentrated toward $t_{\rm ch}$ values of a few tens of days. The $t_{\rm ch}$-specific amplitudes for this class of stars are around a few percent as shown in Fig.~\ref{fig:cov}, in agreement with the theoretical results of \citet{YanFei-2018}. This points to the observed variability in these LBVs associated with the majority of the extracted $t_{\rm ch}$ values as likely corresponding to stochastic microvariability.

On the other hand, despite the lack of color information, the longer tail of the $t_{\rm ch}$ distribution (few hundreds to 1000~days) along with the associated low-amplitude variation (Fig.~\ref{fig:cov}) appear to be in line with the variability shown by weak-active S~Dor variables in the Galaxy and Magellanic Clouds of \citet{Genderen-2001}. Moreover, these authors also found their sample of S~Dor variables spend the majority ($\gtrsim 70\%$) of their lifetime in low(-amplitude) state, which appears to be consistent with our results for the M31 sample.

\subsubsection{Map of $t_{\rm ch}$ for massive stars in M31}\label{sec:map_tch}
Similar to the maps we have constructed for the amplitude of variability, we also map out the characteristic timescale $t_{\rm ch}$ in Fig.~\ref{fig:tch_color}. As mentioned in Sect.~\ref{sec:time}, we obtain multiple $t_{\rm ch}$ values for many of the stars. 
In this figure, we show only one timescale per star corresponding to the maximum power in the wavelet transform of its light curve. As is evident from the plot, the group of red massive stars is characterized by longer $t_{\rm ch}$ than the bluer stars; the latter group shows a comparatively large range of timescales. This result is highlighted even further in Fig.~\ref{fig:tch} (left panel), where we show all extracted timescales against the $B-V$ color, and in the right panel of the same figure, where we show the $t_{\rm ch}$ distribution for the different types of supergiants in M31 with observed variability. The lower (around 10~days) and upper limits (around 1200~days) in $t_{\rm ch}$ in these figures are due to constraints set by the maximum resolution of the Gaussian Process modeling of the light curve (Sect.~\ref{sec:time}) and the baseline of the survey, respectively. 

It can be seen from the right-hand panel of Fig.~\ref{fig:tch} that the LBVs and the A-type supergiants have typical $t_{\rm ch}$ values of a few tens of days, shorter than the yellow (F- and G-types) and red supergiants (K- and M-types), which typically have $t_{\rm ch}$ values of a few hundred days. We find that the WRs, O-, and B-type supergiants in our sample have typical $t_{\rm ch}\gtrsim100$~days; however, we caution that the sample incompleteness for these spectral types could affect our results for them (see also Sect.~\ref{sec:discuss}). For almost all the types, the distribution of $t_{\rm ch}$ values extends to the longest timescale ($>1000$~days) probed in our study, while, as discussed above, the distributions for the bluer stars, i.e., LBVs, O-, B-, and A-types, also extend (close) to the shortest timescale probed here.

For the cool massive stars, long pulsation timescales on the order of a few hundred to a few thousand days are predicted \citep[e.g.,][]{Heger-1997}, and this appears consistent with our observed $t_{\rm ch}$ values for these stars. In Paper~I, we extracted periodic timescales from spectroscopically-confirmed RSGs and derived their period-luminosity relation in M31. We found multiple periodic timescales for many of these stars as evidenced by the multiple peaks in their power spectra ranging from a few hundred to a thousand days---the maximum timescale that could be probed in that study (see Paper~I). The earlier results are thus also in agreement with those from the current study (cf.~Fig.~\ref{fig:tch}).

Furthermore, the observed trend in the timescales for the various types of supergiants (right panel of Fig.~\ref{fig:tch}) broadly agrees with the simple predictions made by \citet{Lovy-1984}. Theoretical results from recent, more sophisticated, 3D hydrodynamical simulations of LBVs by \citet{YanFei-2018} are also in-line with the observations, where the $t_{\rm ch}$ values of the LBVs are concentrated at short timescale values (right panel of Fig.~\ref{fig:tch}; Sect.~\ref{sec:lbvs}). It will be interesting to compare future 3D simulations covering a much longer timescale with our observational results.

\begin{figure*}
\includegraphics[width=60mm]{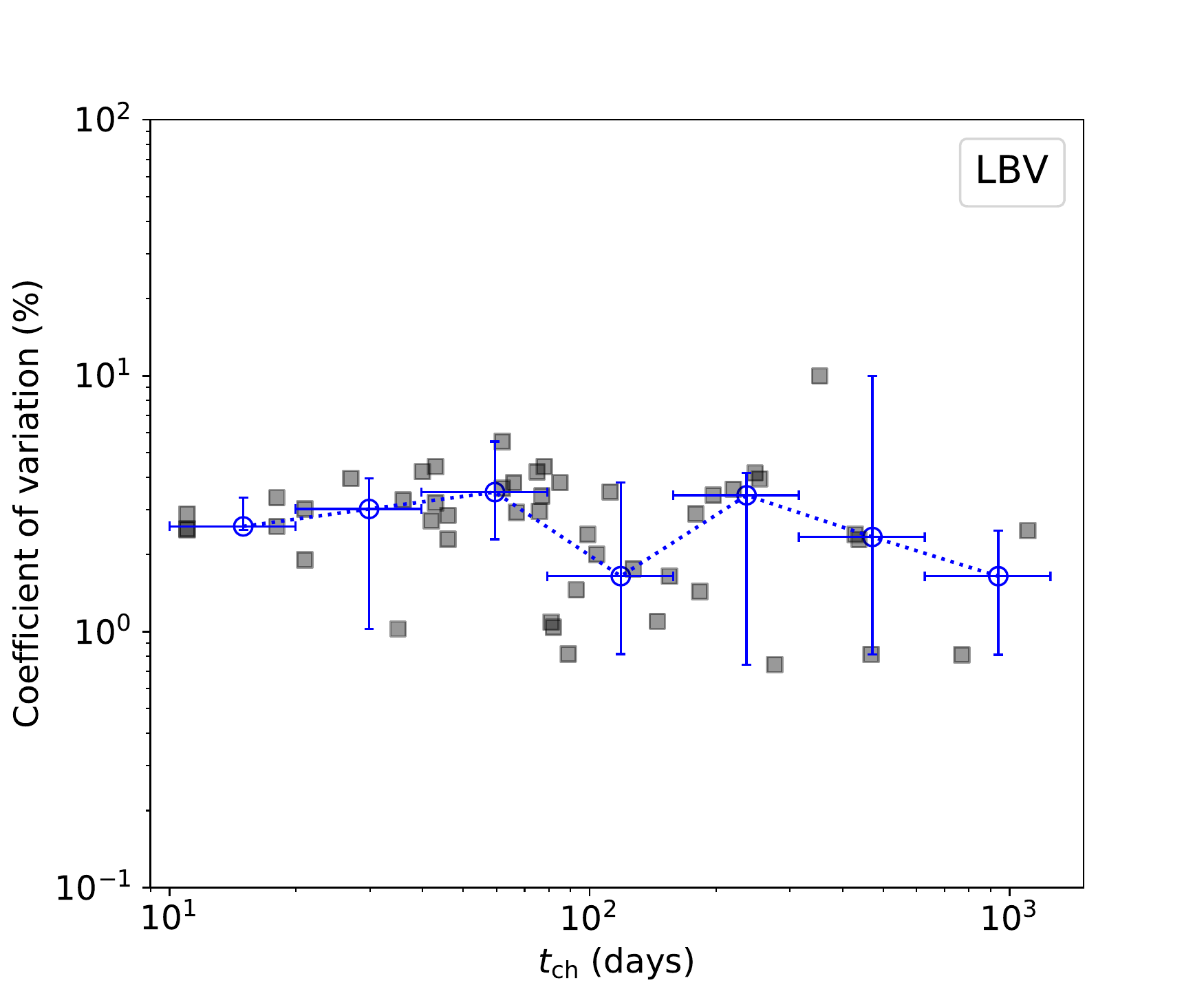}\hfill\includegraphics[width=60mm]{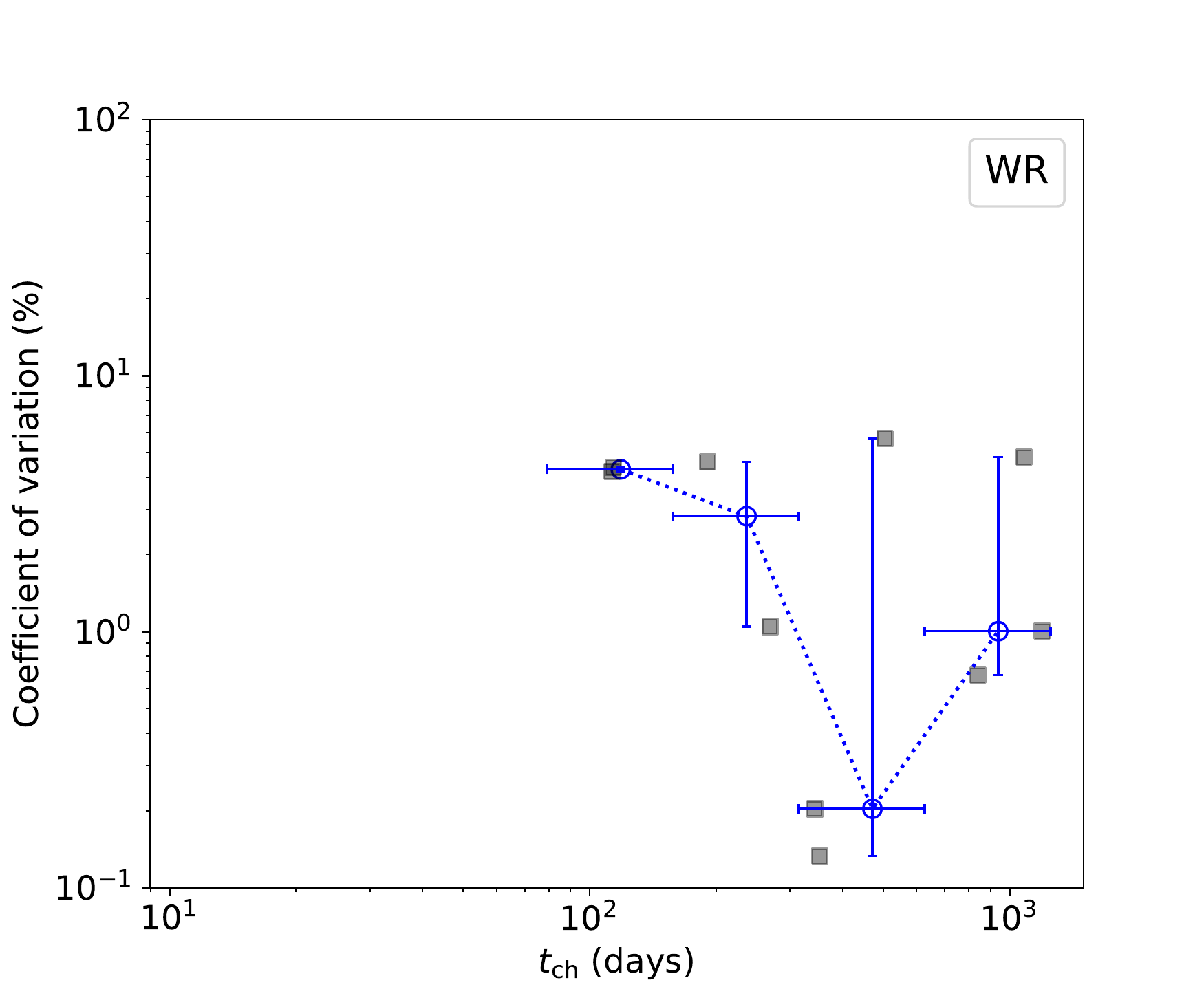}\hfill\includegraphics[width=60mm]{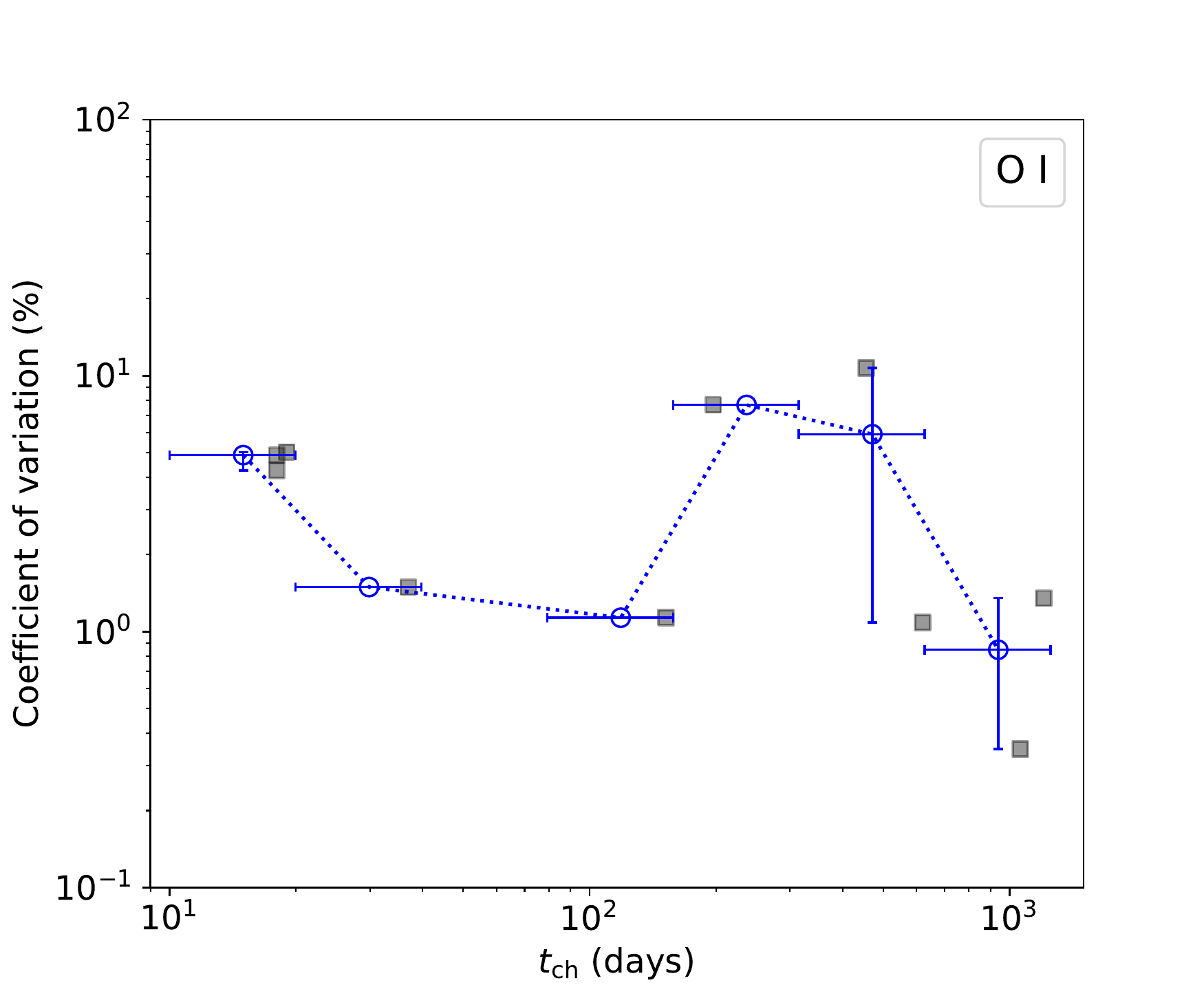}
\includegraphics[width=60mm]{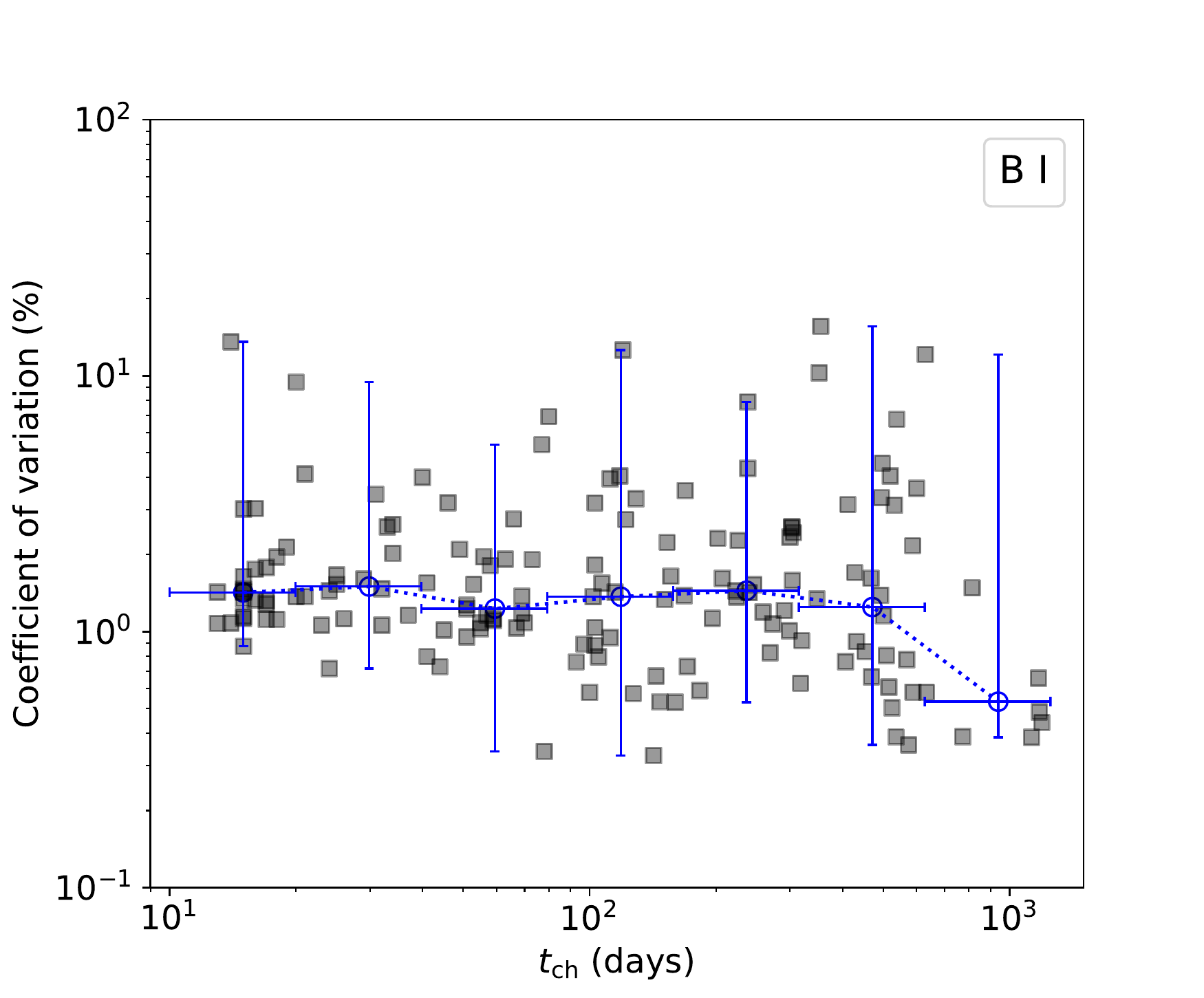}\hfill\includegraphics[width=60mm]{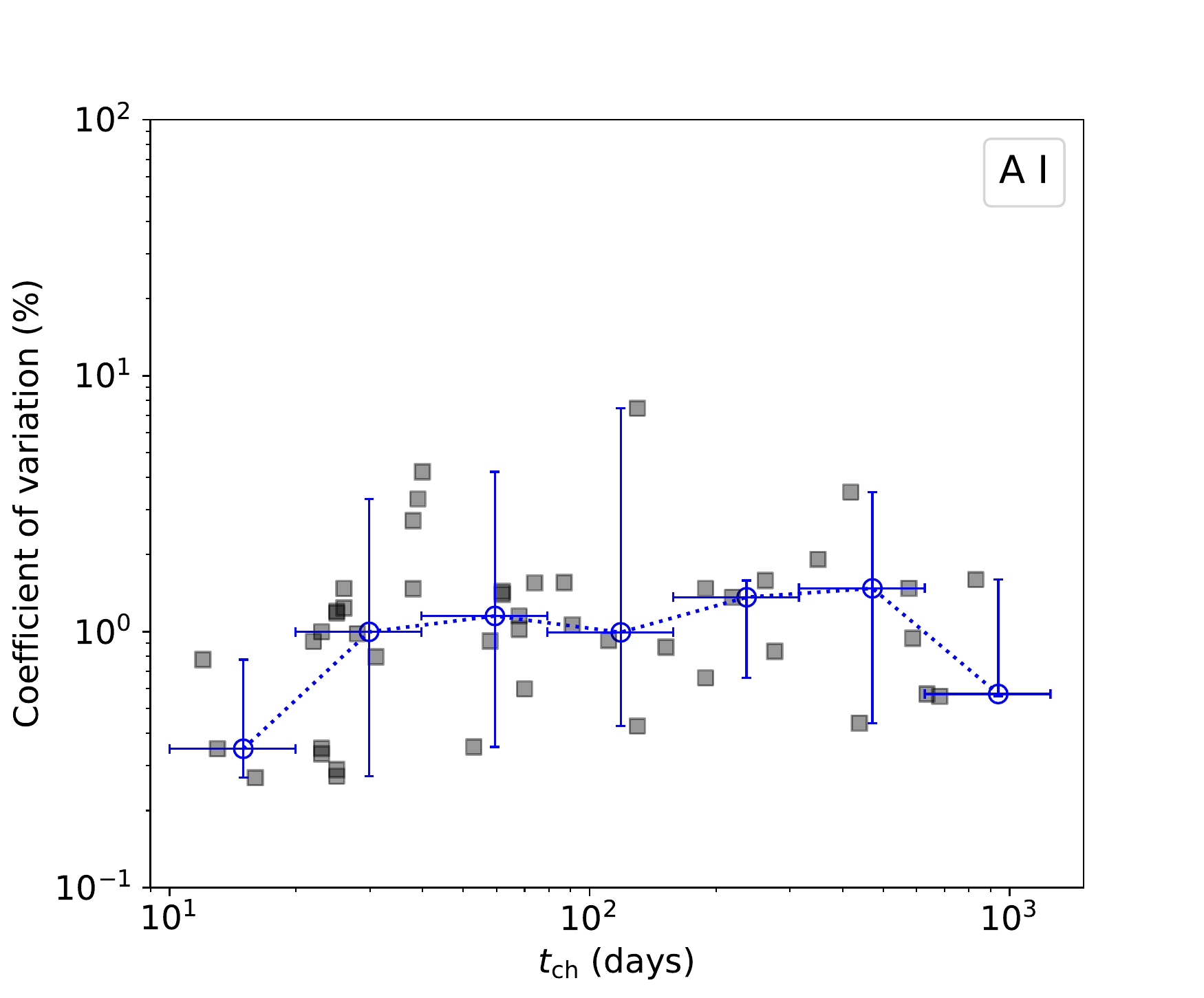}\hfill\includegraphics[width=60mm]{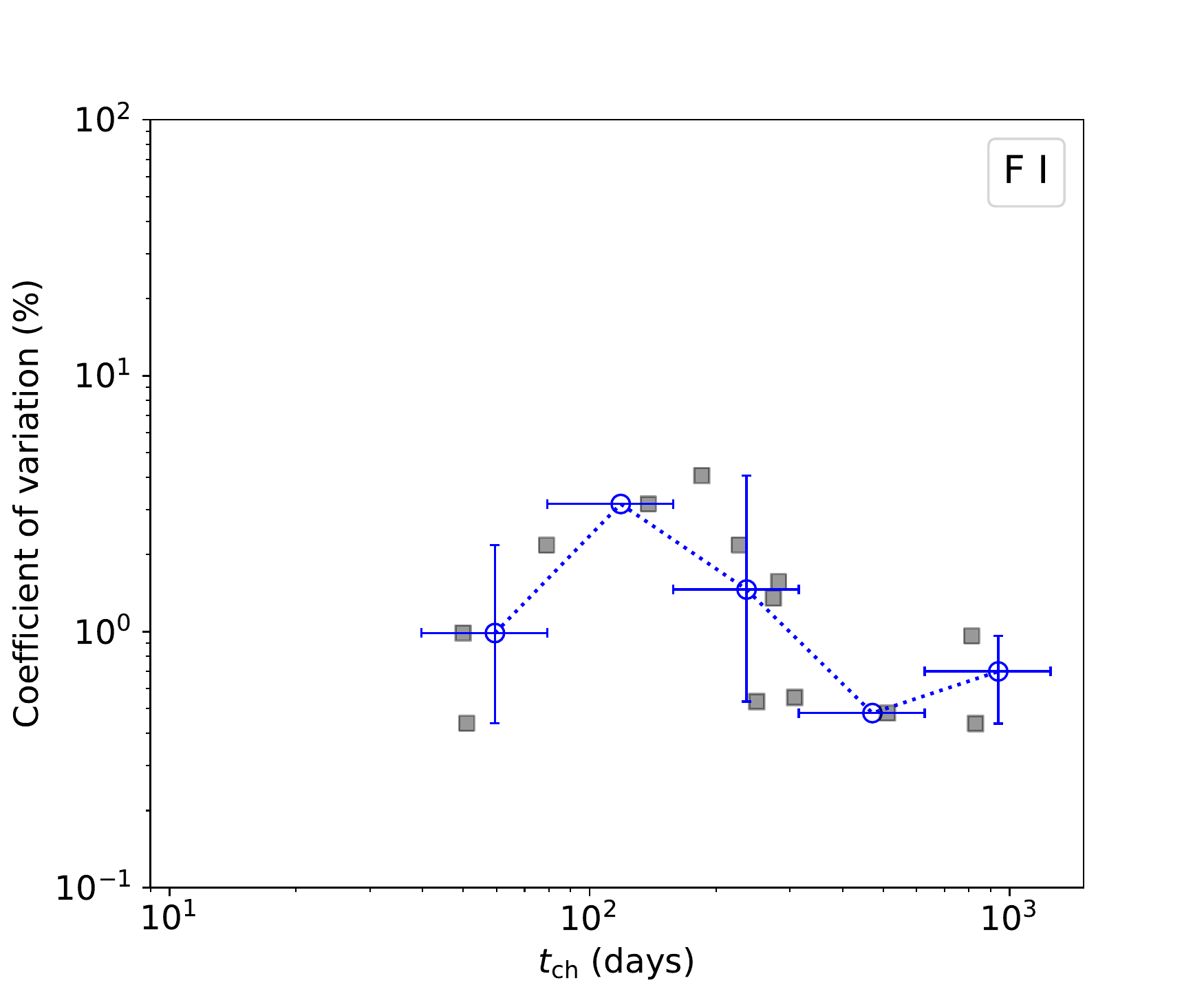}
\includegraphics[width=60mm]{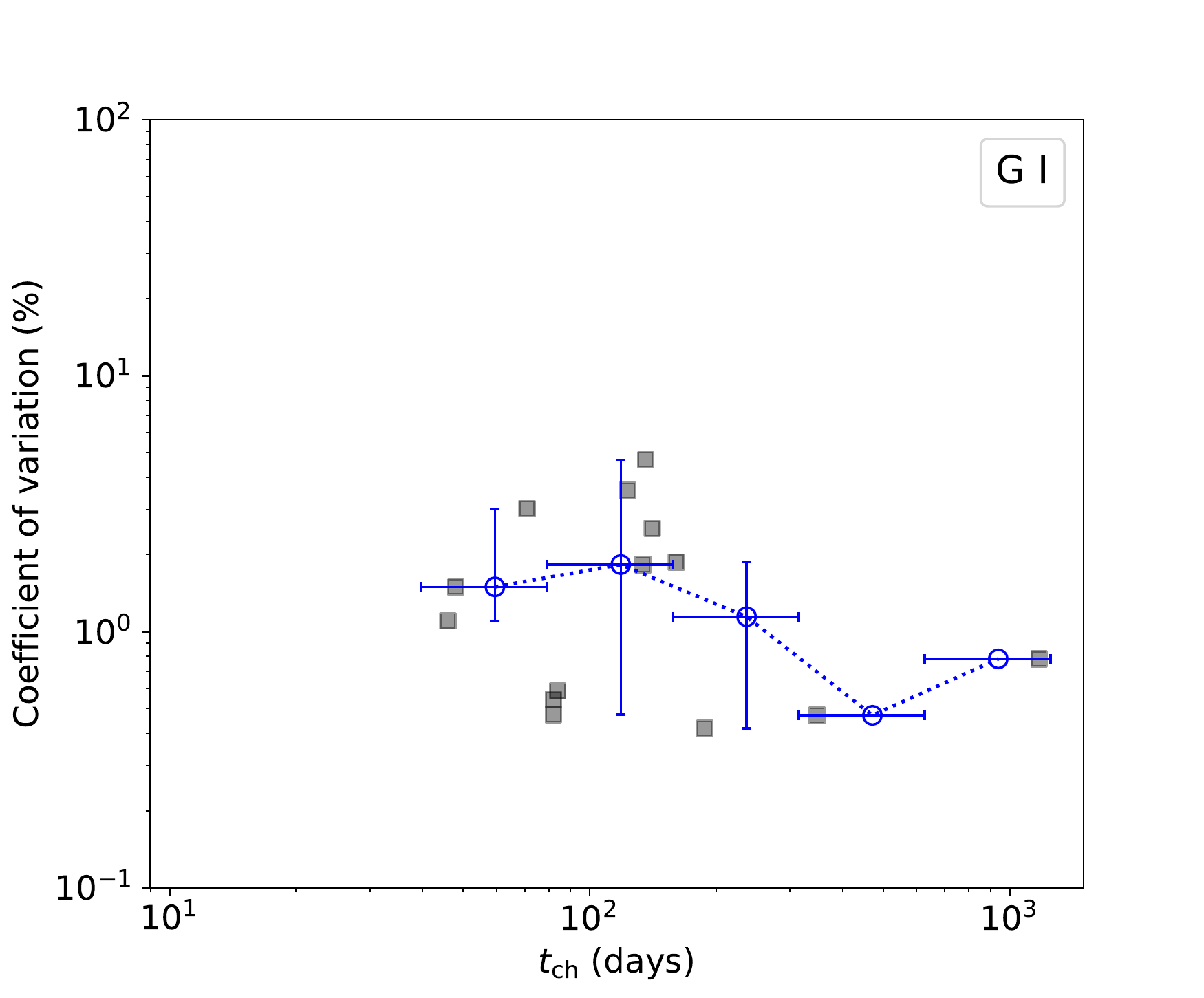}\hfill\includegraphics[width=60mm]{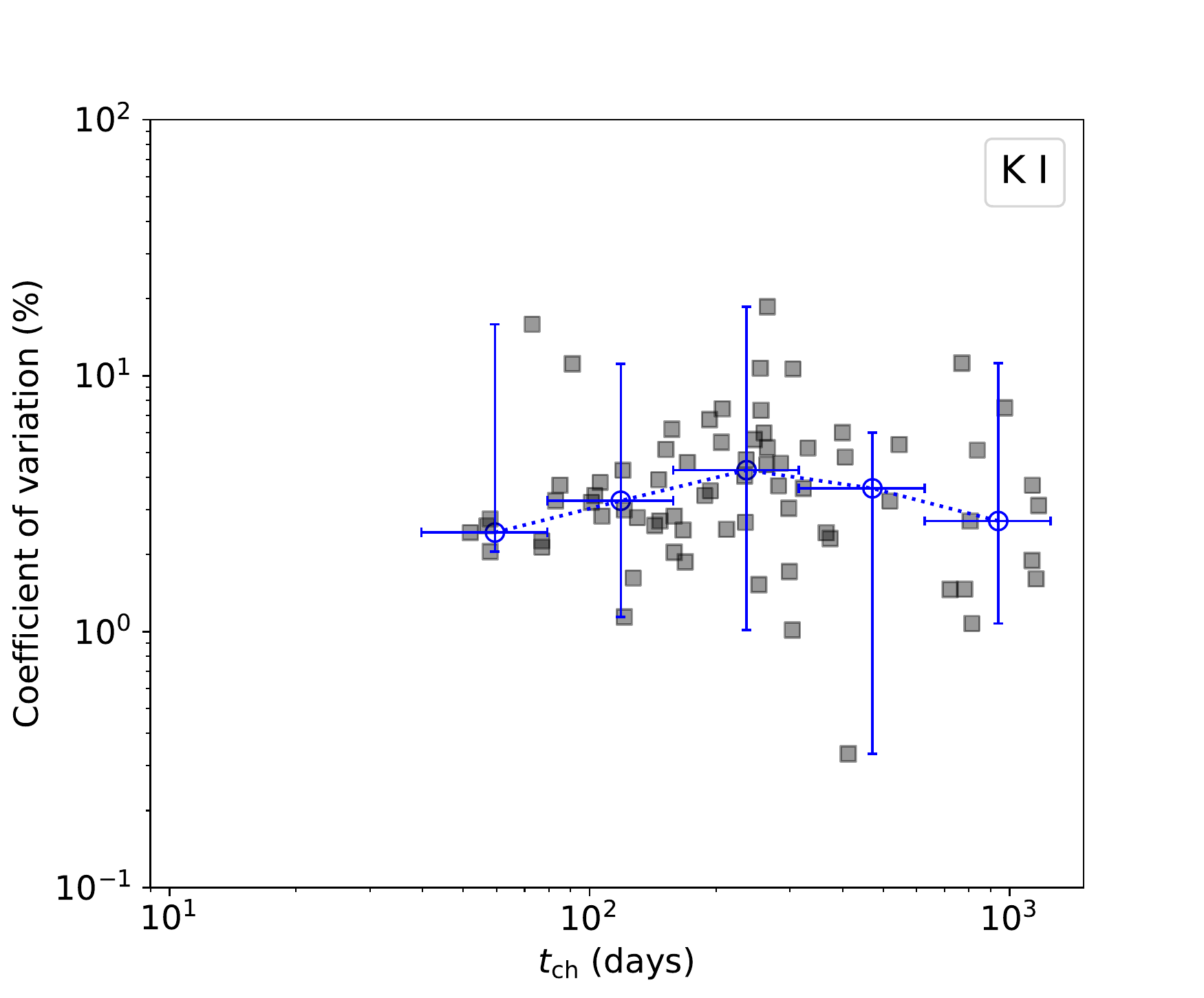}\hfill\includegraphics[width=60mm]{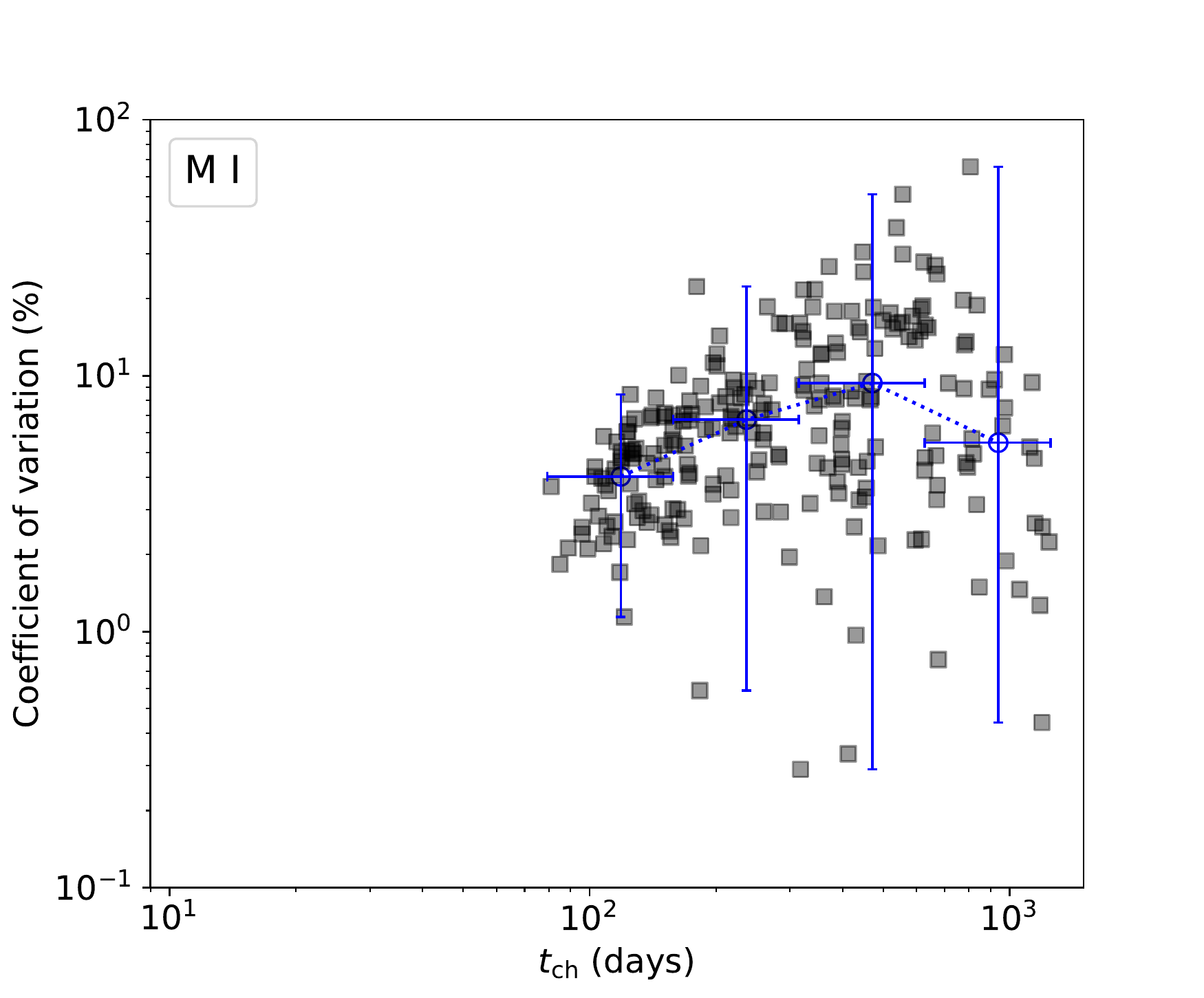}
\caption{Coefficient of variation as a function of characteristic timescales $t_{\rm ch}$ for the different types of supergiants as indicated in the respective legends. The grey squares are the data points, while the blue circles denote the median after binning the data logarithmically into 7 bins between $t_{\rm ch}$ values of 10~days and 1250~days, using the same bins for each panel. The vertical whiskers extend to the minimum and maximum values of the coefficient of variation at a given $t_{\rm ch}$ bin and the horizontal bar shows the bin width.} 
\label{fig:cov}
\end{figure*}

In Fig.~\ref{fig:cov}, we show the amplitude associated with a given $t_{\rm ch}$ relative to the average flux of the star (i.e., coefficient of variation) for the different types of supergiants. 
In almost all cases, the trend between the coefficient of variation and $t_{\rm ch}$ appears flat. However, note that since the noise decreases with increase in timescale probed (cf.~Fig.~\ref{fig:timescale_noise}), we are likely more incomplete at the shorter $t_{\rm ch}$ values than at the longer values. Furthermore, a very low level of variability, especially at short $t_{\rm ch}$ values, will be contributed by the brightest stars.

It can be seen from the figure that the level of variability for the K- and M-type supergiants corresponding to the timescales found for them ($t_{\rm ch}\gtrsim 100$~days) is typically a few percent to tens of percent, and these stars exhibit the highest level of variability amongst all supergiants. For the yellow supergiants, the $t_{\rm ch}$-specific variability is $\approx 0.3\%$ to a few percent, while for the O, B-, and A-types, it ranges between $\approx 0.3\%$ to 10\%. For WRs, the coefficient of variation corresponding to their $t_{\rm ch}\gtrsim100$~days is between 0.1\% to a few percent. For LBVs, the variability level is generally around a few percent.

\section{Discussion} \label{sec:discuss}

The variability characteristics, comprising amplitudes and timescales, for the massive stars in M31 are empirically derived in this work. What is/are then the physical processes responsible for the variability observed in these stars?

Various processes are at work in massive stars. These include rotation, pulsation, convection, binary interaction, variable stellar winds and the consequent mass loss that are common for these massive stars, and possibly all of them influencing each other.

Photometric variability of RSGs has been discussed by \citet[Chapter 7]{Levesque-2017}, which included observation results from the Magellanic Clouds \citep{Yang-2011, Yang-2012} and the Milky Way  (\citealt{Kiss-2006, Stothers-2010}; see also various other references in the Chapter) but not M31, the results of which were presented later in Paper I. As mentioned above, physical mechanisms that can lead to the observed variability in RSGs are diverse, ranging from radial pulsations \citep{Stothers-1969, Heger-1997, Guo-2002}, convection-driven surface effects \citep[e.g.,][]{Chiavassa-2009, Stothers-2010},  variable mass loss \citep{Yoon-2010} to binary effects (e.g., the eclipsing binaries VV Cep and $\zeta$~Aur \citep{Bennett-1996, Wright-1977}) . The binary fraction for RSGs is not yet well-established. \citet{Patrick-2019} estimated a binary fraction of 30\% for these stars -- \citet{Neugent-2019} found 63 RSG+B star binaries in M31 and M33 out of a sample size of 149 (one of their stars J004327.01+412808.7 typed as BI in MNS16 is included in our sample and we find a $t_{\rm ch}$ of around 150~days for this star; cf.~Fig.~\ref{fig:append}). Because of their large physical size, however, the binary companions of these stars are expected at large orbital periods -- thousands of days, which are much longer than their extracted timescales from the PTF data. RSGs also rotate very slowly---the projected rotational velocity mapped for Betelgeuse based on ALMA observations by \citet{Karvella-2018} is around 5~km/s and thus the rotation period is on the order of decades, again much longer than the baseline of the data studied here. Some of them exhibit a long secondary period $>1000$~days \citep[e.g.,][]{Kiss-2006, Yang-2011, Yang-2012}, which has been suggested, among others, to be related to the turnover timescale of surface convection cells \citep[e.g.,][]{Stothers-1971, Stothers-2010}. These timescales are again much longer than the ones probed in this work. Multi-epoch spectroscopic data on a long baseline are required to investigate the variable mass-loss scenario, which are not available for these stars.  
On the other hand, in Paper~I, we found that the extracted periodic timescales for these stars are consistent with the fundamental and first overtone modes of radial pulsations, based on theoretical models computed with MESA \citep{Paxton-2015} coupled with the linear asteroseismology code \texttt{GYRE} \citep{Townsend-2013}. The $t_{\rm ch}$ values for these red massive stars in the present study are similar to those found in Paper~I and hence, are likely associated with pulsation as well.

Like for the RSGs, the binary fraction of the rarer yellow supergiants (F- and G-types) is also highly uncertain. For a few cases, orbital periods of around a few hundred days have been determined \citep{Prieto-2008, Sperauskas-2014}. Besides, when (some of) these stars go through the classic instability strip, they will undergo pulsation as Cepheids and their variants such as the double-mode Cepheids \citep{Bono-1999a} with periods on the order of a few days to tens of days, and even reaching over 100 days for the ultra-long period Cepheids \citep{Bird-2009}. 
\citet{Genderen-2004} presented observation results of optical photometric variability for F-type supergiants in the LMC and found a maximum variability amplitude of $\approx0.2$~mag and timescales on the order of  100-200~days, which are consistent with our results (see Figs.~\ref{fig:sg_rms} and \ref{fig:tch}). 
The observed variability in these stars (see e.g., Fig.~\ref{fig:append}) likely includes pulsation, but disentangling various other processes, like the presence of a binary companion, is impossible without complementary information from multi-epoch long-baseline spectroscopy and comprehensive theoretical models.

Earlier work in the literature, before the era of large-scale time-domain surveys, on observed photometric variability of O-, B-, A-type supergiants in the Galaxy includes \citet{Maeder-1980, Burki-1978, Rufener-1978, Schild-1983} and references therein (though not exhaustive). \citet{Maeder-1980} found, using the catalog of stars in the Geneva Observatory photometric system \citep{Rufener-1976}, the optical variability amplitude to increase with the luminosity in a given spectral type, for example Ia versus Iab, and Ib, and the RMS amplitude, typically at the level of a hundredth of a magnitude, increases toward later spectral types in the most luminous supergiants (i.e, Ia) with the largest amplitude for the M-type supergiants and a local maximum around early B-type stars. This is based on approximately only 7 data points per star measured at different times over a baseline of 20 years. Further, based on four months of high-cadence data for seven supergiants,  \citet{Burki-1978} found timescales on the order of a few tens of days for B-type supergiants increasing to more than 80~days for the G-type, which was discussed to be related to non-radial gravity mode oscillations by \citet{Maeder-1980}. \citet{Schild-1983} also found variability in the OB-type supergiants, but with typical amplitudes on the order of 0.1~mag and timescales between weeks and 1000~days.  

We do not make the distinction between the luminosity subtypes of supergiants in the present work, and from Fig.~\ref{fig:sg_rms} the typical amplitudes for OBA stars appear similar at around 0.06~mag (with roughly similar typical noise floor). Although this amplitude level differs from the old work on Galactic supergiants, the trend of the late M supergiants reaching the largest amplitude is consistent with the earlier results. Our extracted  timescales $t_{\rm ch}$ (Fig.~\ref{fig:tch}) for the supergiants cover the timescales quoted by the earlier work, but detailed comparison is not possible given the disparate nature of the data and their  analysis.

For the hot luminous stars, the binary fraction is better determined, which is around 50\%--70\% \citep[e.g.,][]{Sana-2012,Dunstall-2015} including systems with long orbital periods extending to around 10~years, much longer than the baseline of our data. However, the period distribution for these stars indicate that orbital periods of tens to hundreds of days, i.e., the timescales covered in our study, are well populated, with the O-types preferring short periods \citep{Sana-2012} and the distribution of B-types being flat \citep{Dunstall-2015}. \citet{Aerts-2018} performed a study of the blue supergiant $\rho$~Leo using {\it K2} photometric data and multi-epoch HERMES spectroscopic data, and found dominant variability at the level of 8~mmag with a periodic timescale of a few tens of days, which they attributed to rotation. The shorter $t_{\rm ch}$ values in Fig.~\ref{fig:cov} and their corresponding amplitudes for the O and B stars may indeed be connected to rotation. Coherent pulsations triggered by the $\kappa$ effect of the Fe opacity bump are known to occur in these stars \citep[e.g.,][]{Cox-1992, Dziembowski-1993}. Given the high metallicity of M31, opacity-driven pulsations are certainly operating in these hot stars. Thus, some, if not all, of the variability we have observed for these stars may be coherent pulsations.

\citet{Pedersen-2019} performed a classification of the photometric variability in a large sample (over 150) of O- and B-type stars, which included rotating, eclipsing, and pulsating stars,  based on {\it TESS} data and visually examining the light curves and their discrete Fourier transforms. Many of the variable stars in the Pedersen et al.\ sample show simultaneous modulations from the different phenomena, e.g., eclipsing light curves with rotational modulation and/or coherent pulsation. 
Further, a number of blue supergiants in their sample show stochastic low-frequency variability, similar to that found by \citet{Bowman-2019} in their analysis of more than 150 hot luminous stars in the ecliptic and the LMC using {\it K2} and {\it TESS} data. The characteristic amplitudes of this variability are less than a few millimags (see \citealt{Bowman-2019}). Such variability could be caused by internal gravity waves excited by turbulent core convection \citep{Bowman-2019} or sub-surface convection triggered by local opacity enhancements associated with Fe and He \citep[e.g.,][]{Cantiello-2009, Cantiello-2019}. Being a low-frequency phenomenon, it should manifest at long timescales. Thus, the long $t_{\rm ch}$ values of around hundreds to a thousand days that we find for these stars in M31 and the corresponding amplitude of $<1\%$ (Fig.~\ref{fig:cov}) may be related to such stochastic variability.

For WRs, the close binary fraction is around 30-40\% \citep[similar to that of the OB stars]{Neugent-2014}, with corresponding orbital period $<100$~days \citep{Neugent-2019}, while our extracted $t_{\rm ch}$ for the WRs (Fig.~\ref{fig:tch}) are typically on the order of a few hundred days. Short-timescale low-level ($<0.1$~mag) photometric variability on order of hours to a few days, attributed to different mechanisms such as non-radial \citep{Vreux-1985} or radial \citep{Maeder-1985} pulsation, rotation \citep{Poe-1989} and wind instabilities \citep[e.g.,][]{Moffat-1991}, has been observed for many WRs. We did not find such short timescales for the WRs in our sample (Sect.~\ref{sec:short_tch}), probably due to the larger noise of the iPTF data for WRs (see Fig.~\ref{fig:sg_rms}). \citet{Genderen-2013} performed an analysis of the WC-type WR star WR~103 using a long-baseline (11 years) dataset and found stochastic variability with amplitude in the visual band on the order of 0.1~mag. This is consistent with our result where we find the typical RMS amplitude for WRs to be on the order of 0.1~mag (Fig.~\ref{fig:sg_rms}). The long $t_{\rm ch}$ values are consistent with the stochastic nature of the variability in a way similar to that discussed above for the OB-supergiants, i.e., the stochasticity is concentrated at low frequencies. However, we caution that the majority of the WRs in our sample are of the WN type.

A consistent treatment and prediction of observables accounting for the various physical processes, e.g, interaction between pulsation and convection, mass loss, binary effects, is currently lacking to interpret the wealth of observational results. Studies like ours, however, will provide important constraints in modeling the poorly understood physical processes in the evolution of massive stars.

\section{Conclusions} \label{sec:conclude}
By mining the well-sampled, long-baseline iPTF time-domain data of M31, we have mapped the variability of stars in the upper part of the HR diagram. The earlier work of MNS16 and also \citet{Massey-2016b} in M31 provided the identification of the massive stars, including their spectral types. These stars exhibit a wide variety of light curve shapes, encoding the varied physical phenomena modulating their observed radiation fields. 
In agreement with \citet{Conroy-2018}, who studied the variability of the stellar populations in M51, we find that (photometric) variability is widespread in the upper parts of the CMD of M31 with the observed variability fraction increasing for the later spectral types toward close to 100\% for the cool supergiants. The incompleteness of the spectral catalog, however, likely affects the result for the early-type supergiants.
In the observed variability maps, the cooler stars also show larger variability amplitudes than the bluer counterparts.

Further, using the powerful signal reconstruction tool of \citet{Oppermann-2013}, we are able to extract characteristic timescales $t_{\rm ch}$ of variability for these stars that are both localized in time (e.g., in the case of irregular and semi-regular variables) and unlocalized (i.e., periodicity). For the first time, we are thus able to map out the $t_{\rm ch}$ values for the massive star population characterized by diverse variability behavior.

Using a block of the time-series data straddling two nights with a high cadence of $\approx2$~minutes, we find significant $t_{\rm ch}$ in the range 0.1--10~days for 13 stars in our sample. This is in agreement with recent results from space-based data, for example from \citet{Wallenstein-2019} that also found such short timescales in a smaller sample of evolved stars.

Using the long baseline PTF light curve for probing $t_{\rm ch}\gtrsim10$~days, we find that the cool supergiants have longer $t_{\rm ch}$ (hundreds of days and more) relative to the hotter stars. The $t_{\rm ch}$ values of the latter cover a larger range, exhibiting variability typically on short timescales of tens of days and extending to the larger timescale domain of the cooler stars. These observations are in general agreement with the theoretical predictions of pulsation in massive stars, especially for the cooler supergiants. For the hot luminous stars, a myriad of effects including rotational modulation, pulsation, binary companions, and perhaps an interplay among them can result in photometric and/or spectroscopic variability. Contemporaneous multi-epoch spectra will greatly complement studies of variability in massive stars. On the theoretical side, various uncertainties in the treatment of physical processes operating in the massive stars' envelopes (interplay of convection and pulsation, stellar wind, companion effect, etc.) in the models prevent a more detailed comparison with observations.

The maps of the variability characteristics such as those presented in this paper will serve as a powerful tool to explore the phenomena themselves, but also in investigating their relation to the host galaxy environment and consequently galaxy evolution, given the important role that these massive stars play. To this end, a large statistical sample of the maps of variability characteristics of stellar populations in different host environments covering a large range of metallicity, star formation rate, etc., will be invaluable. Fortunately, this is a task that is now possible to accomplish with the large amount of archival time-domain data that have become available from the present generation of wide-field, high-cadence, and long-baseline optical surveys.

\acknowledgments{Acknowledgments:}
MDS is supported by the Illinois Survey Science Fellowship of the Center for Astrophysical Surveys at the University of Illinois at Urbana-Champaign. 
This research was supported in part by the National Science Foundation through grant PHY-1748958 at the KITP and benefited from interactions that were funded by the Gordon and Betty Moore Foundation through Grant GBMF5076. 
MDS thanks Chien-Hsiu Lee for helpful discussions on observational studies of stellar variability as well as Niels Oppermann, on proper applications of signal analysis tools. We also thank Charlie Conroy and Yan-Fei Jiang for discussions, and the latter for providing his simulation data. MRD acknowledges support from the Dunlap Institute at the University of Toronto and the Canadian Institute for Advanced Research (CIFAR). This work was based on observations obtained with the 48-inch Samuel Oschin Telescope at the Palomar Observatory as part of the Palomar Transient Factory project. Operations were conducted by Caltech Optical Observatories and data processing by IPAC. 
We thank Xiaodian Chen for helpful comments on the period-luminosity relation of contact binaries.

%\facilities{}
\software{\texttt{numpy} \citep{numpy}, \texttt{scipy} \citep{scipy}, \texttt{astropy} \citep{astropy}, \texttt{pyfits}, \texttt{pandas} \citep{pandas}, \texttt{mpi4py} \citep{mpi4py}, \texttt{matplotlib} \citep{matplotlib}, \texttt{DAOPHOT} \citep{Stetson}, \texttt{DAOGROW} \citep{Stetson-1990},\texttt{NIFTy} \citep{Selig-2013}.}

\begin{table*}[ht]
\caption{Variability characteristics of massive stars in M31}\label{tab:all}
\renewcommand\arraystretch{1.0}
\renewcommand{\tabcolsep}{22pt}
\centering
\begin{tabularx}{1.0\textwidth}{*{4}l*{1}{X}}
\hline
ID\tablenotemark{a}	
&SpT\tablenotemark{b}		
&$\left<m_{R}\right>$
&$\Delta m_{R}$\tablenotemark
&$t_{\rm ch}$ (days)\tablenotemark{c}\\
\hline
J004314.06+415301.8           &B8I           &17.75           &0.05           &223, 407\\
J004353.34+414638.9           &WN7           &18.28           &0.12           &-99\\
J004043.95+405901.6           &G8I           &17.34           &0.06           &71\\
J004028.48+404440.2           &B9I           &17.23           &0.03           &37, 41, 112, 773\\
J004157.56+410753.3           &A4I           &17.68           &0.05           &62, 189, 830\\
J004213.75+412524.7           &M3 I           &18.57           &0.13           &144, 362, 846\\
J004313.31+413329.1           &F0I           &18.48           &0.10           &274, 812\\
J004252.10+414516.4           &K5 I           &19.02           &0.13           &235, 781\\
J004621.08+421308.2           &cLBV           &17.76           &0.05           &36, 75, 183\\
J004331.17+411203.5           &WC8           &19.08           &0.17           &-99\\
J004130.30+411603.8           &A7I           &17.29           &0.11           &25, 58, 68, 276, 682\\
J004009.43+405932.3           &ON9.7Iab           &18.93           &0.17           &18, 19, 1206\\
J003953.55+402827.7           &YSG           &17.28           &0.21           &170, 1130, 1131\\
J004158.87+405316.7           &O9.5I           &17.98           &0.06           &37, 152\\

\hline
\end{tabularx}
\begin{flushleft}
\tablenotetext{a}{LGGS ID of star as given in MNS16.}
\tablenotetext{b}{Spectral type of the star from MNS16.}
\tablenotetext{c}{Measurements not available are indicated -99.}
\tablenotetext{}{(This is only a part of the table; the full version is available in machine-readable form online.)}
\end{flushleft}
\end{table*}

\bibliographystyle{aasjournal}
\bibliography{references}

\appendix

\section{W~Ursa Majoris-type contact binary candidate}\label{append:wuma}
\begin{figure*}[h]
    \centering
    \includegraphics[width=60mm]{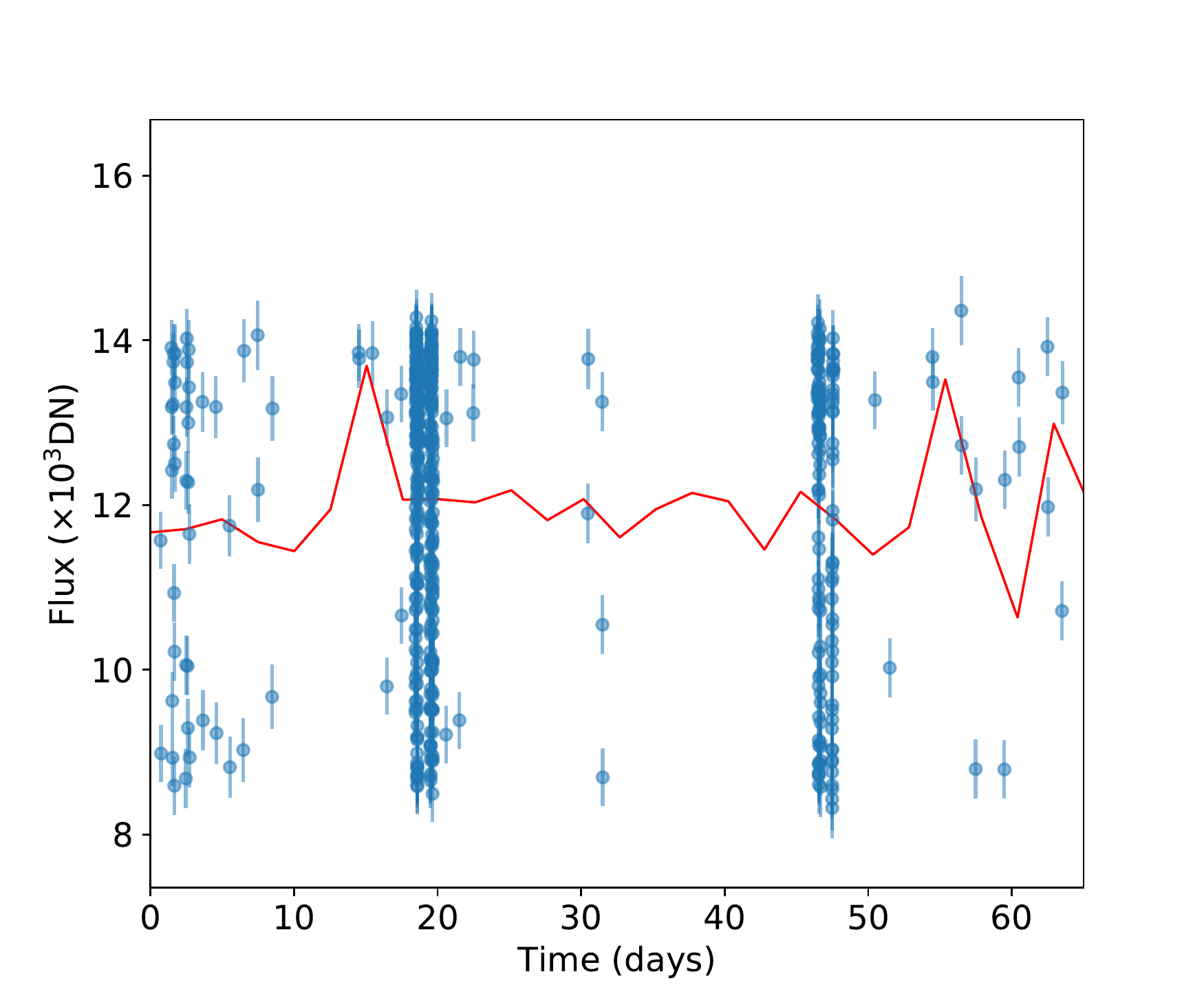}\hfill\includegraphics[width=60mm]{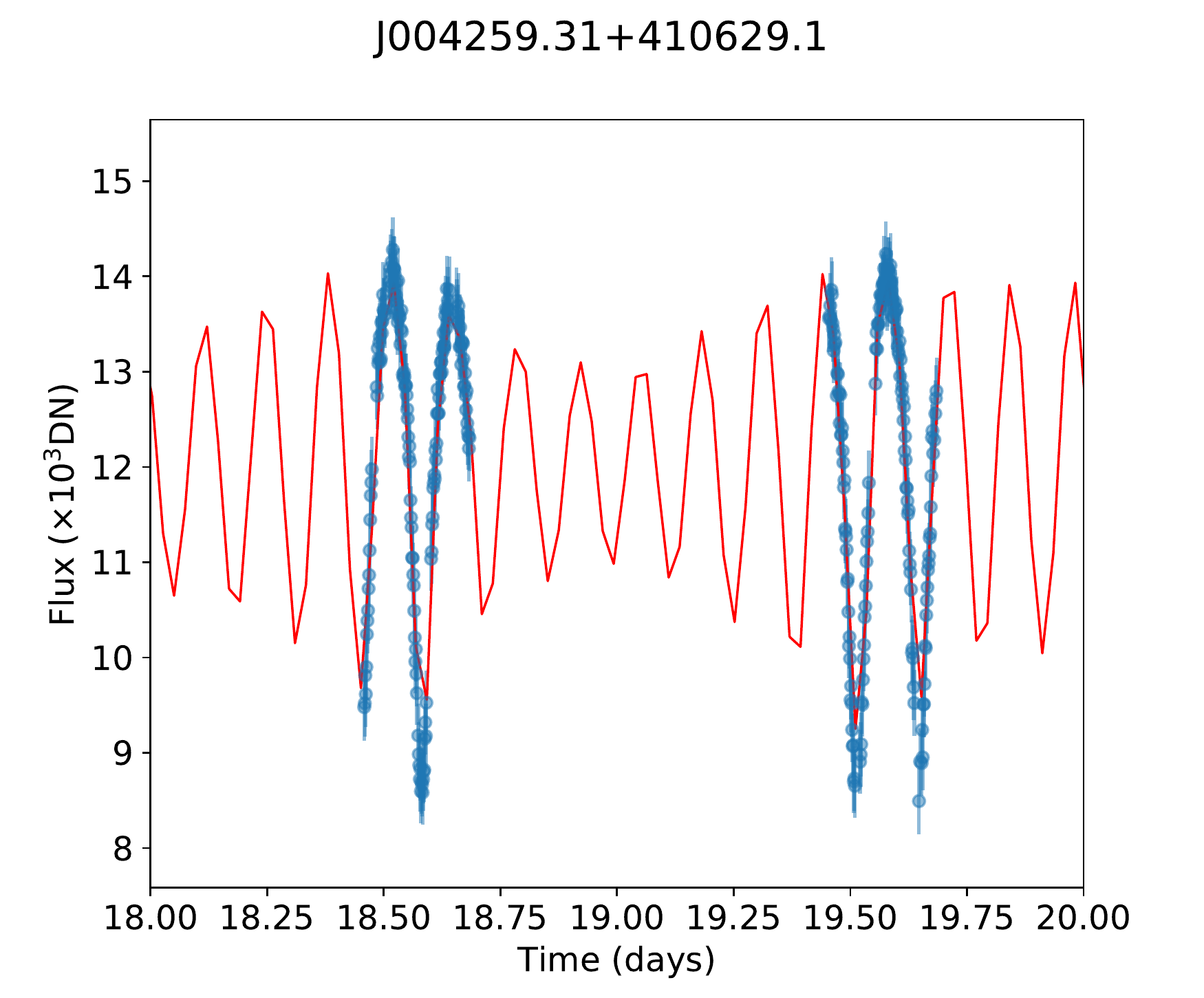}\hfill\includegraphics[width=60mm]{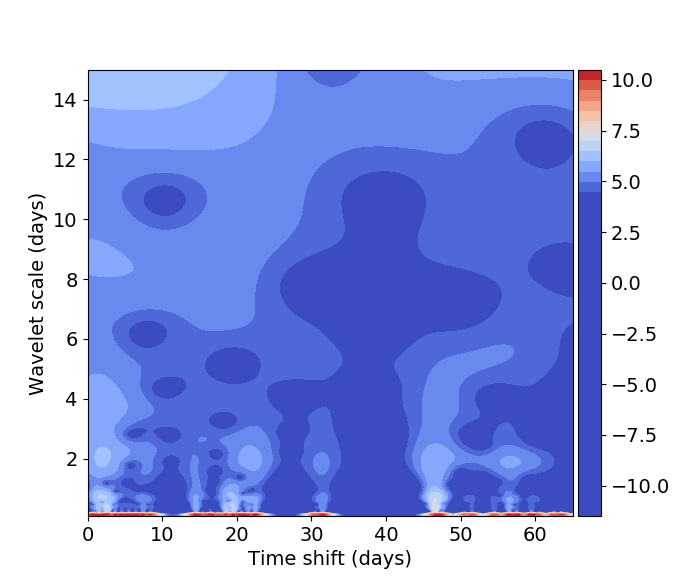}
    \caption{iPTF light curve for the W~UMa candidate along with its wavelet transform based on the high-cadence block ({\it right}). The reconstructed signals are shown in red for both the lower and higher resolution in the left and middle panels, respectively, and the wavelet transform corresponds to the signal in the middle panel.}
    \label{fig:wuma}
\end{figure*}

For the 11 stars belonging to clusters and HII regions, and without proper spectral labels, which we have dropped in Sect.~\ref{sec:var}, we also perform both the high- and low-resolution reconstruction of their light curves to determine $t_{\rm ch}$. Interestingly for one of the stars, J004259.31+410629.1, belonging to an HII region and with Mm of 1 in the MNS16 catalog (Sect.~\ref{sec:spectra}), we obtain $t_{\rm ch}$ of 0.14~day based on its high-cadence block---its low-resolution reconstruction does not represent the data well (see Fig.~\ref{fig:wuma}). Very short periods (less than around 0.5~day) with sinusoidal light curve shape are typical characteristics of W~UMa-type contact binaries---both features exhibited by J004259.31+410629.1, such that it may be indeed a W~UMa-type contact binary.

The mean magnitude of the iPTF light curve of this star is $m_{R}\approx17.0$. Applying the optical $R$ band period-luminosity (PL) relation for W~UMa-type contact binaries from \citet{wuma-2018}, we find its absolute magnitude $M_{R}\approx4.65$, and accordingly a distance estimate of around 2.9~kpc. Note that we use $2\times t_{\rm ch}=0.28$~day in the PL relation since the full period of contact binaries contains two maxima and minima (Chen, private communication). We find a source in {\it Gaia}-DR2 \citep{Gaia-dist} within $0.27''$ of J004259.31+410629.1, with measured parallax of 0.527~mas. Its estimated distance based on the {\it Gaia} parallax is around 2~kpc \citep{Bailer-2018}. There is around 0.9~kpc difference between the two distance estimates. Nevertheless, given the number of uncertainties (e.g., extinction), the discrepancy may not be surprising. It thus appears likely that J004259.31+410629.1 is in the foreground and a contact binary.

\section{Light curves and their wavelet transform}\label{append}
Examples of iPTF light curves for stars belonging to different spectral types are shown in Fig.~\ref{fig:append}, while light curves for all (candidate) LBVs in M31 with detected photometric variability from iPTF are shown in Fig.~\ref{fig:lbvs}. Light curves of the 13 stars with significant $t_{\rm ch}$ in the high-cadence block are also shown in Fig.~\ref{fig:small_tch} along with their corresponding wavelet transform maps.

\begin{figure*}[h]
\includegraphics[width=60mm]{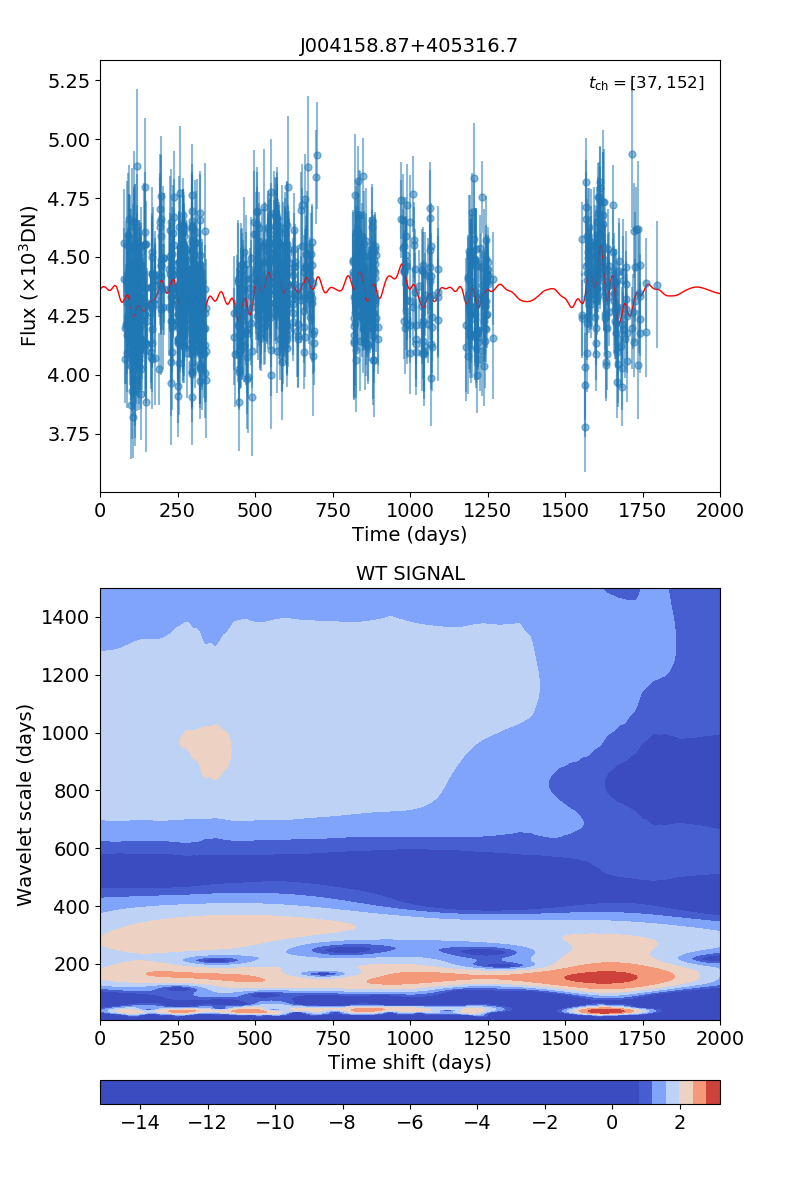}\hfill\includegraphics[width=60mm]{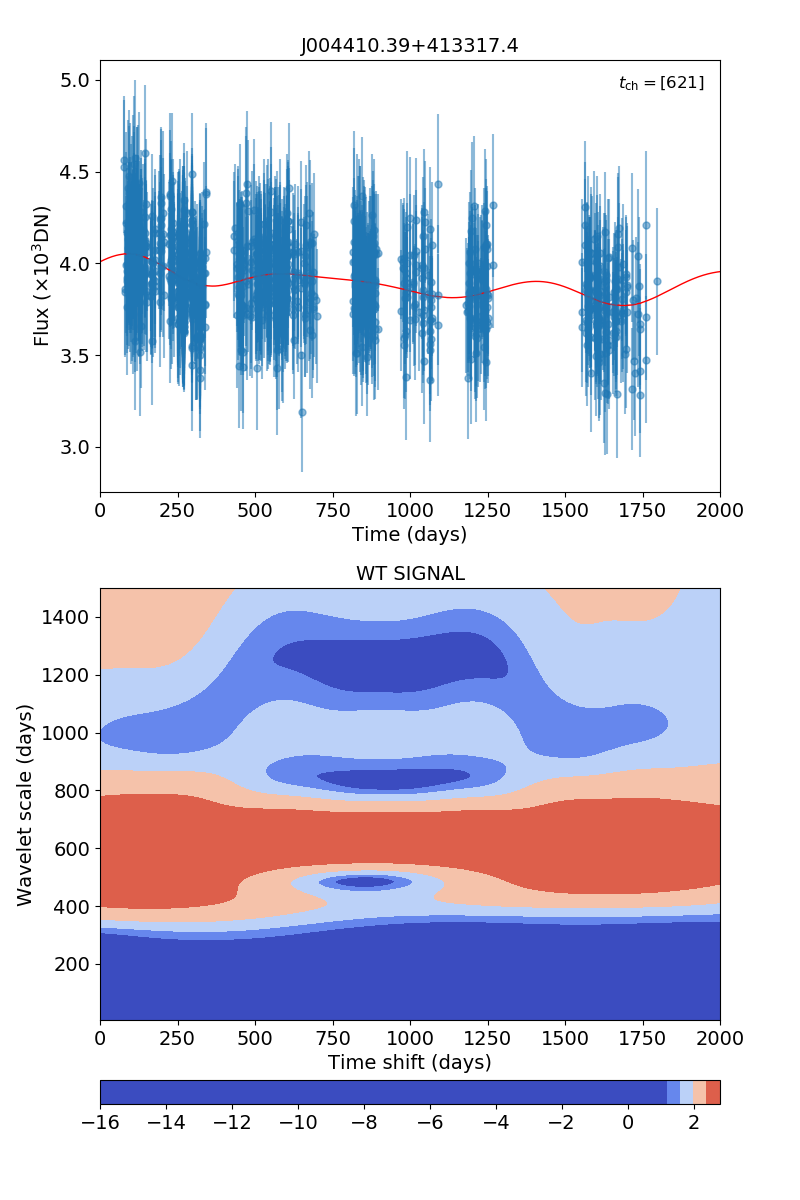}\hfill\includegraphics[width=60mm]{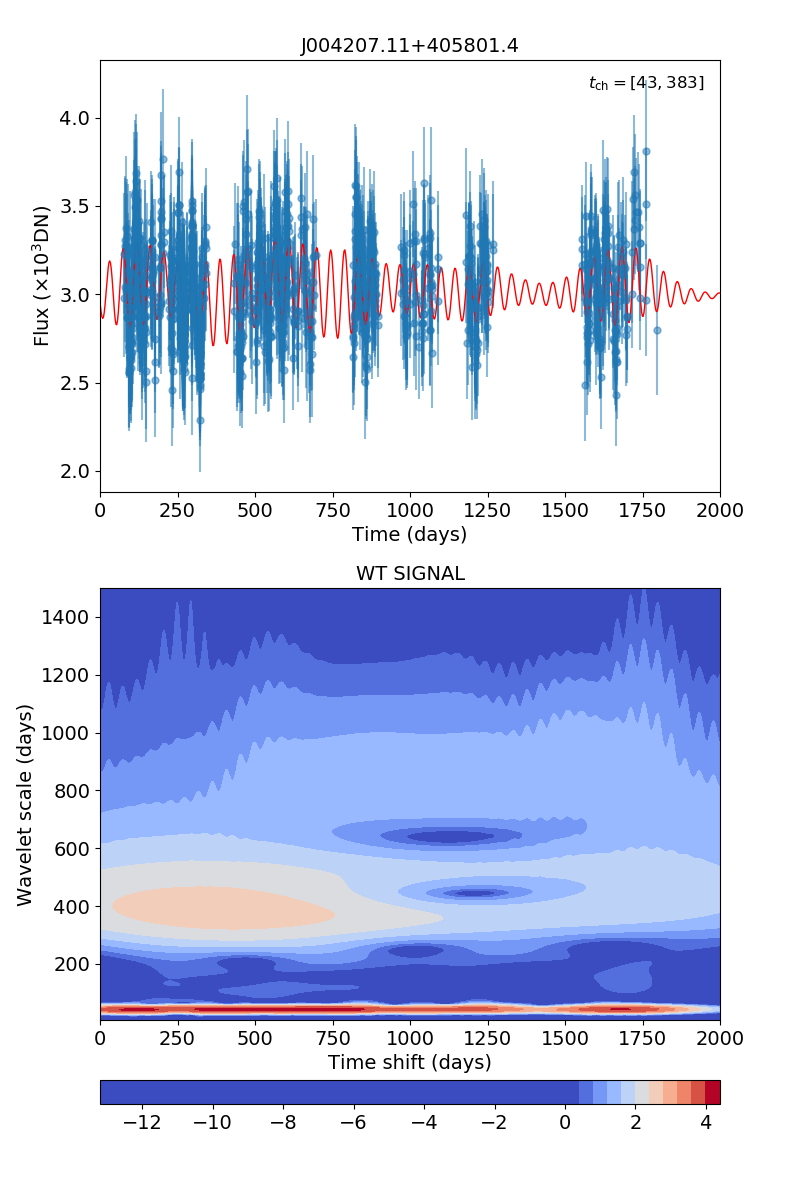}
\includegraphics[width=60mm]{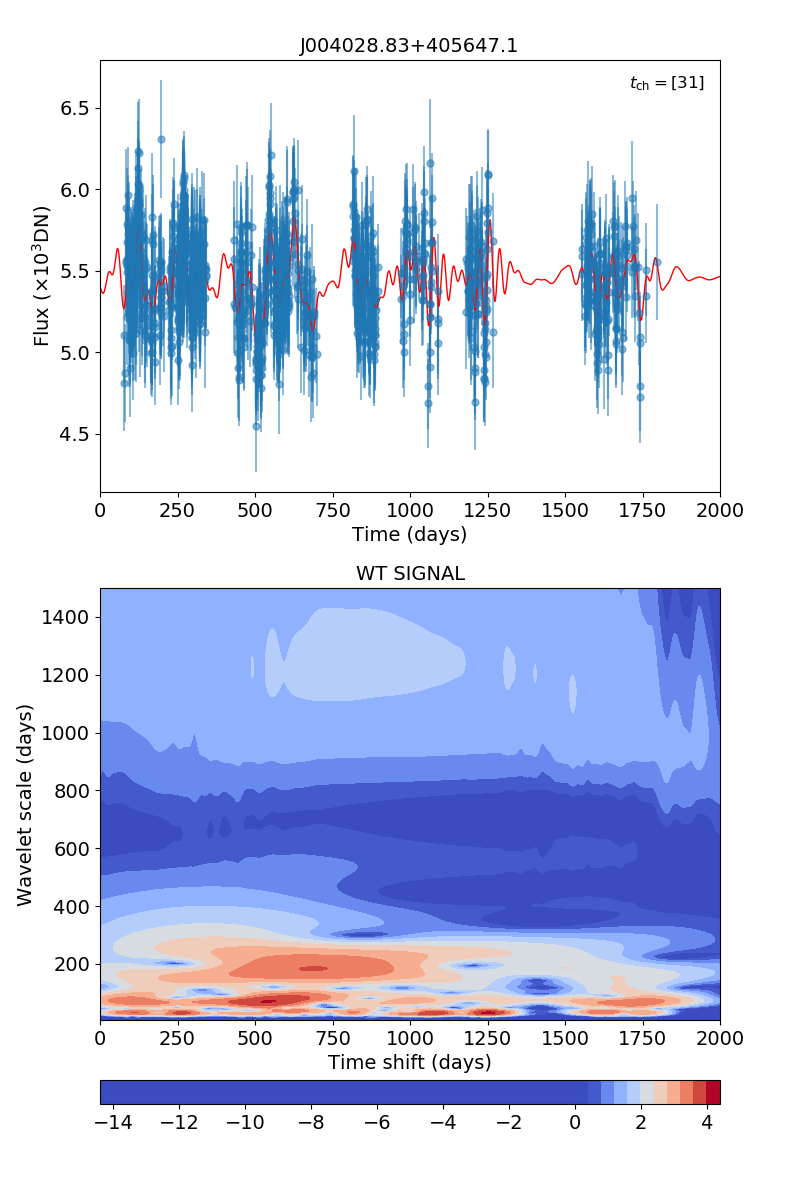}\hfill\includegraphics[width=60mm]{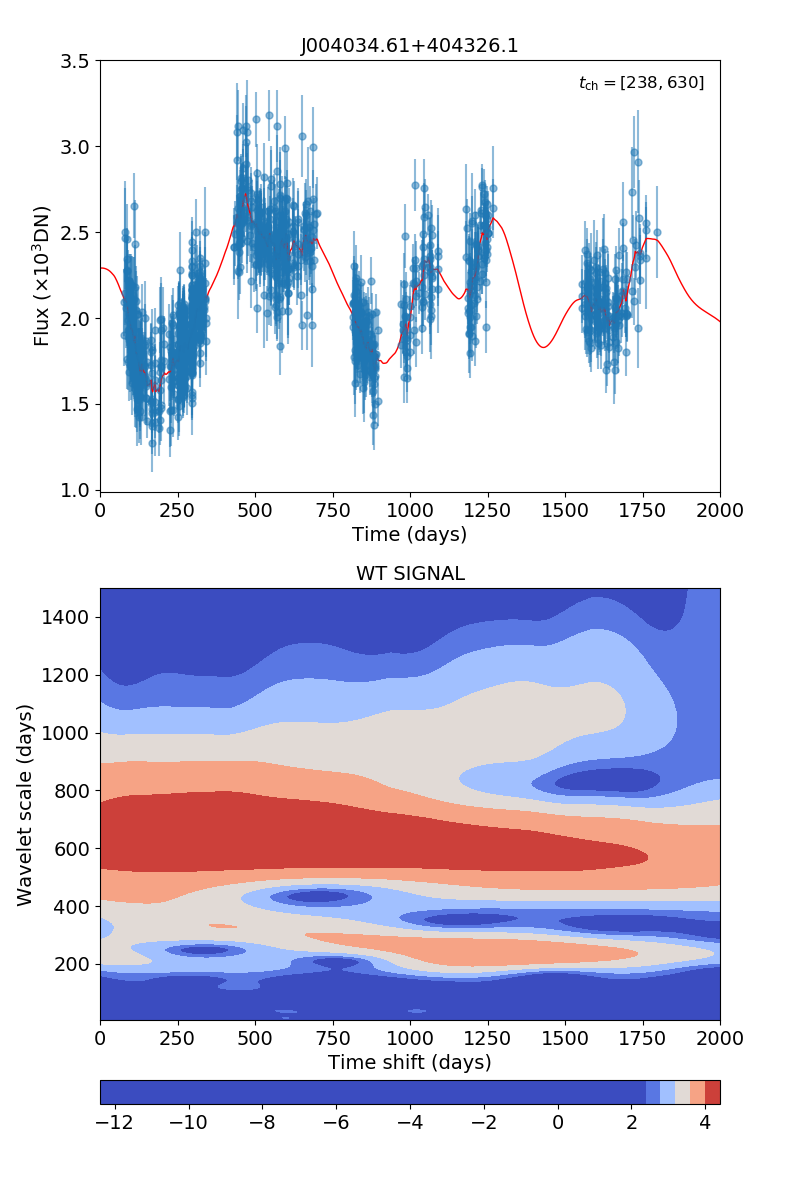}\hfill\includegraphics[width=60mm]{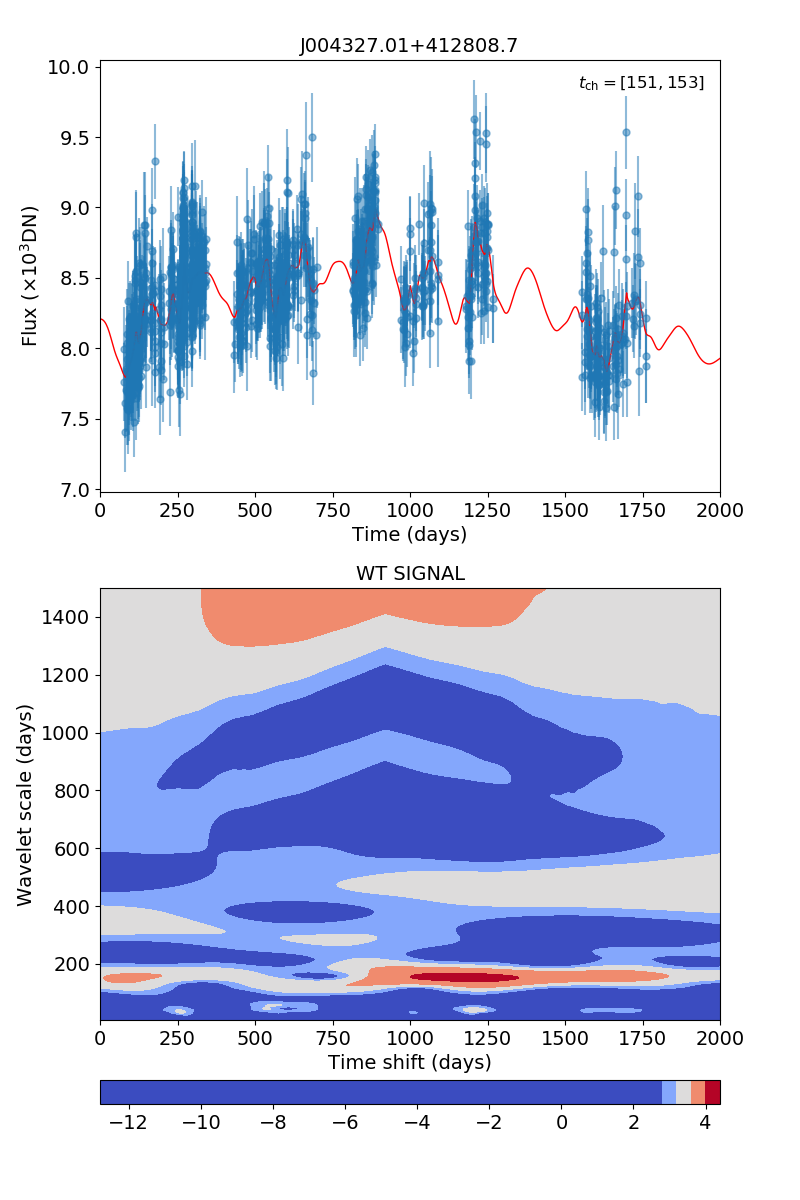}
\caption{iPTF light curves along with the corresponding wavelet transform maps for O stars ({\it top}) and B stars ({\it bottom}). The red curve shows the reconstructed signal, and the ID of the star from MNS16 is indicated on top of the plot.}
\label{fig:append}
\end{figure*}

\begin{figure*}
\ContinuedFloat
\includegraphics[width=60mm]{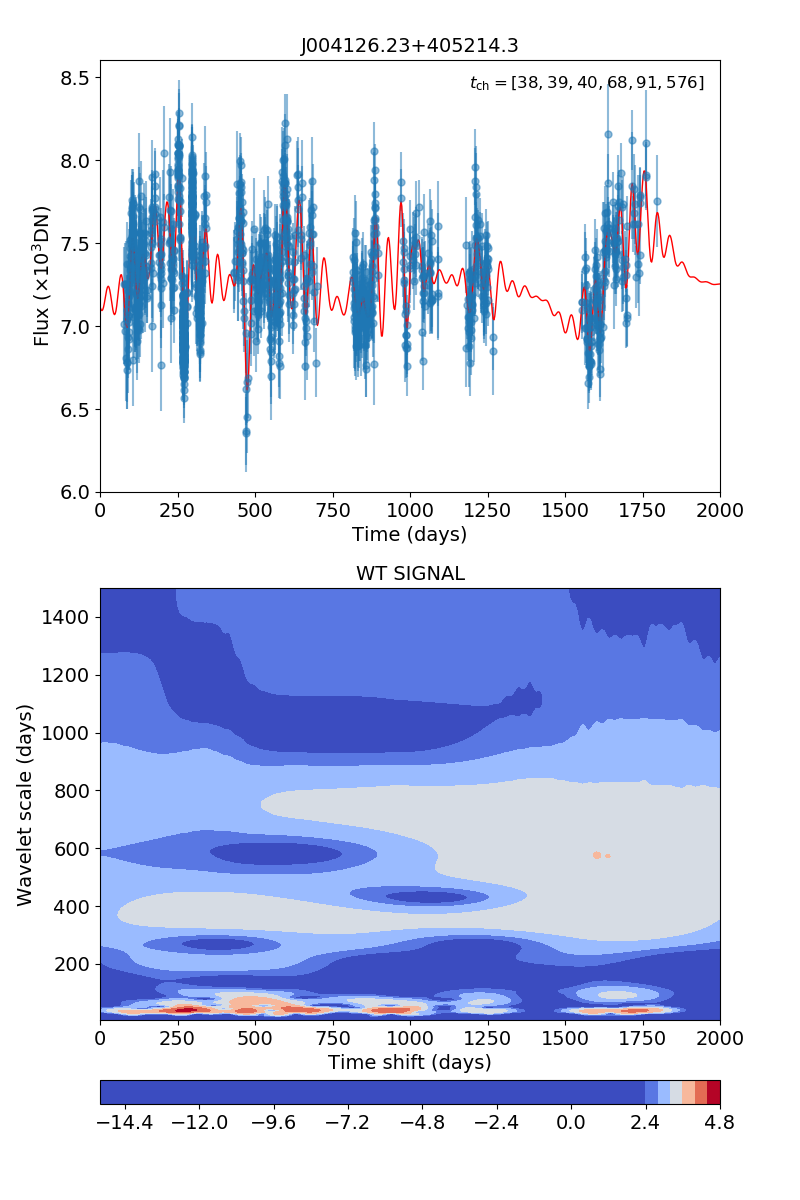}\hfill\includegraphics[width=60mm]{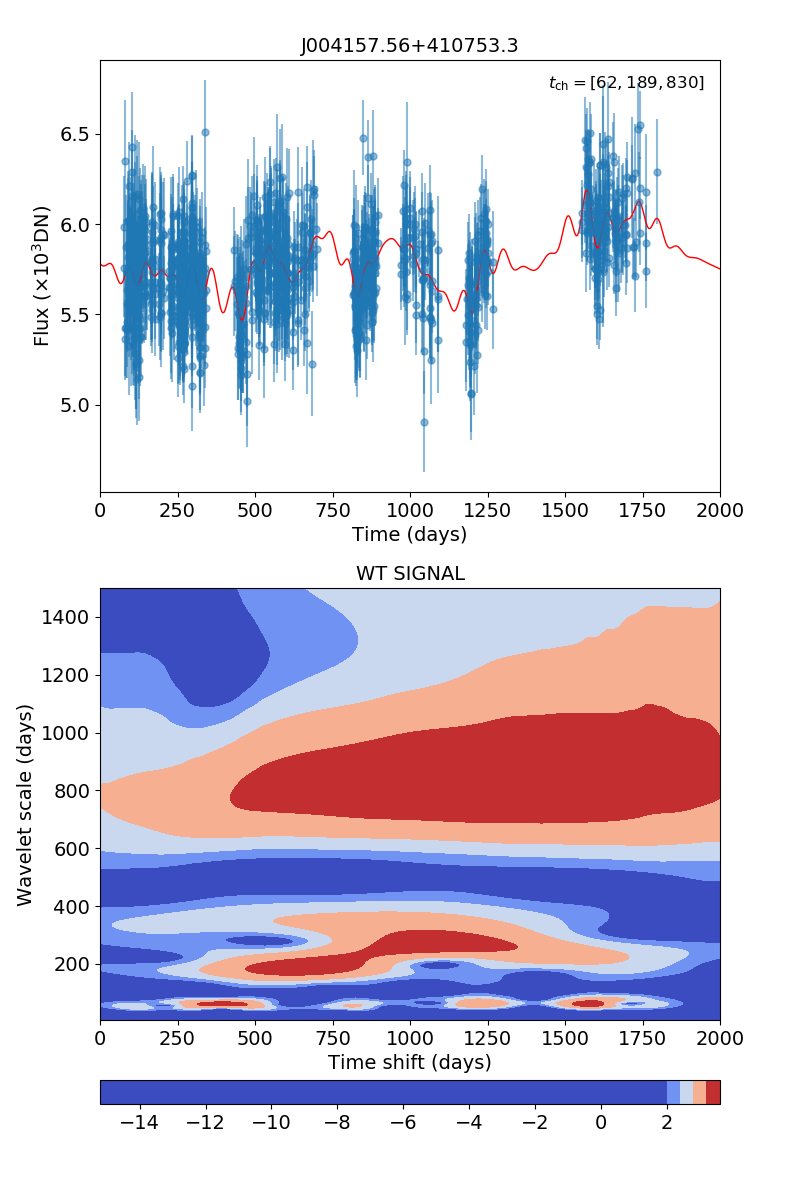}\hfill\includegraphics[width=60mm]{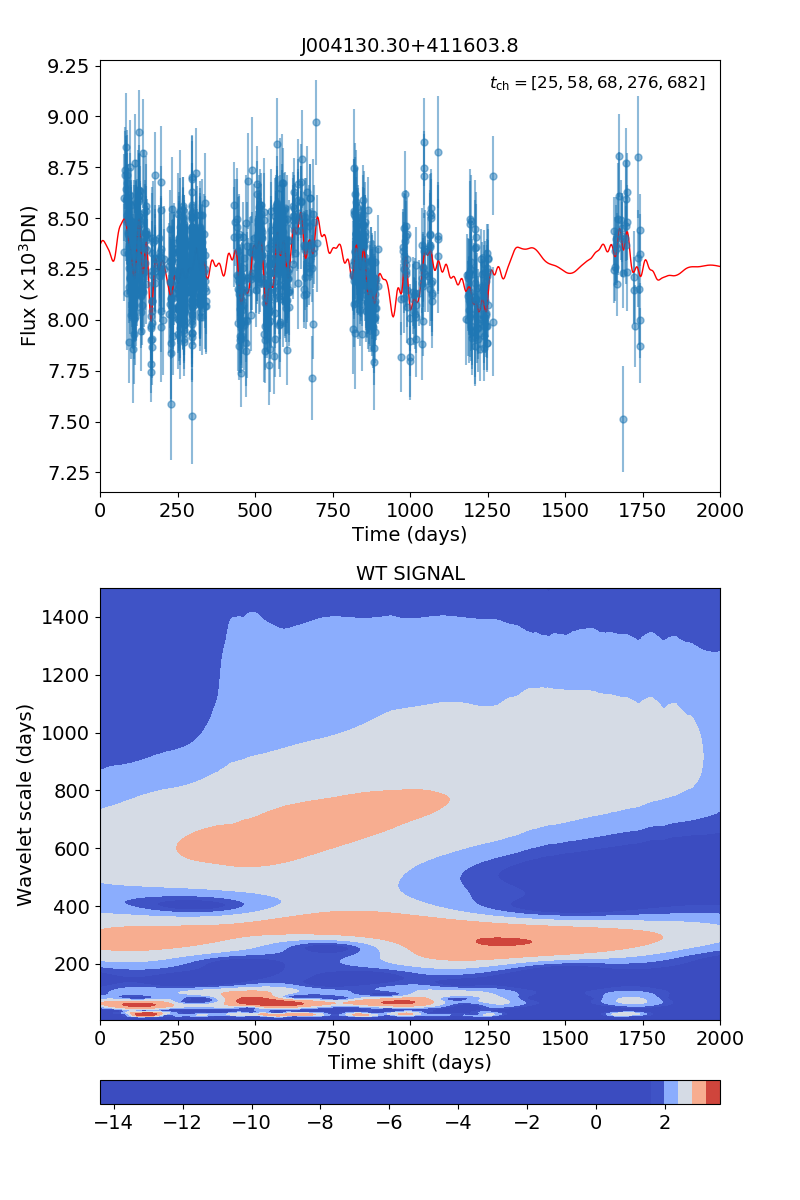}
\includegraphics[width=60mm]{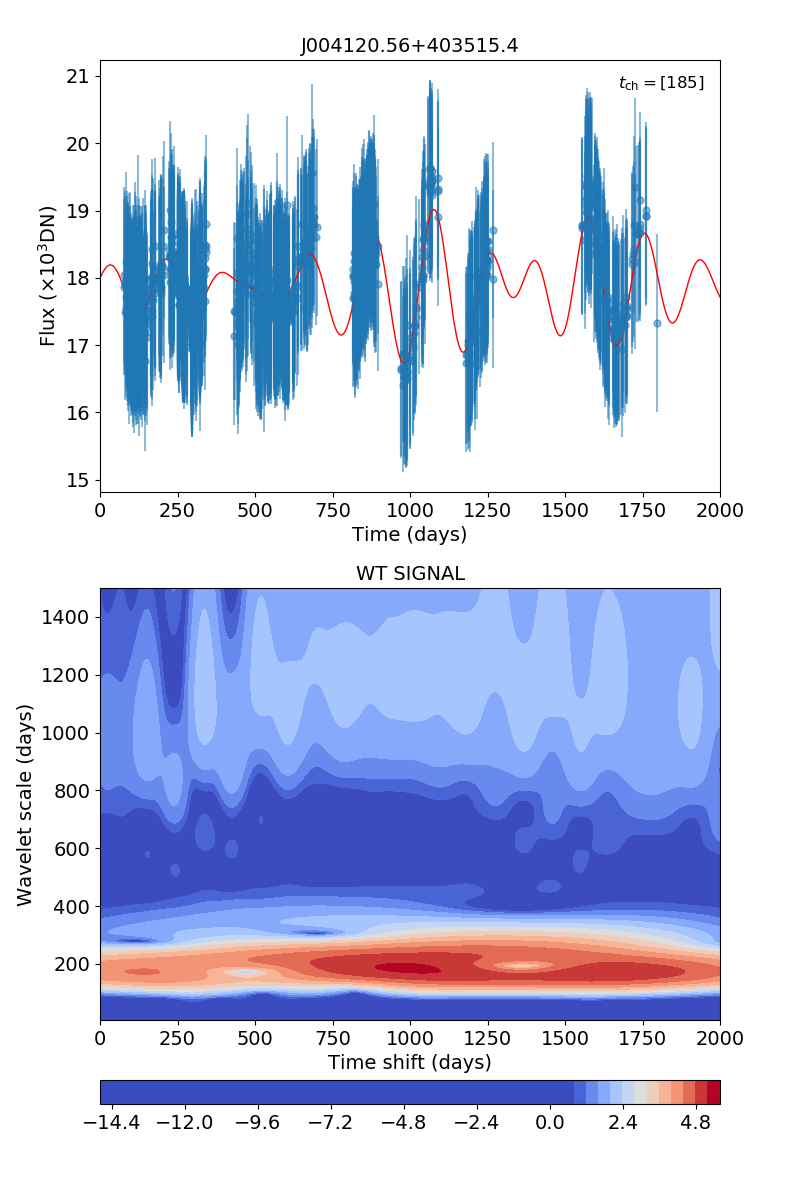}\hfill\includegraphics[width=60mm]{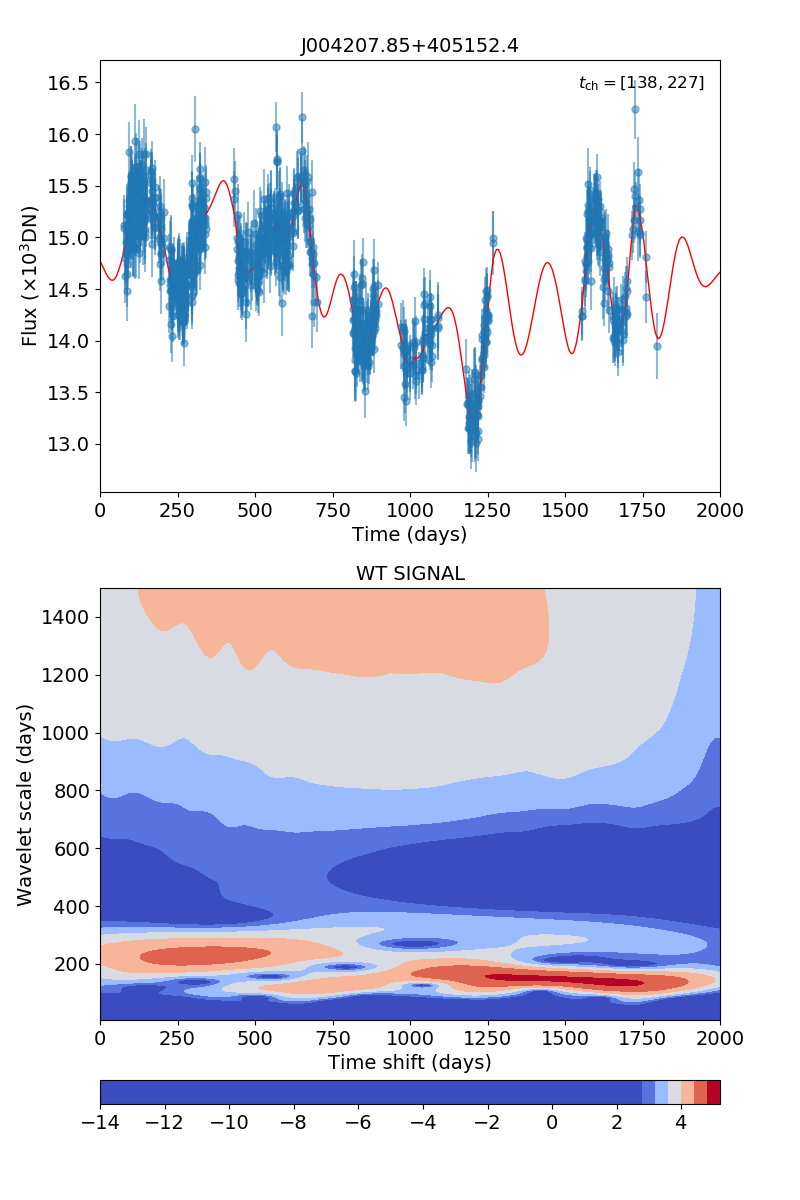}\hfill\includegraphics[width=60mm]{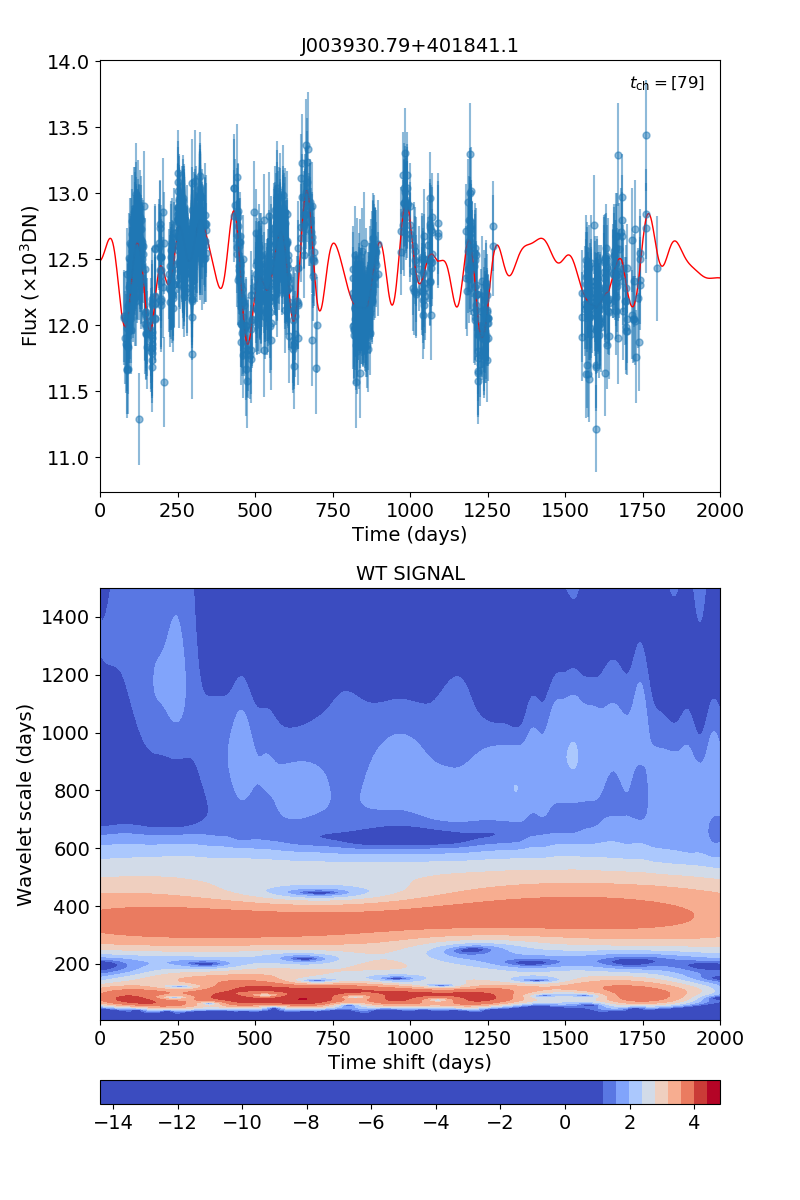}
\caption{Contd. for AI stars ({\it top}) and FI stars ({\it bottom}).}
\end{figure*}

\begin{figure*}
\ContinuedFloat
\includegraphics[width=60mm]{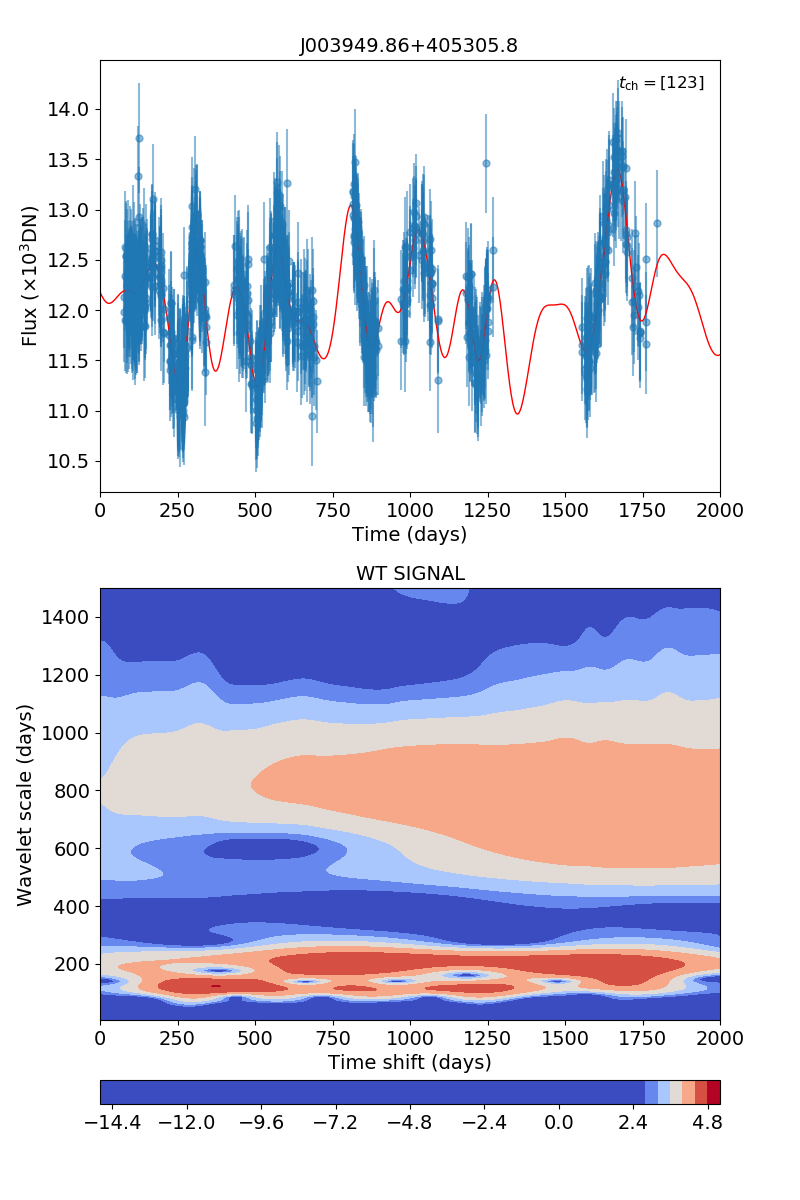}\hfill\includegraphics[width=60mm]{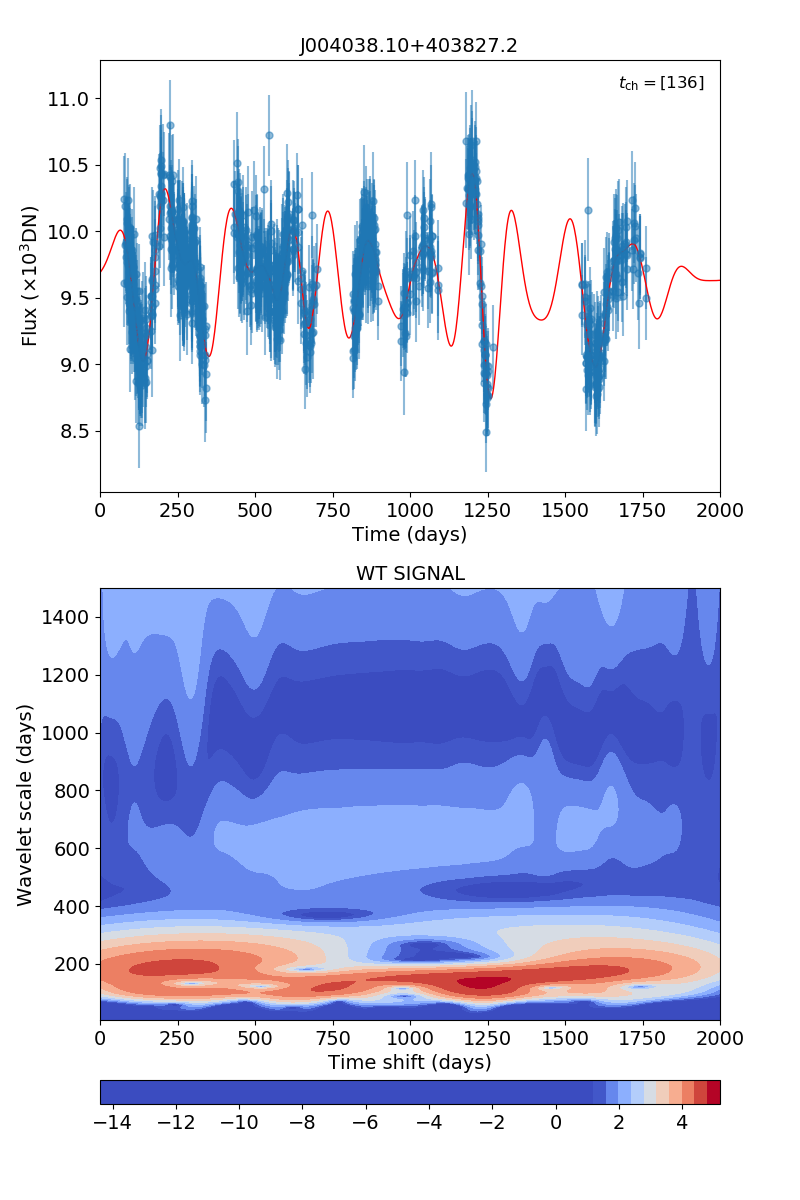}\hfill\includegraphics[width=60mm]{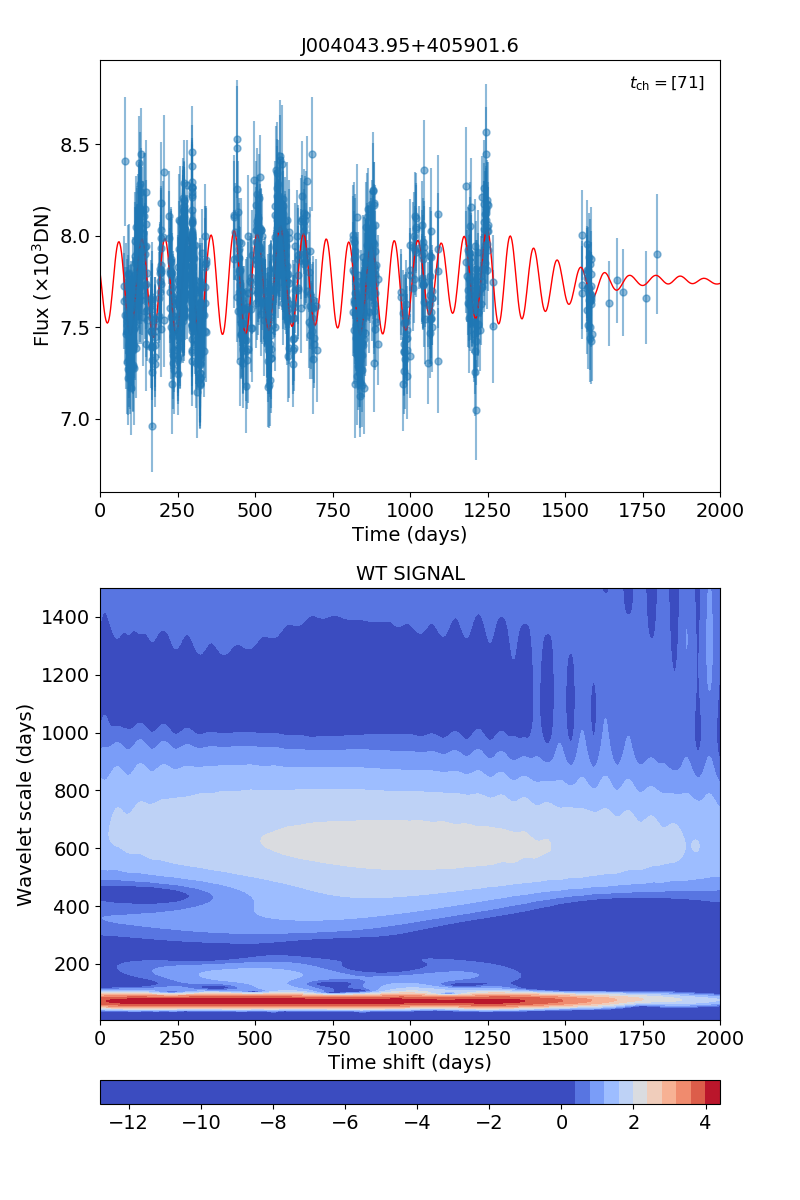}
\includegraphics[width=60mm]{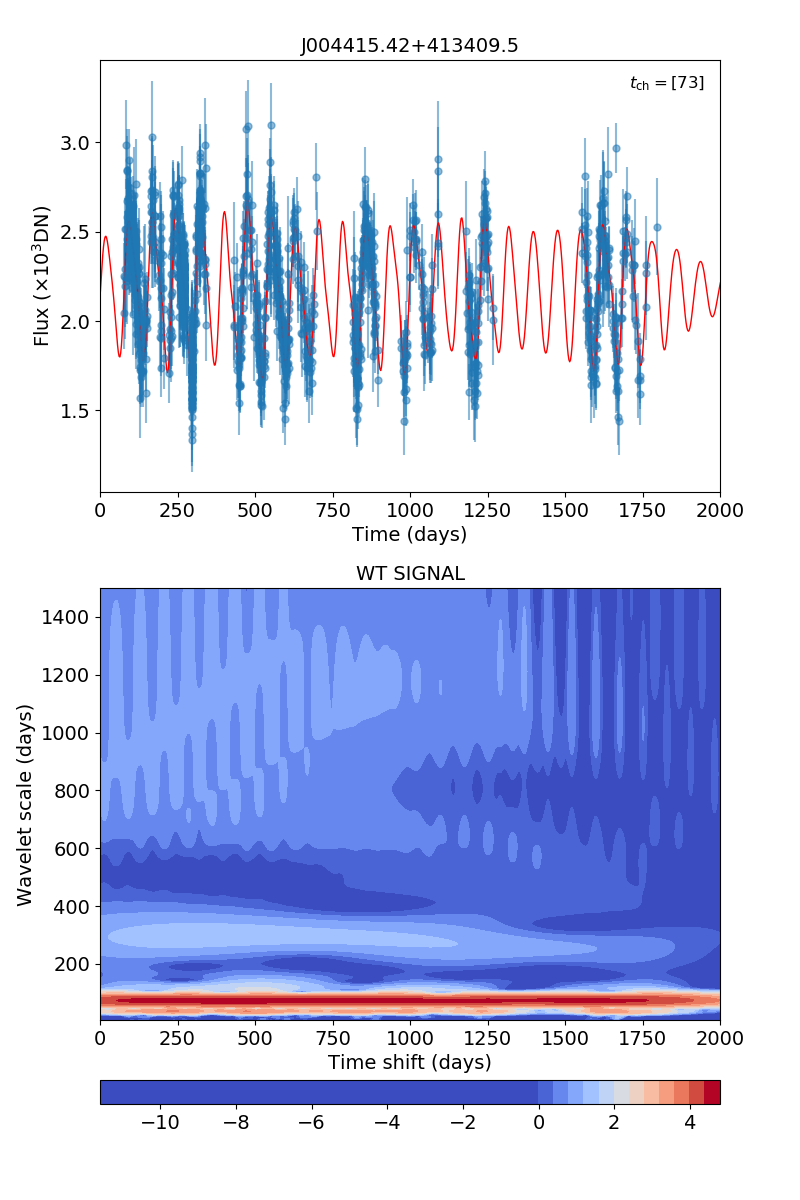}\hfill\includegraphics[width=60mm]{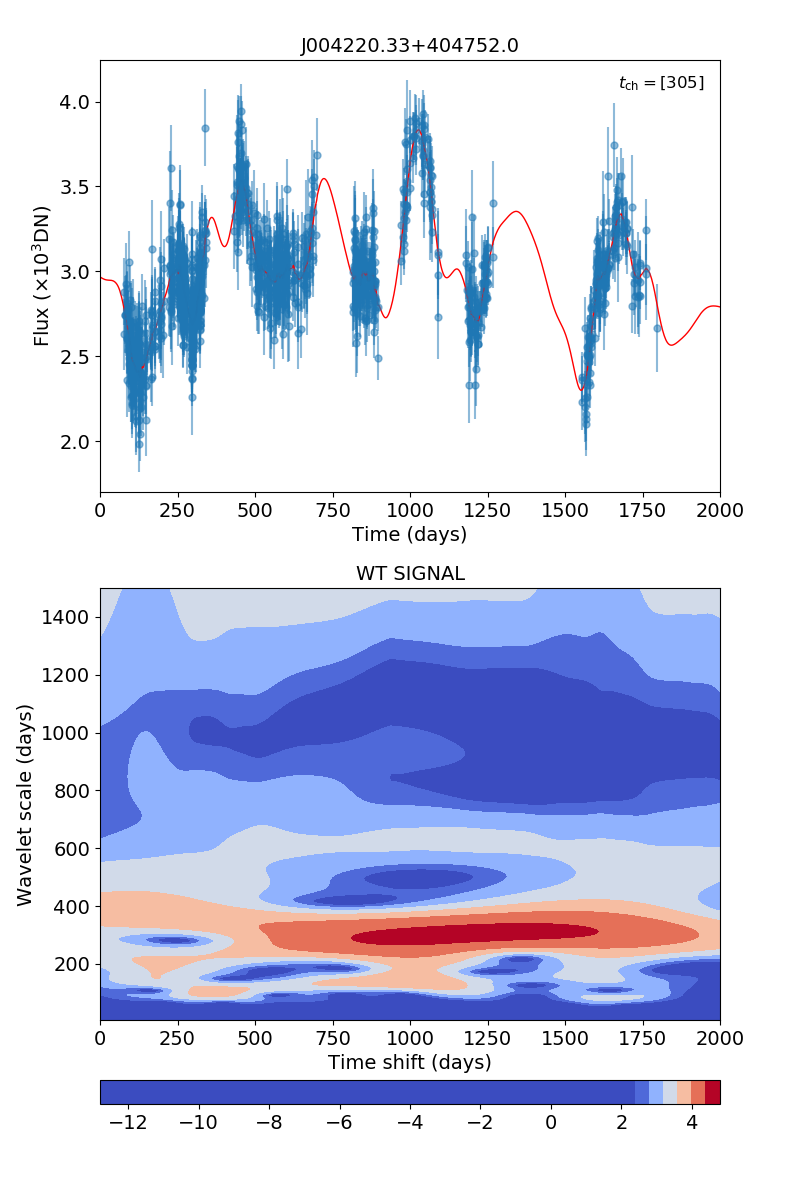}\hfill\includegraphics[width=60mm]{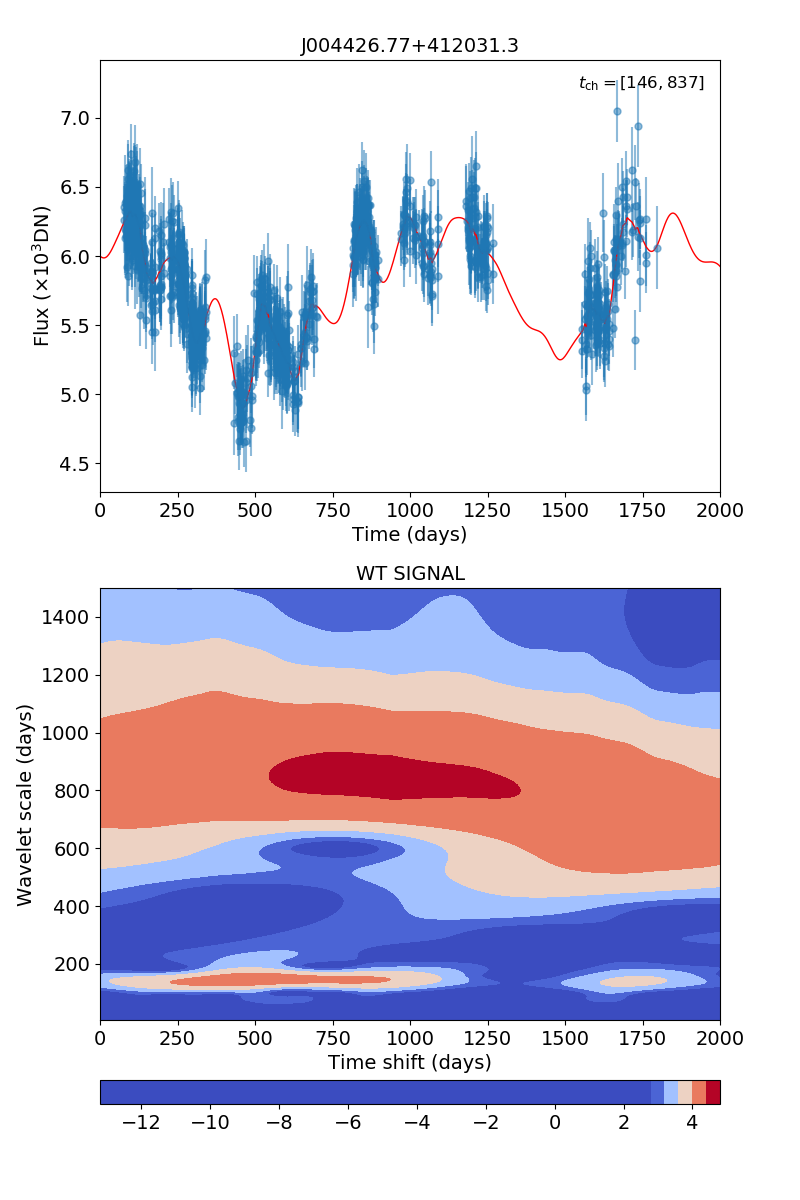}
\caption{Contd. for GI stars ({\it top}) and KI stars ({\it bottom}).}
\end{figure*}

\begin{figure*}[h]
\ContinuedFloat
\includegraphics[width=60mm]{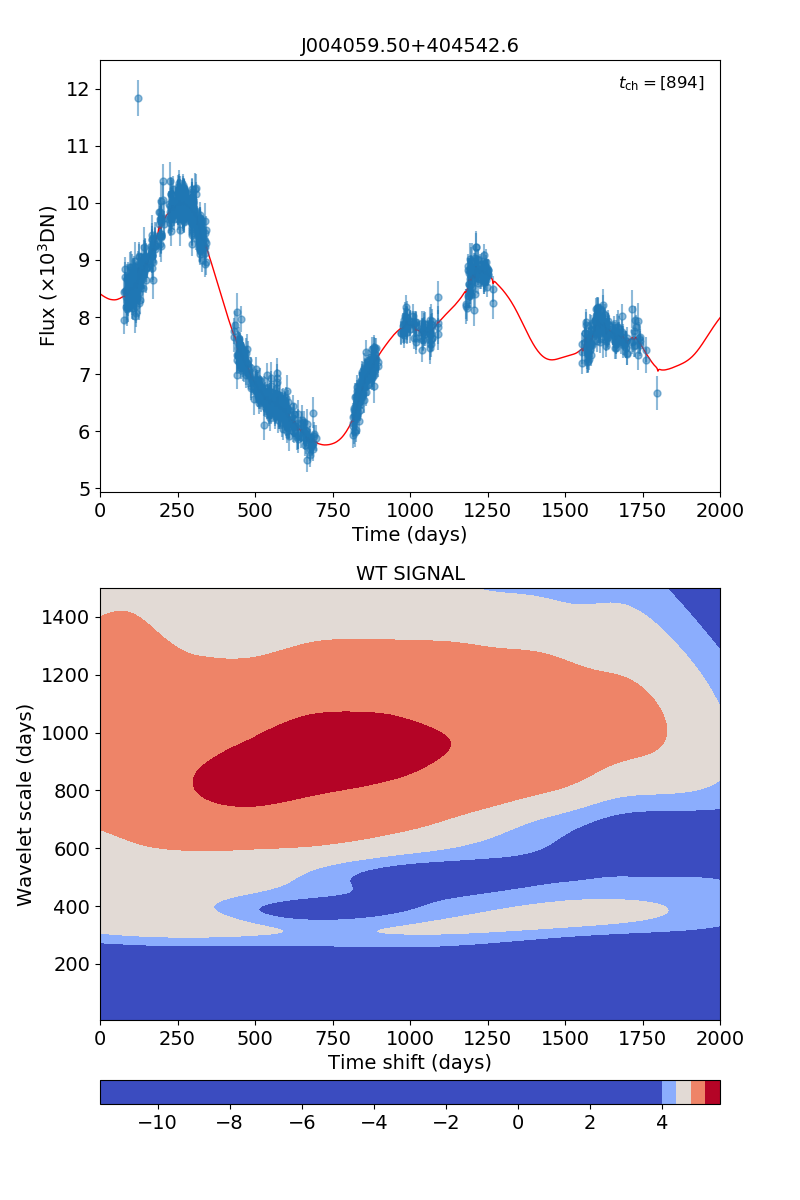}\hfill\includegraphics[width=60mm]{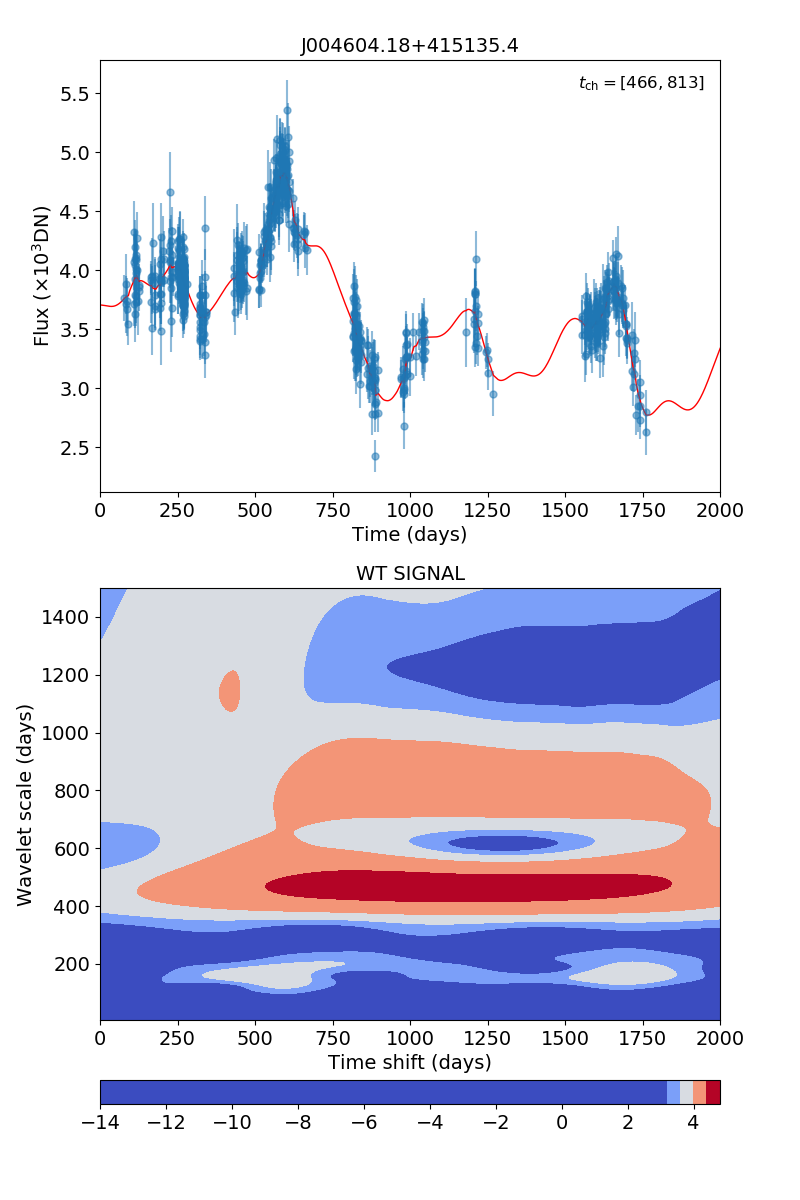}\hfill\includegraphics[width=60mm]{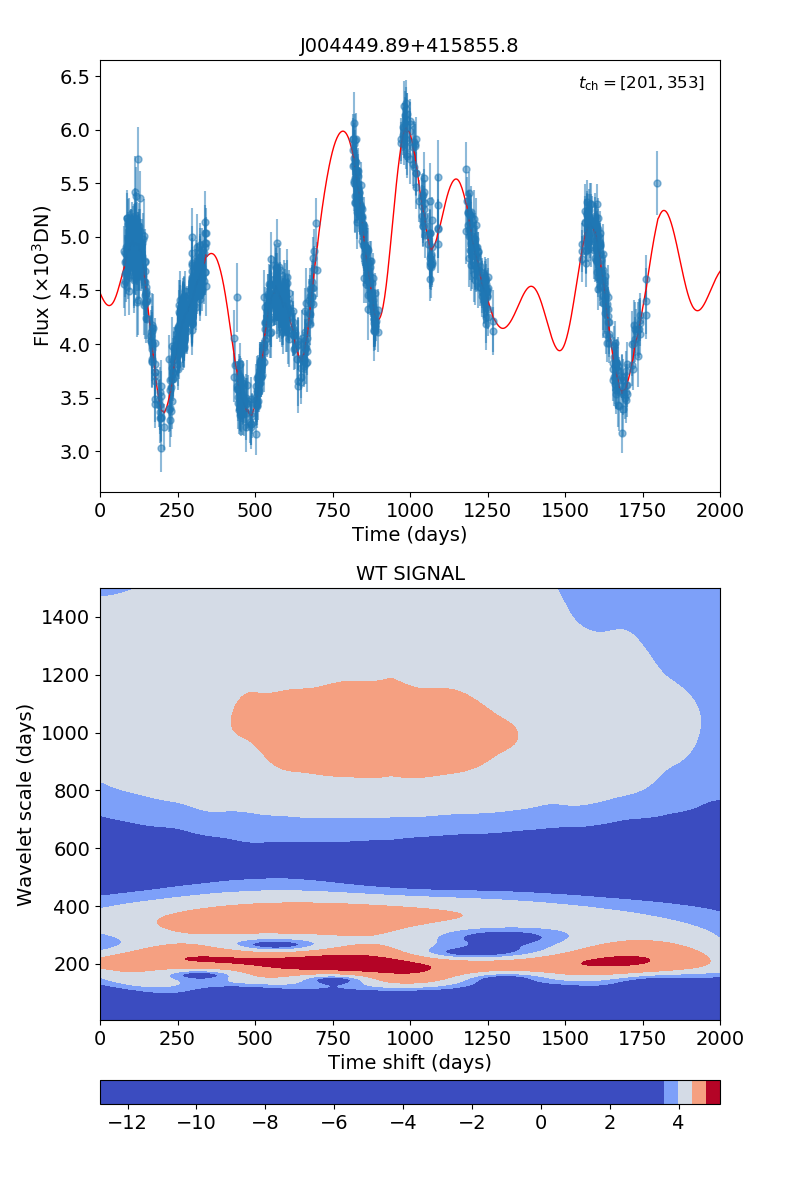}
\includegraphics[width=60mm]{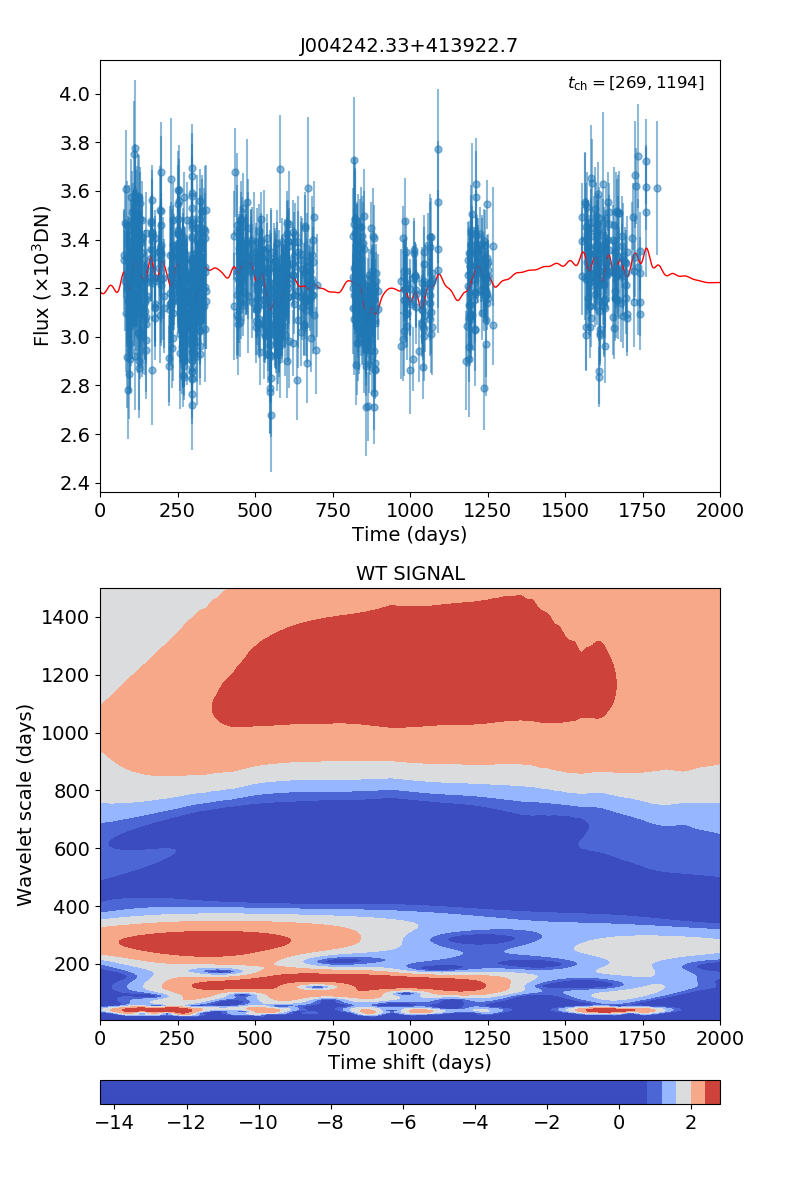}\hfill\includegraphics[width=60mm]{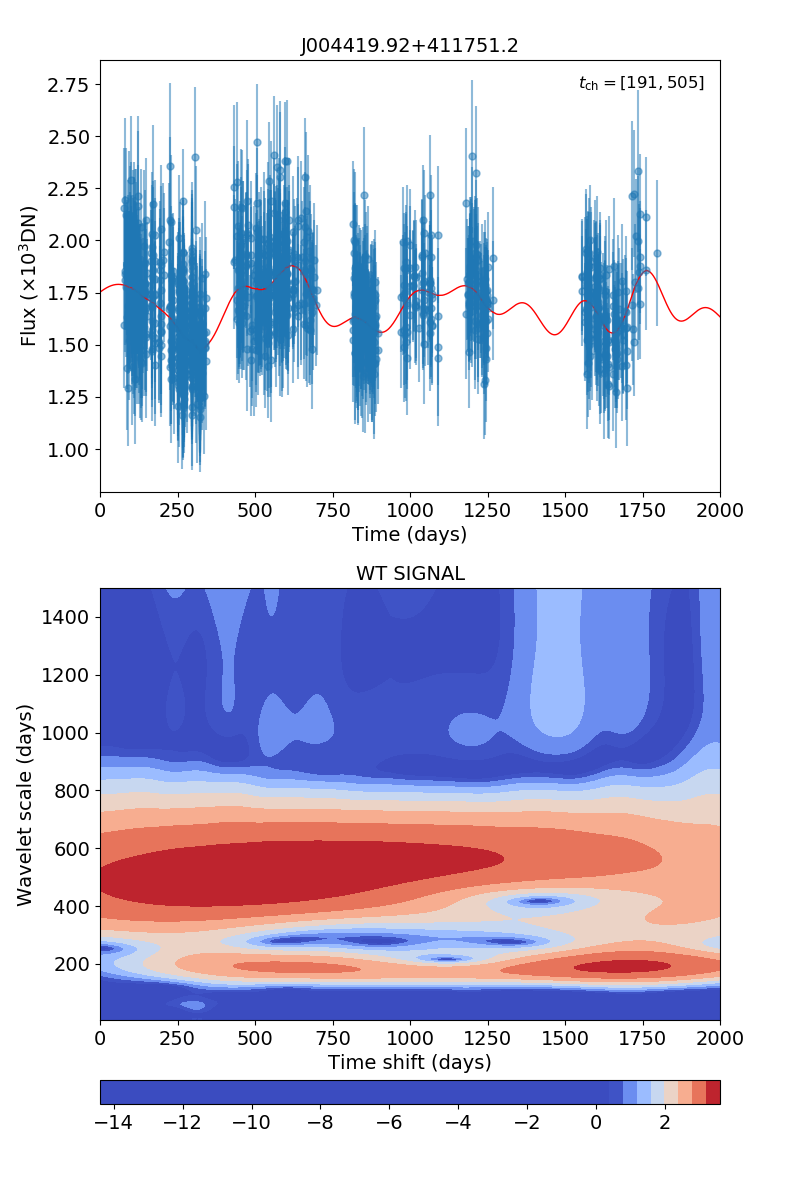}\hfill\includegraphics[width=60mm]{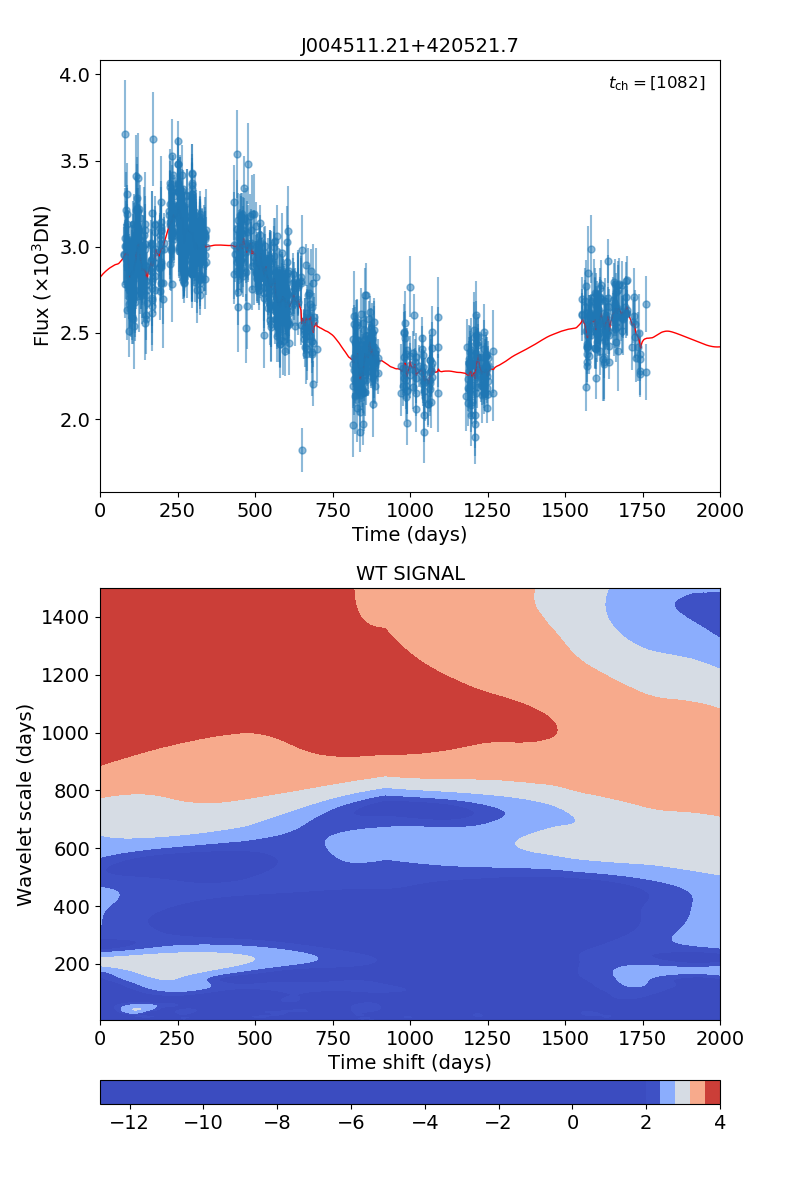}
\caption{Contd. for M stars ({\it top}) and WR stars ({\it bottom}).}
\end{figure*}

\begin{figure*}
\includegraphics[width=60mm]{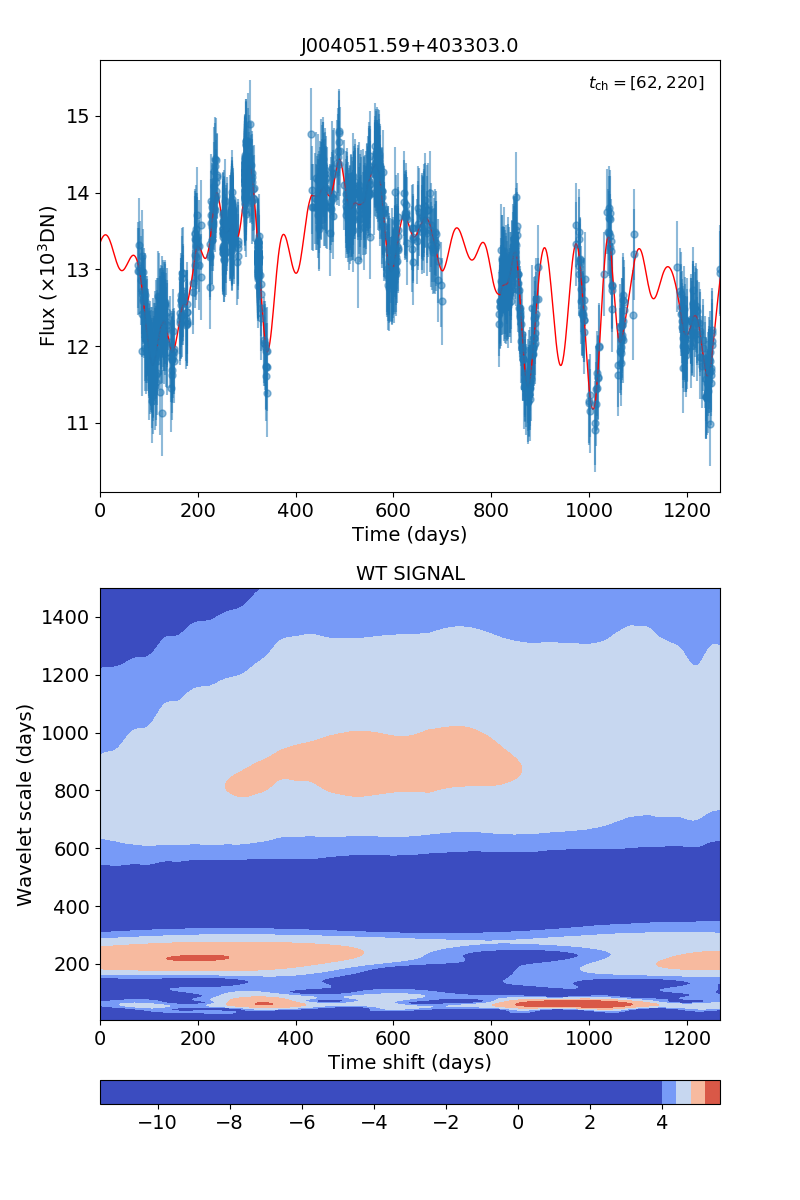}\hfill\includegraphics[width=60mm]{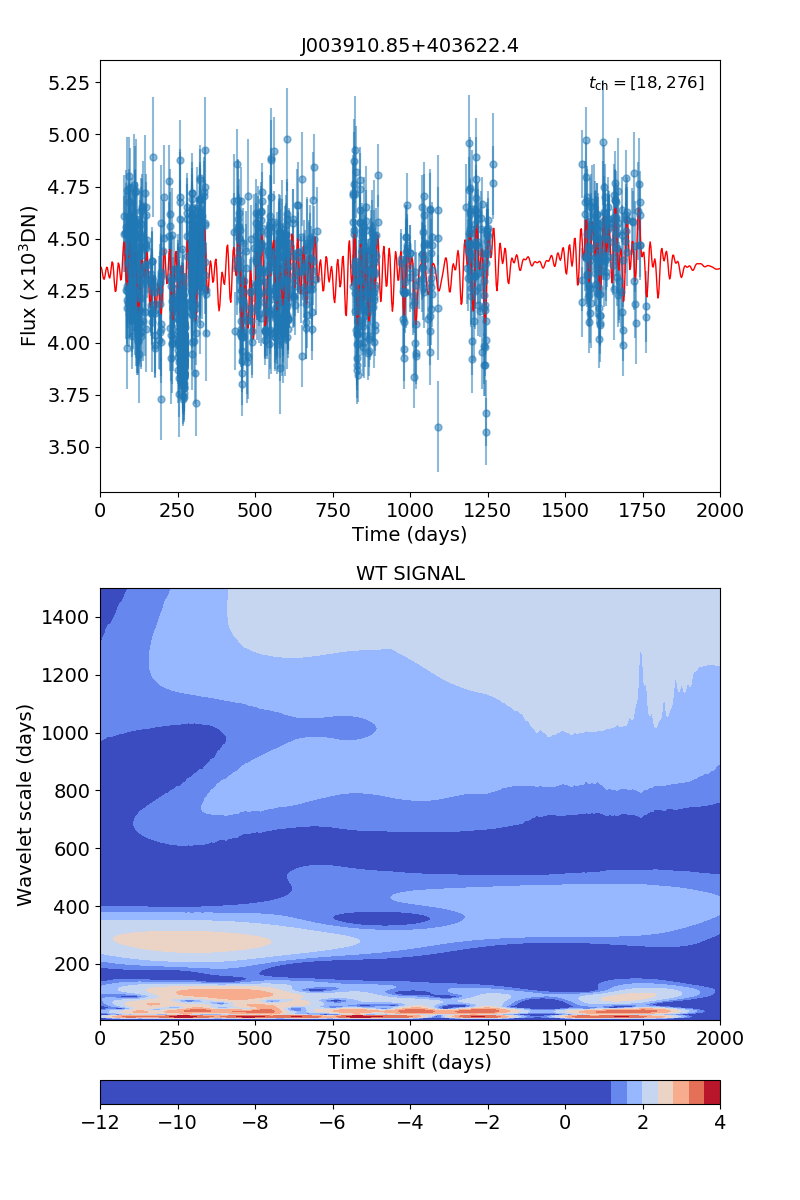}\hfill\includegraphics[width=60mm]{{LBVs/J004302.52+414912.4_lc_wt}.png}
\includegraphics[width=60mm]{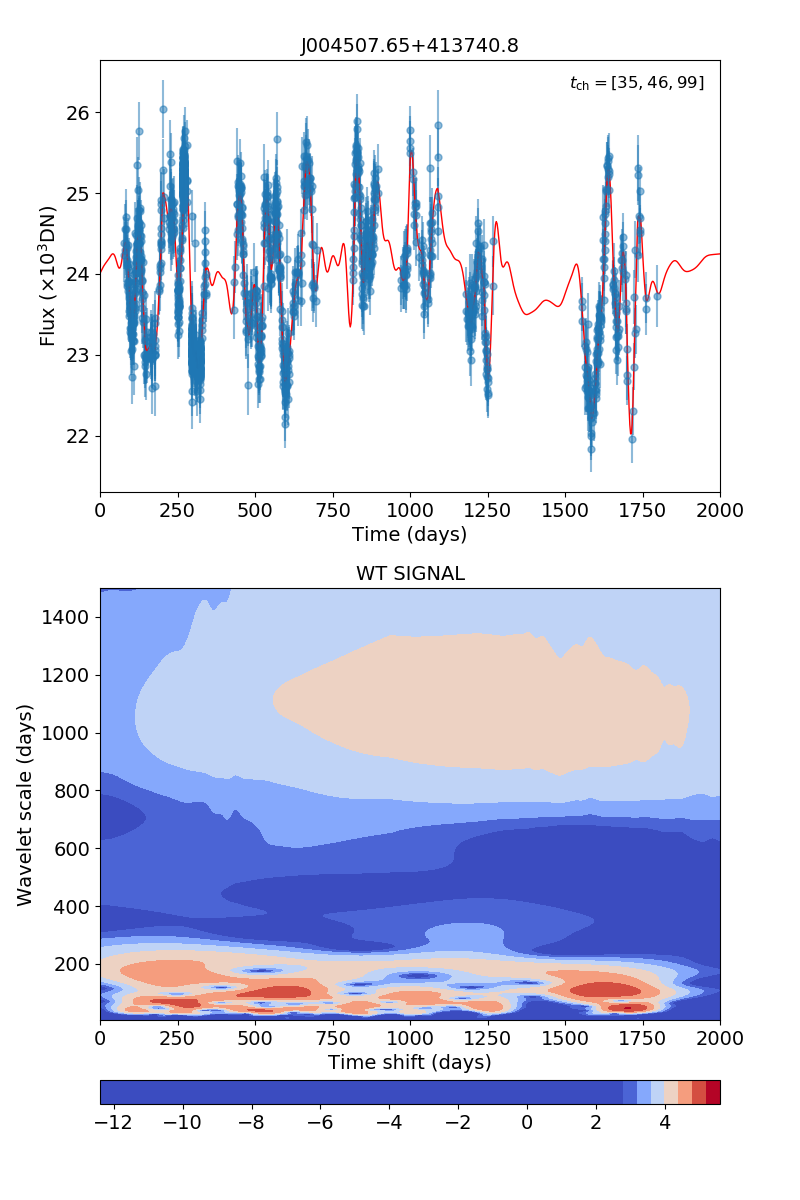}\hfill\includegraphics[width=60mm]{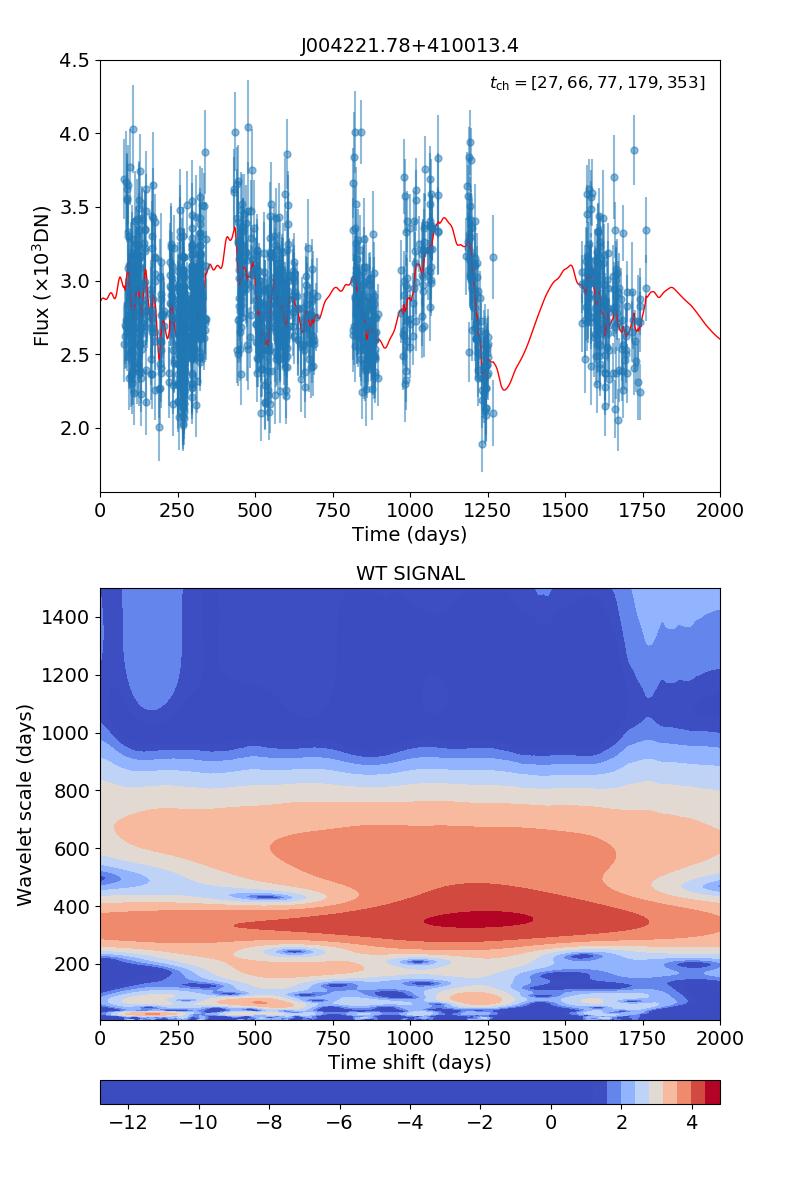}\hfill\includegraphics[width=60mm]{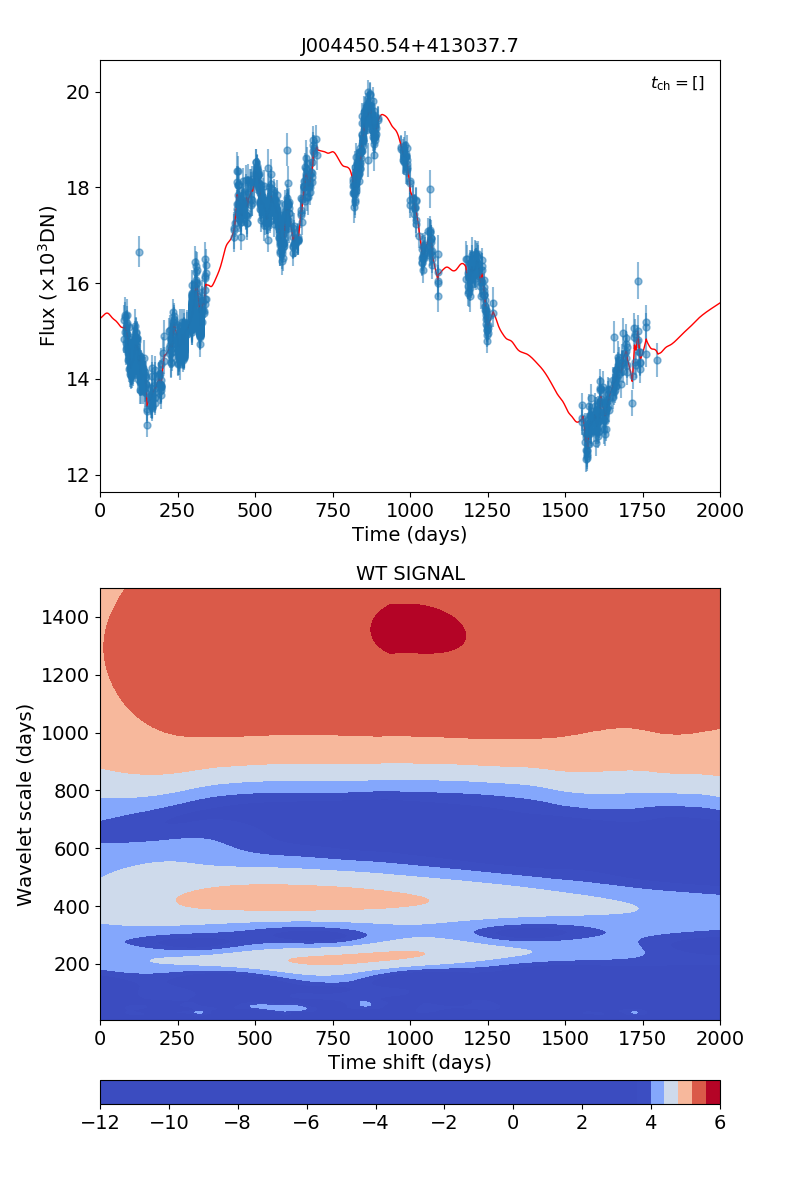}
\caption{iPTF light curves along with the corresponding wavelet transform maps for known and candidate LBVs in M31 from MNS16. The red curve in each panel with the observed light curve shows the reconstruction (see text). The ID of the star from MNS16 is shown on top of each plot. The time axis in the light curve plots is with respect to a reference value of MJD 56000.} 
\label{fig:lbvs}
\end{figure*}

\begin{figure*}
\ContinuedFloat
\includegraphics[width=60mm]{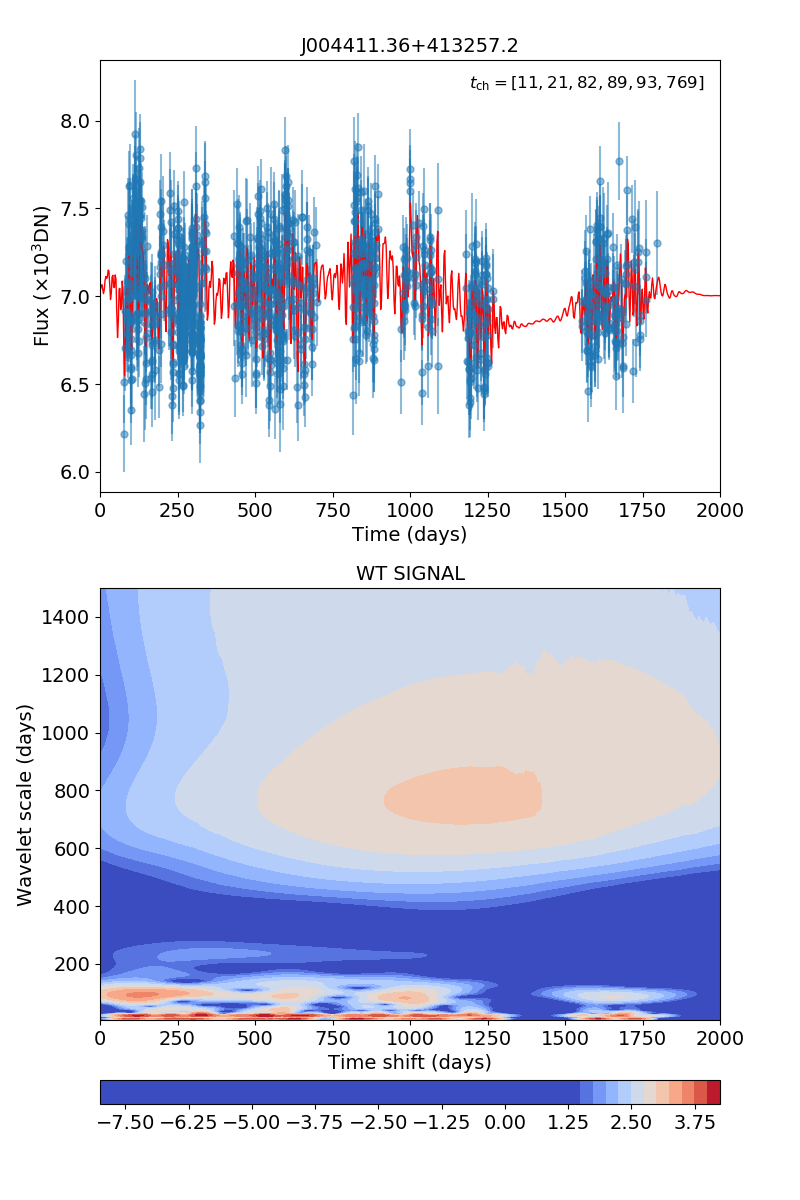}\hfill\includegraphics[width=60mm]{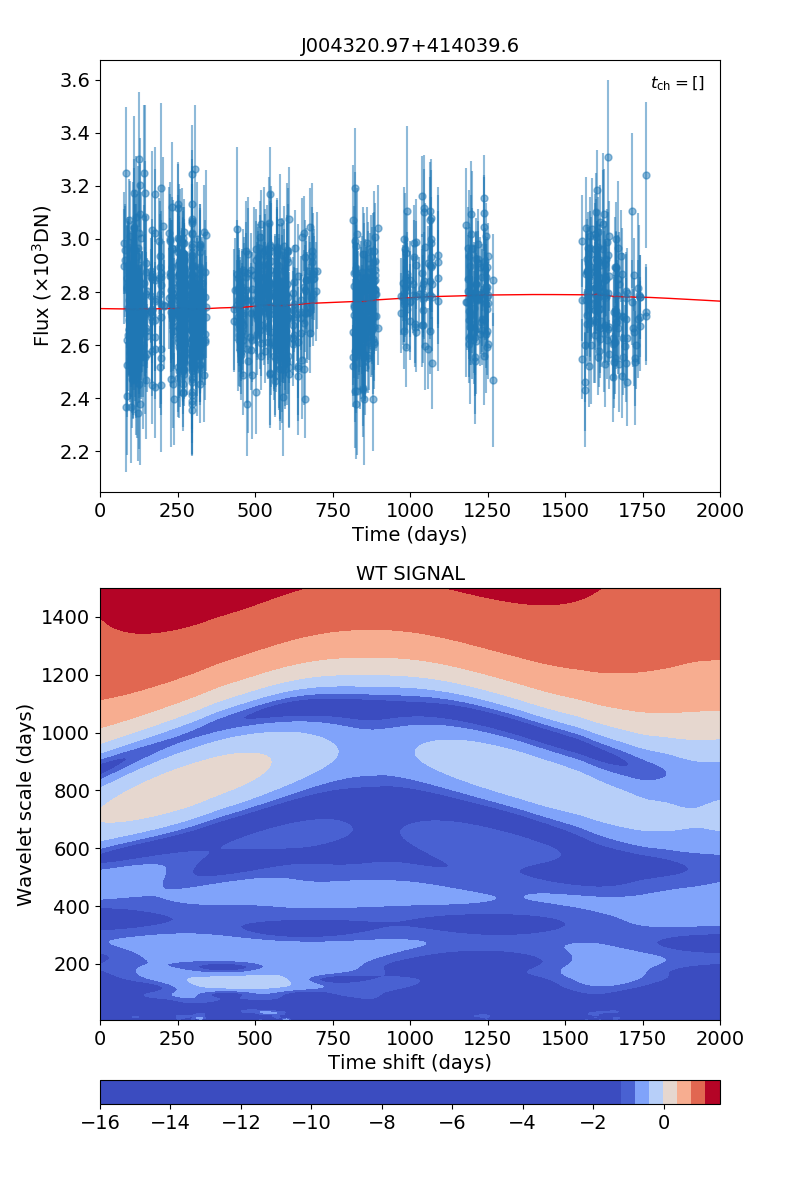}\hfill\includegraphics[width=60mm]{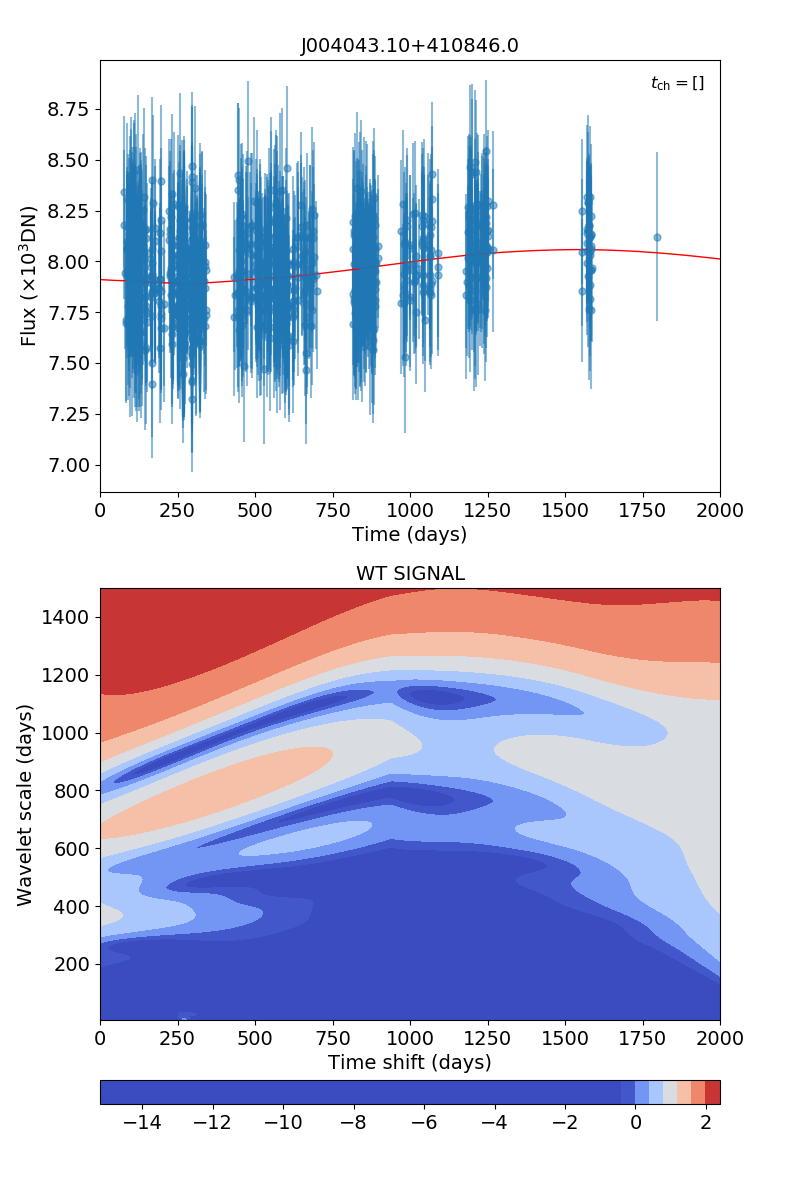}
\includegraphics[width=60mm]{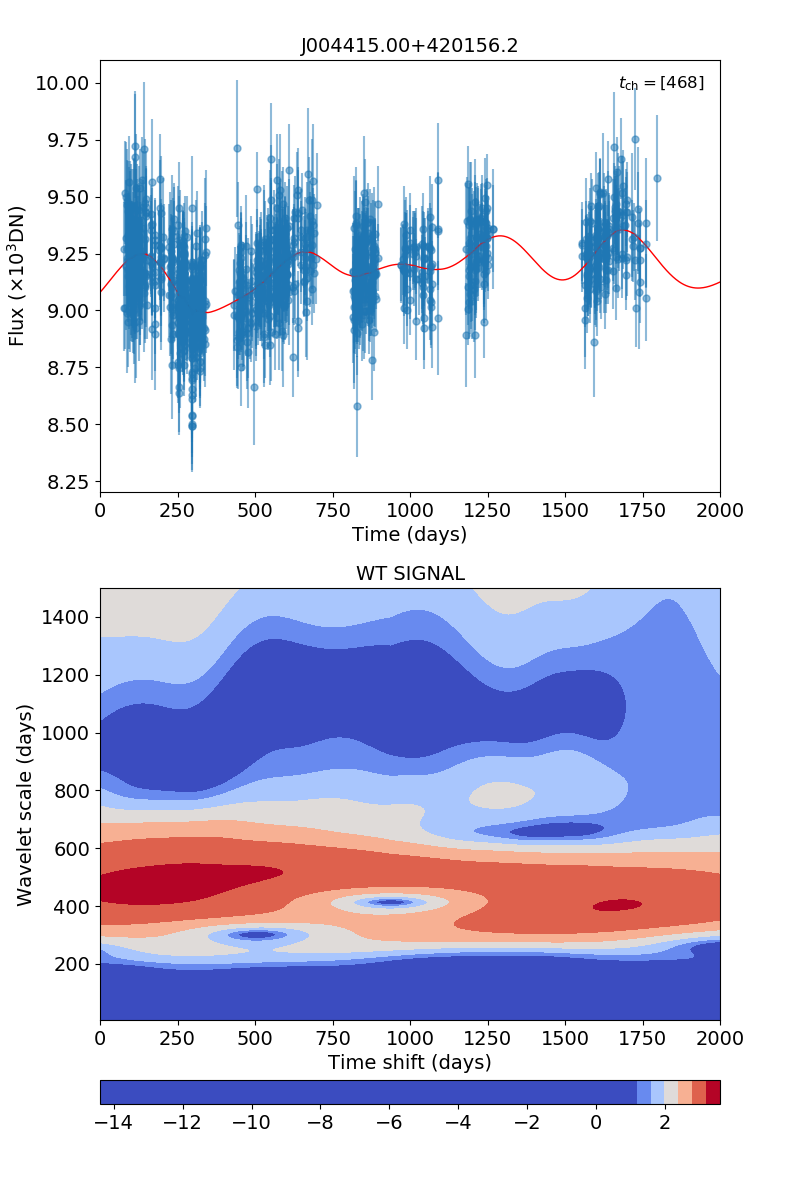}\hfill\includegraphics[width=60mm]{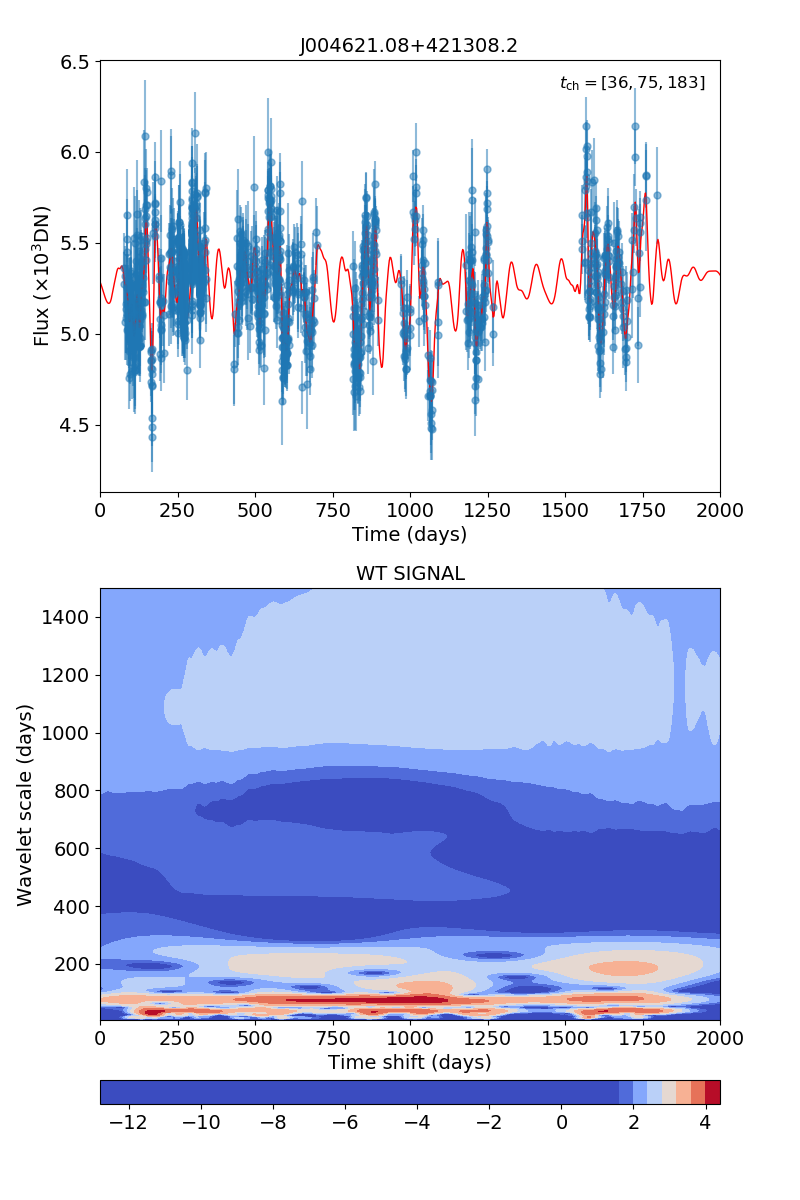}\hfill\includegraphics[width=60mm]{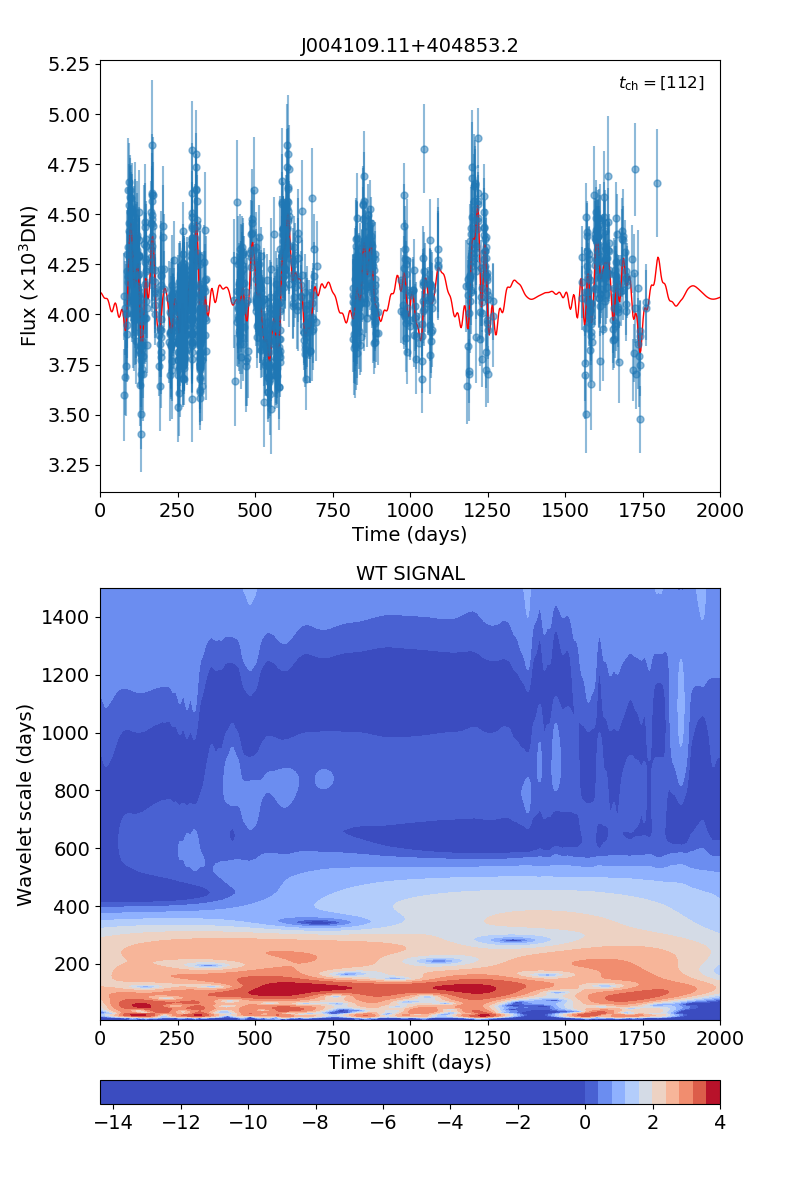}
\caption{Contd.} 
\end{figure*}

\begin{figure*}
\ContinuedFloat
\includegraphics[width=60mm]{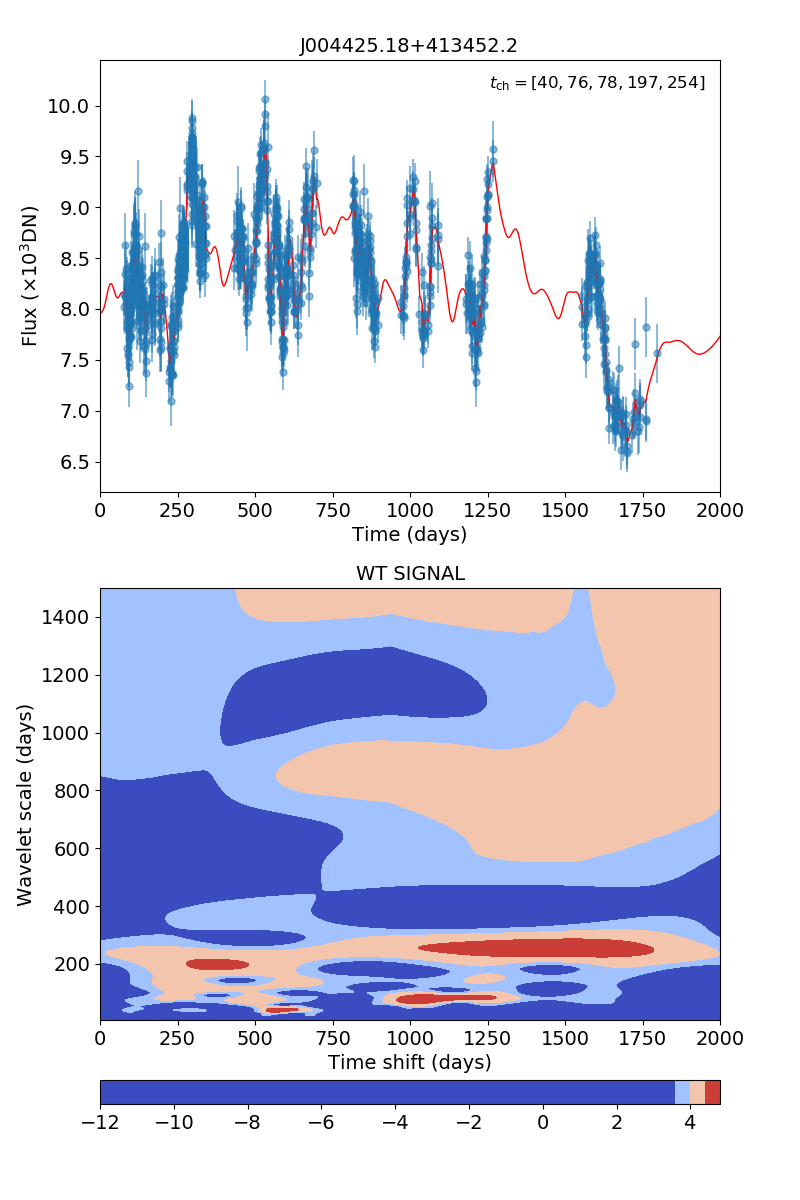}\hfill\includegraphics[width=60mm]{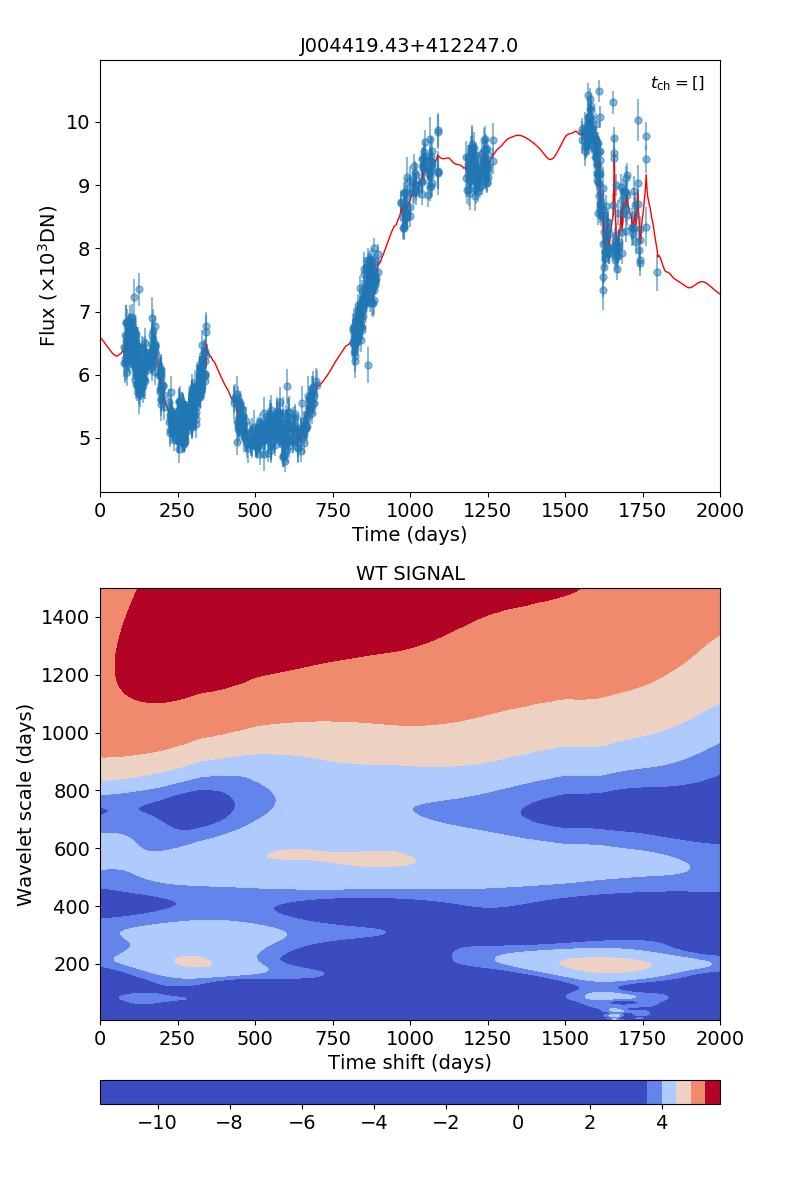}\hfill\includegraphics[width=60mm]{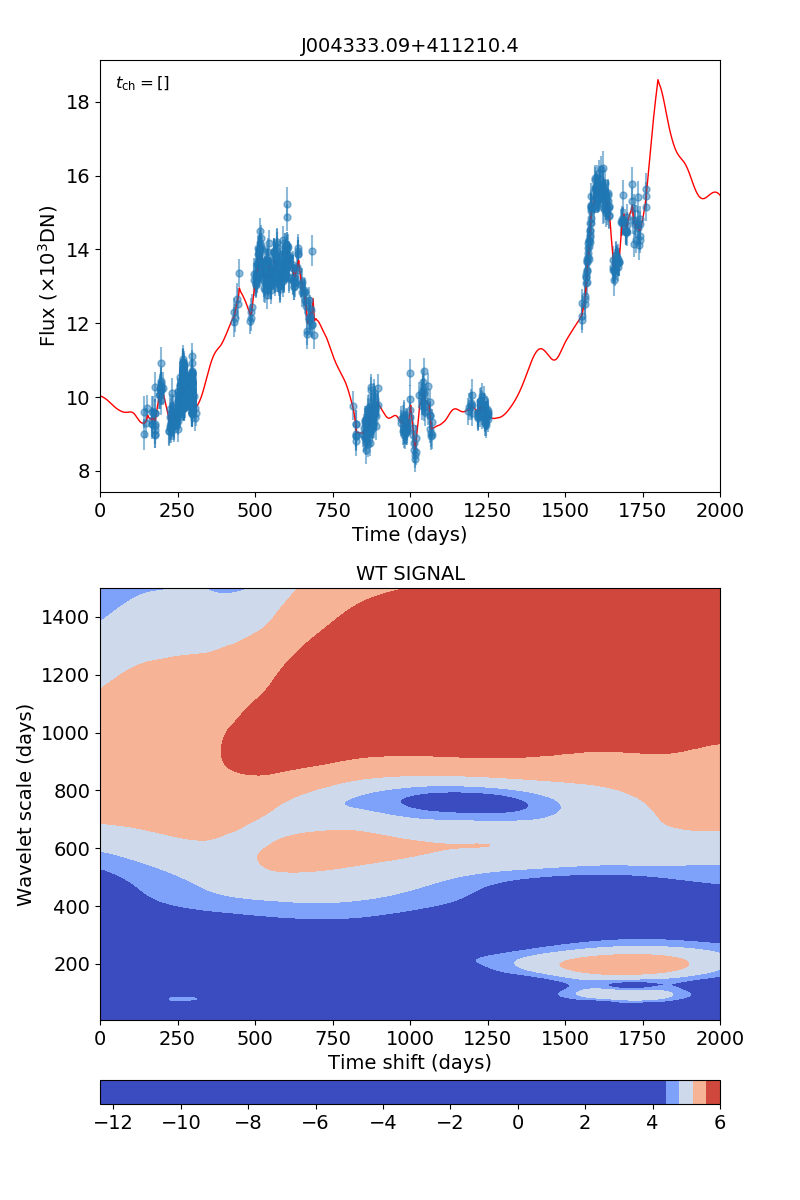}
\includegraphics[width=60mm]{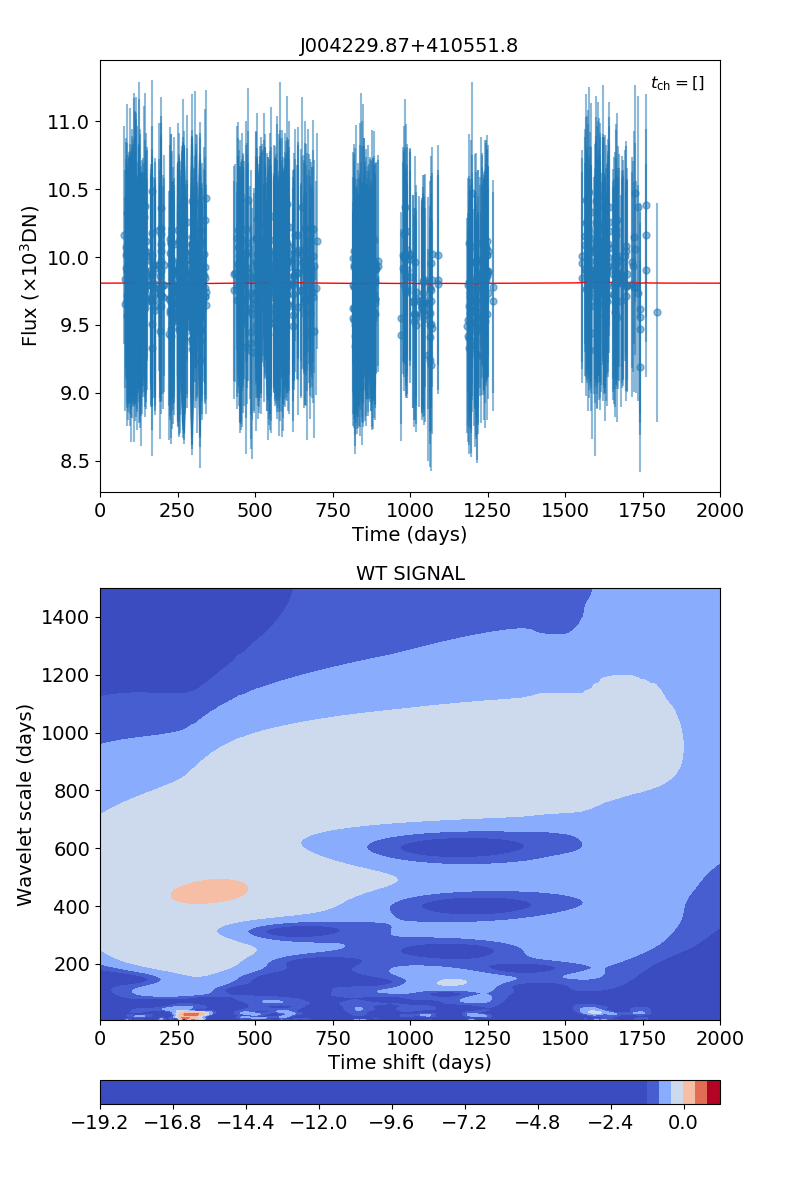}\hfill\includegraphics[width=60mm]{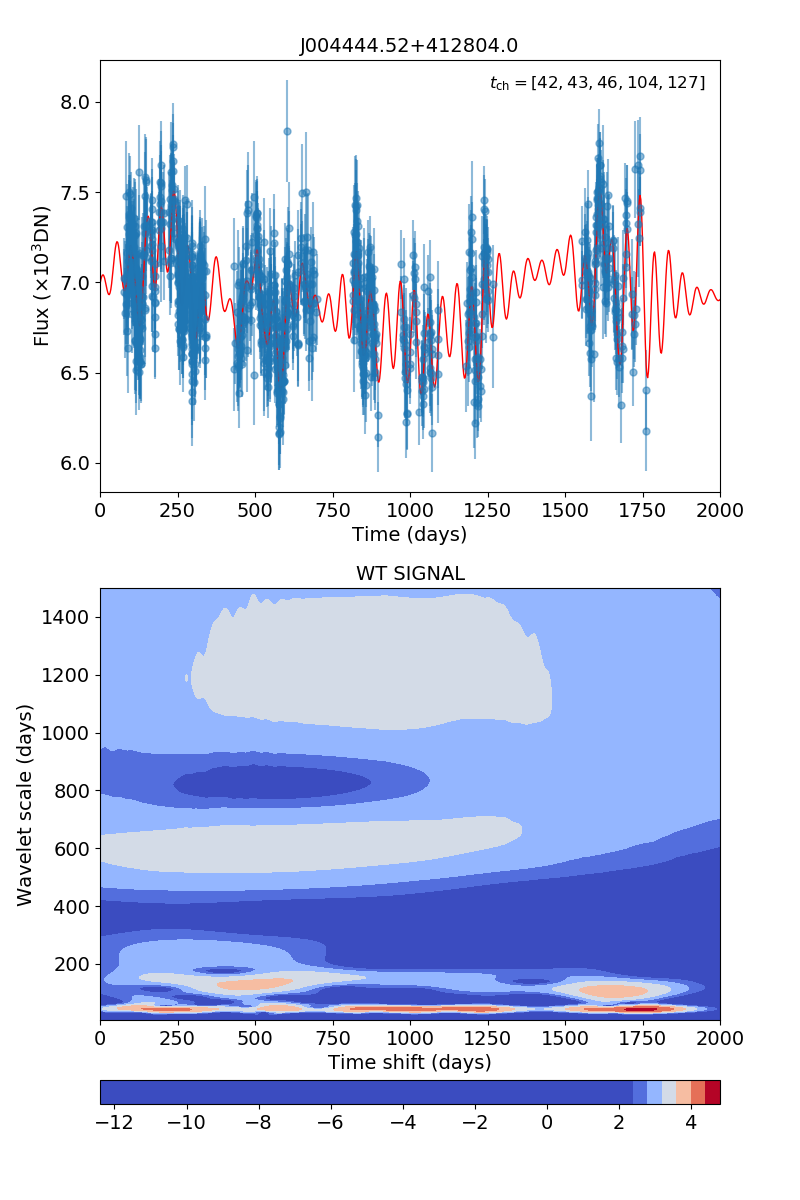}\hfill\includegraphics[width=60mm]{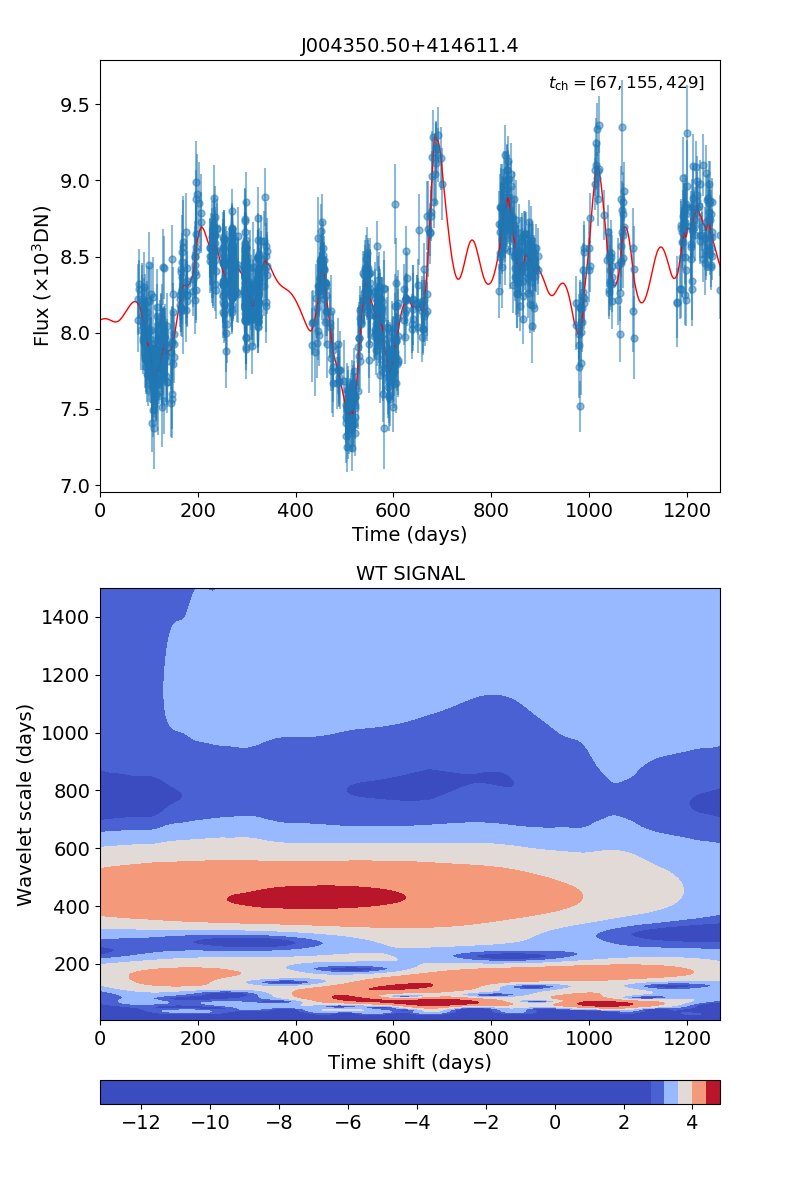}
\caption{Contd.} 
\end{figure*}

\begin{figure*}
\ContinuedFloat
\includegraphics[width=60mm]{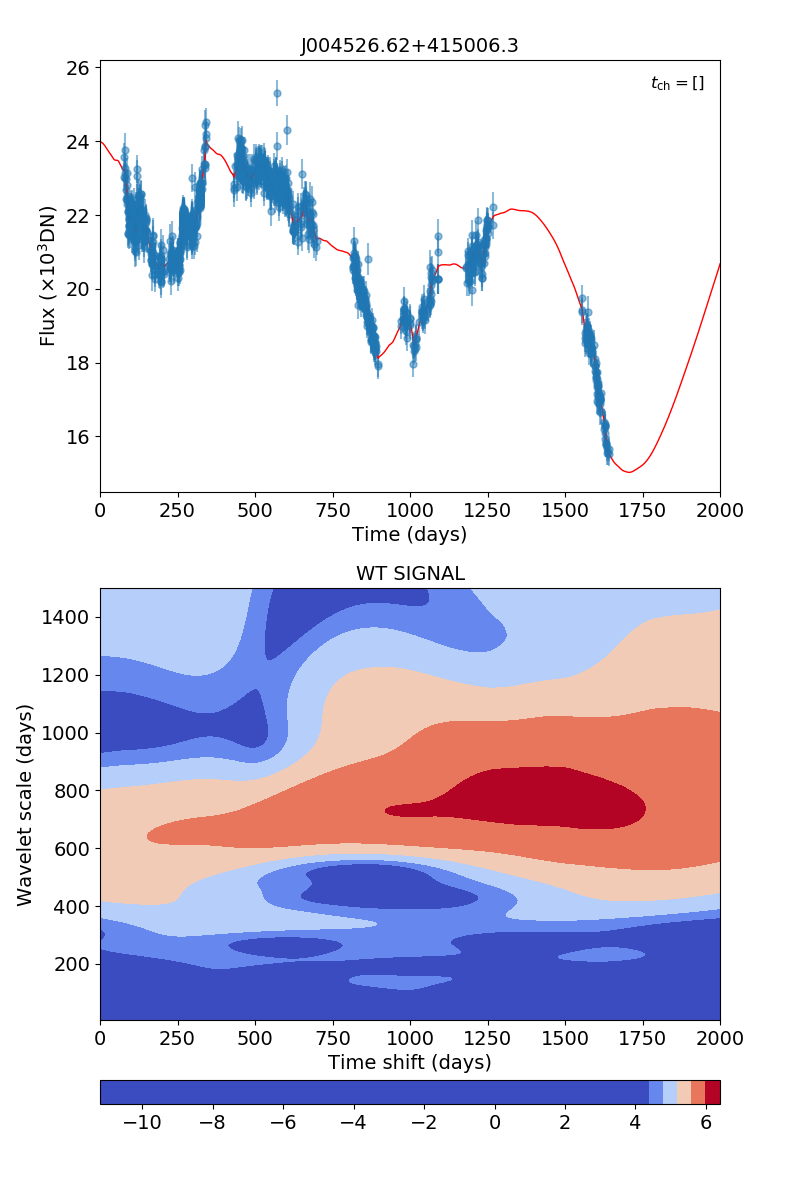}\hfill\includegraphics[width=60mm]{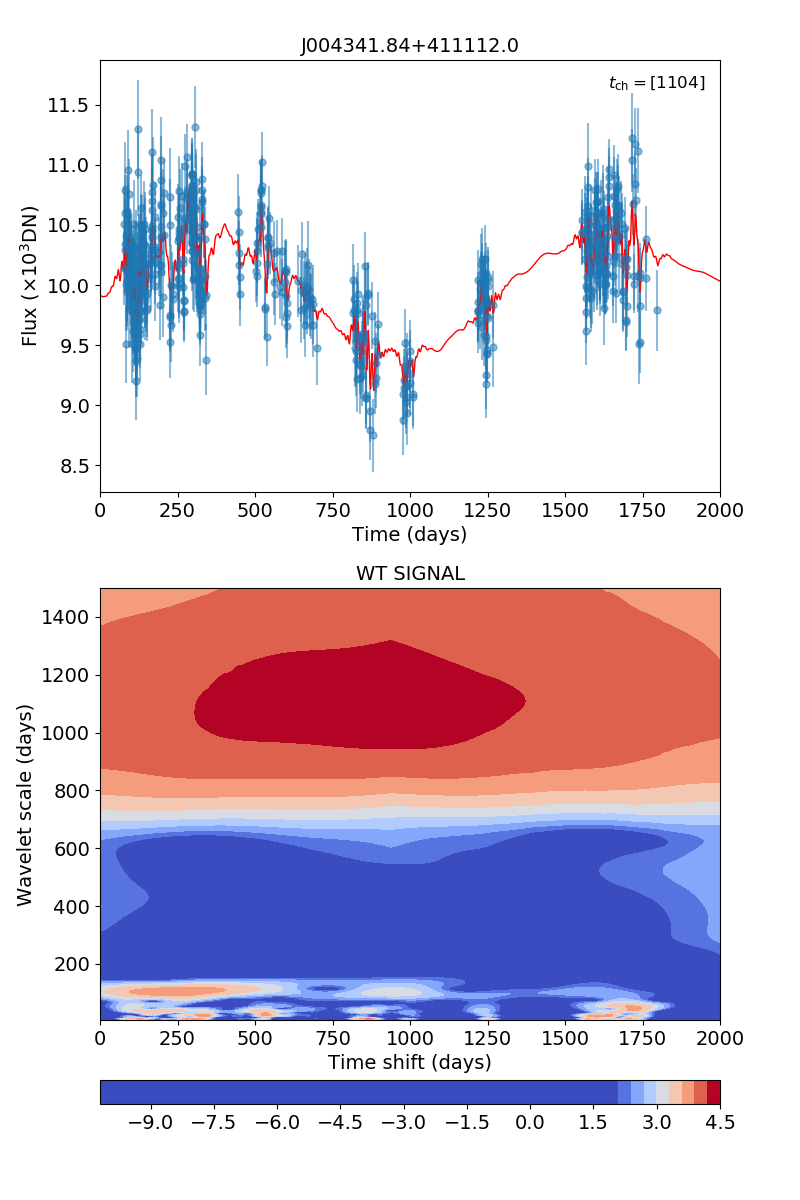}\hfill\includegraphics[width=60mm]{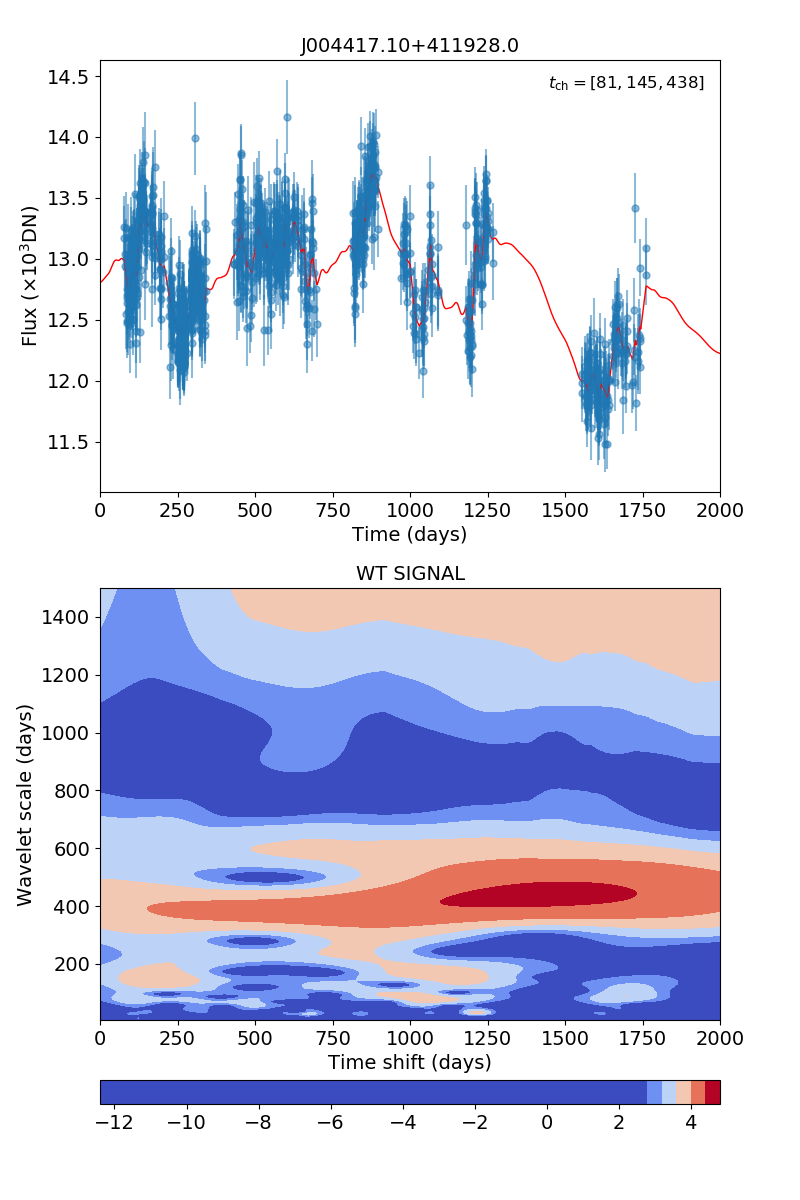}
\caption{Contd.} 
\end{figure*}

\begin{figure*}
\includegraphics[width=60mm]{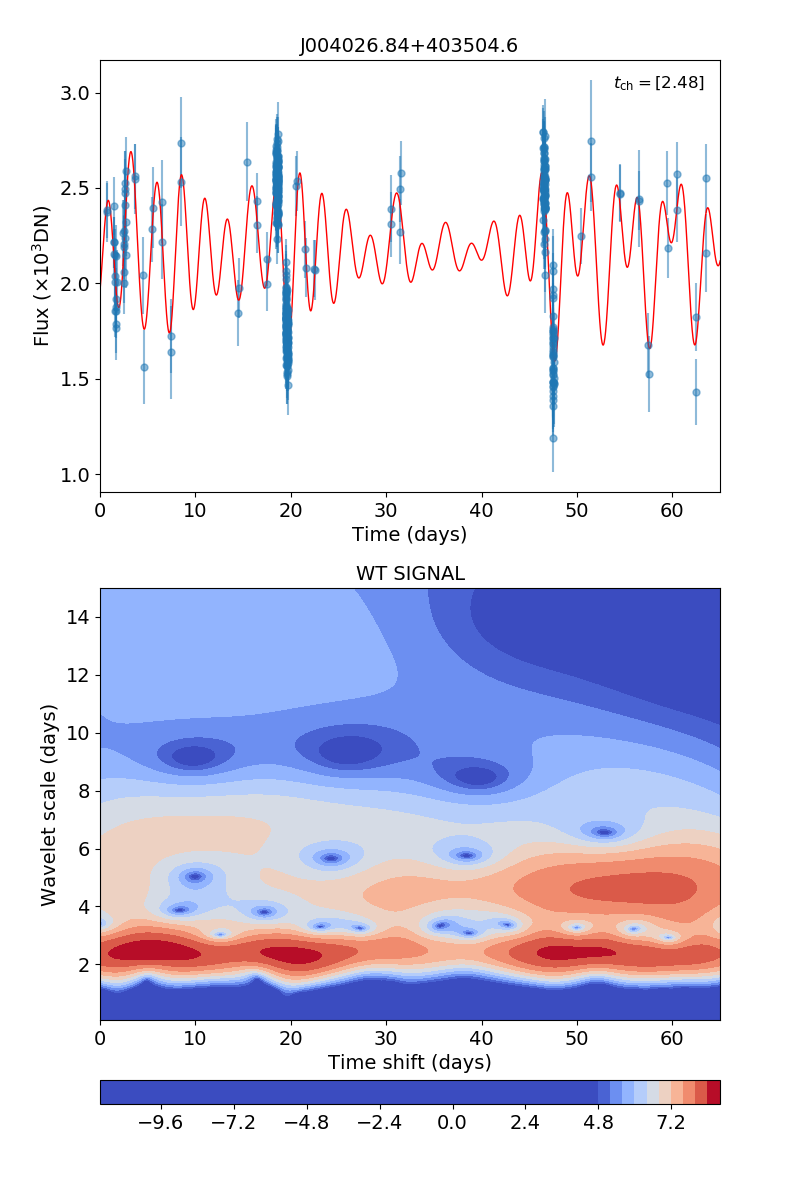}\hfill\includegraphics[width=60mm]{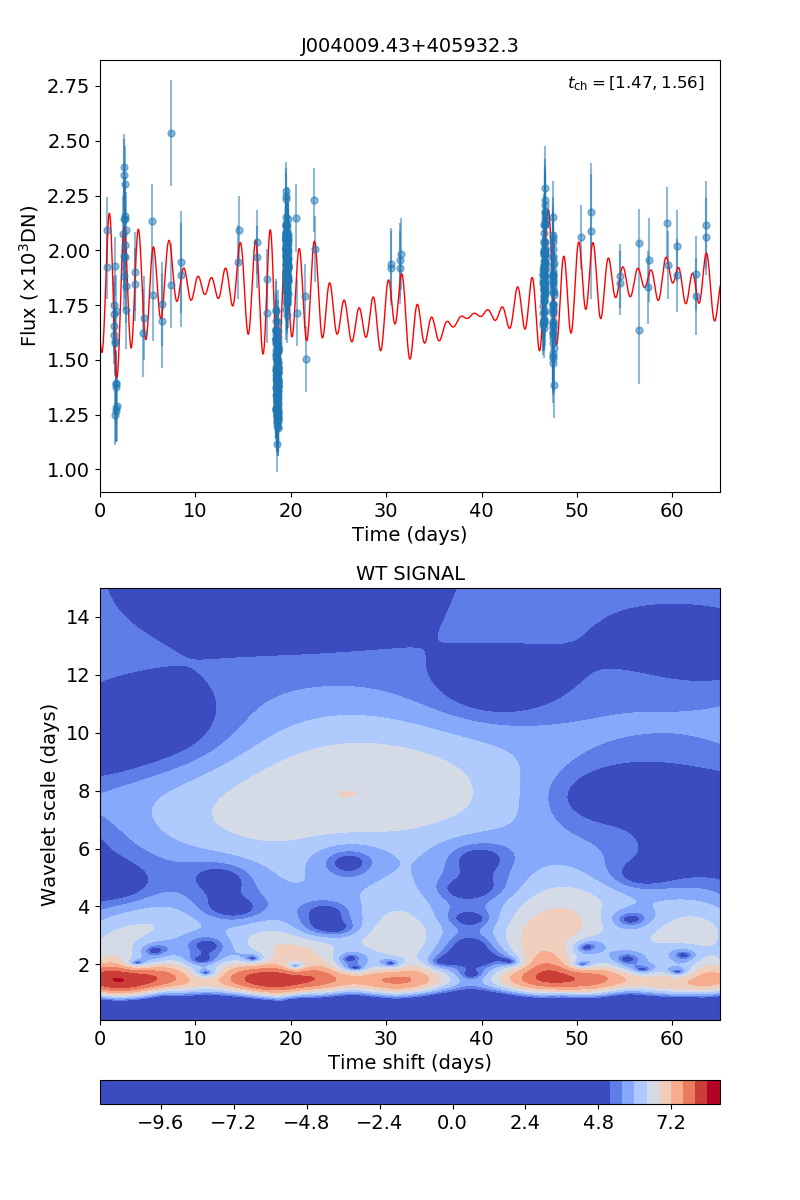}\hfill\includegraphics[width=60mm]{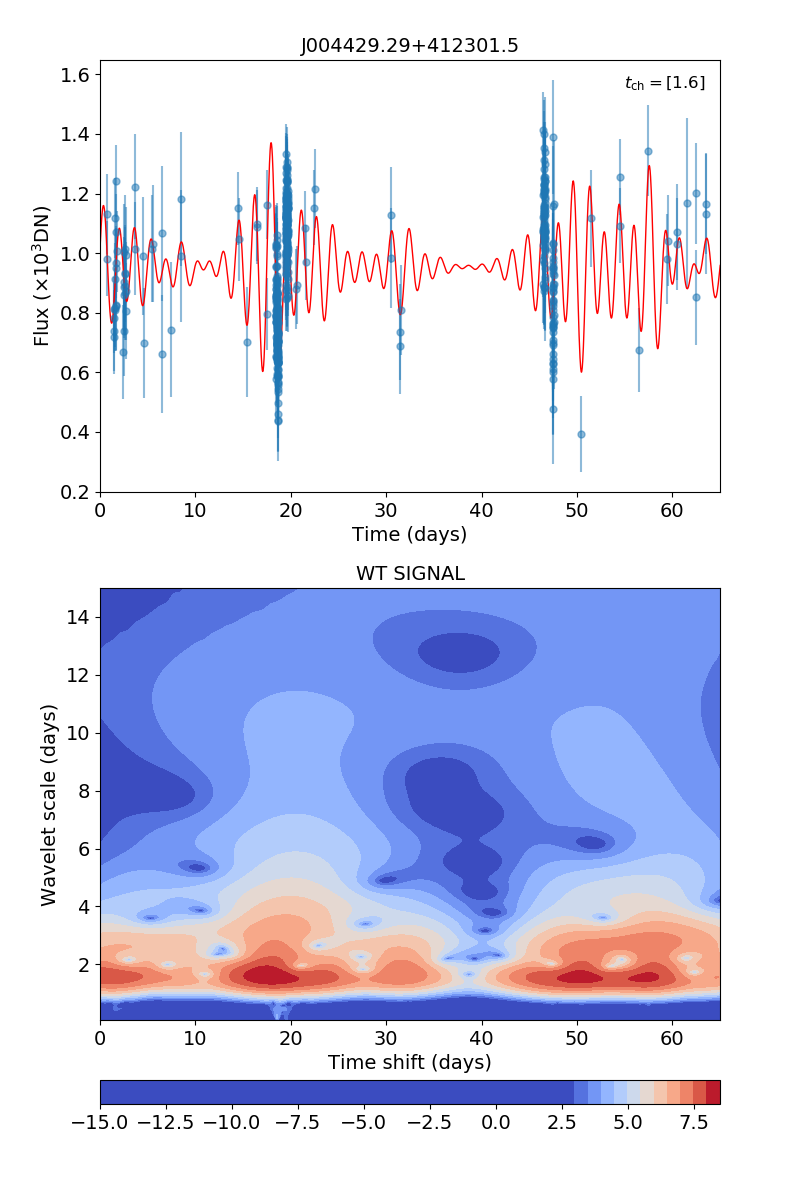}
\caption{iPTF light curves along with the corresponding wavelet transform maps for stars with significant timescales in the high-cadence block. The ID of the star from MNS16 is shown on top of each plot and the spectral types of the stars (going from left to right) are O7+O9f:, ON9.7Iab, and O8V. The time axis in the light curve plots is with respect to a reference value of MJD 56250.621783. The wavelet transform power is shown in log scale.} 
\label{fig:small_tch}
\end{figure*}

\begin{figure*}
\centering
\ContinuedFloat
\includegraphics[width=60mm]{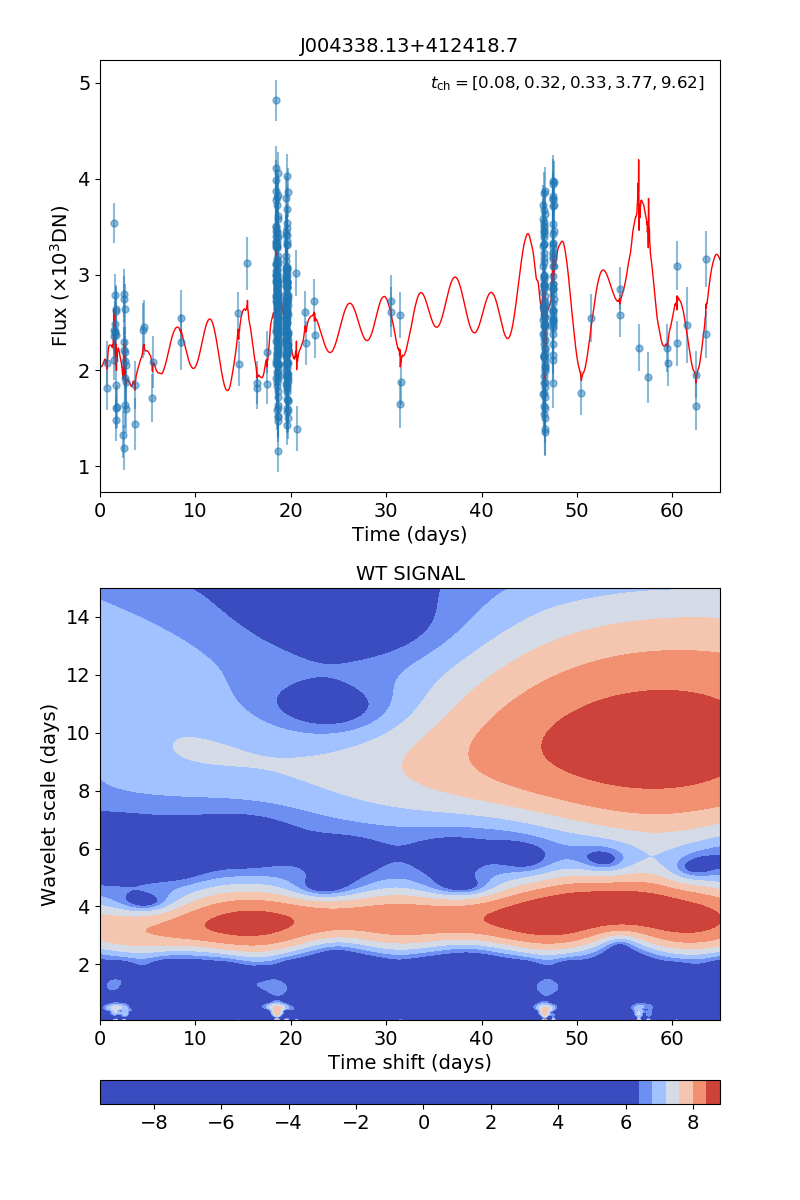}\hfill\includegraphics[width=60mm]{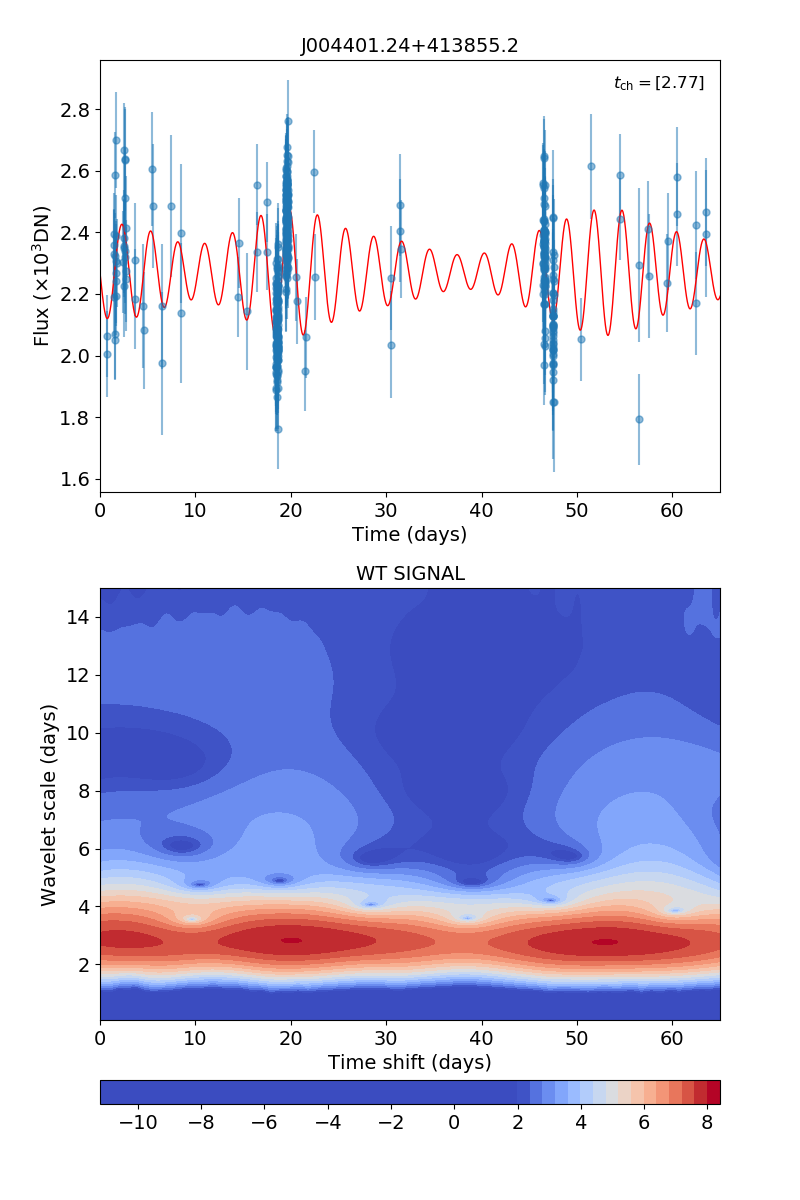}\hfill\includegraphics[width=60mm]{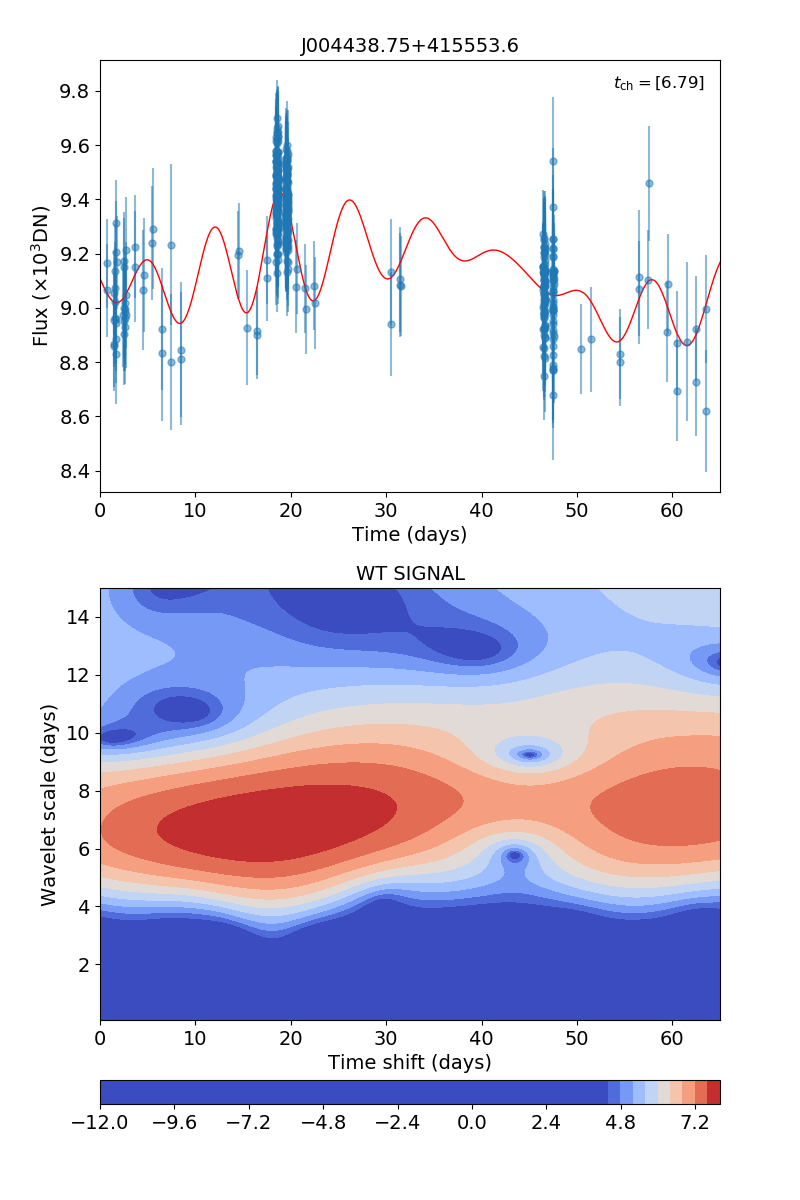}
\includegraphics[width=60mm]{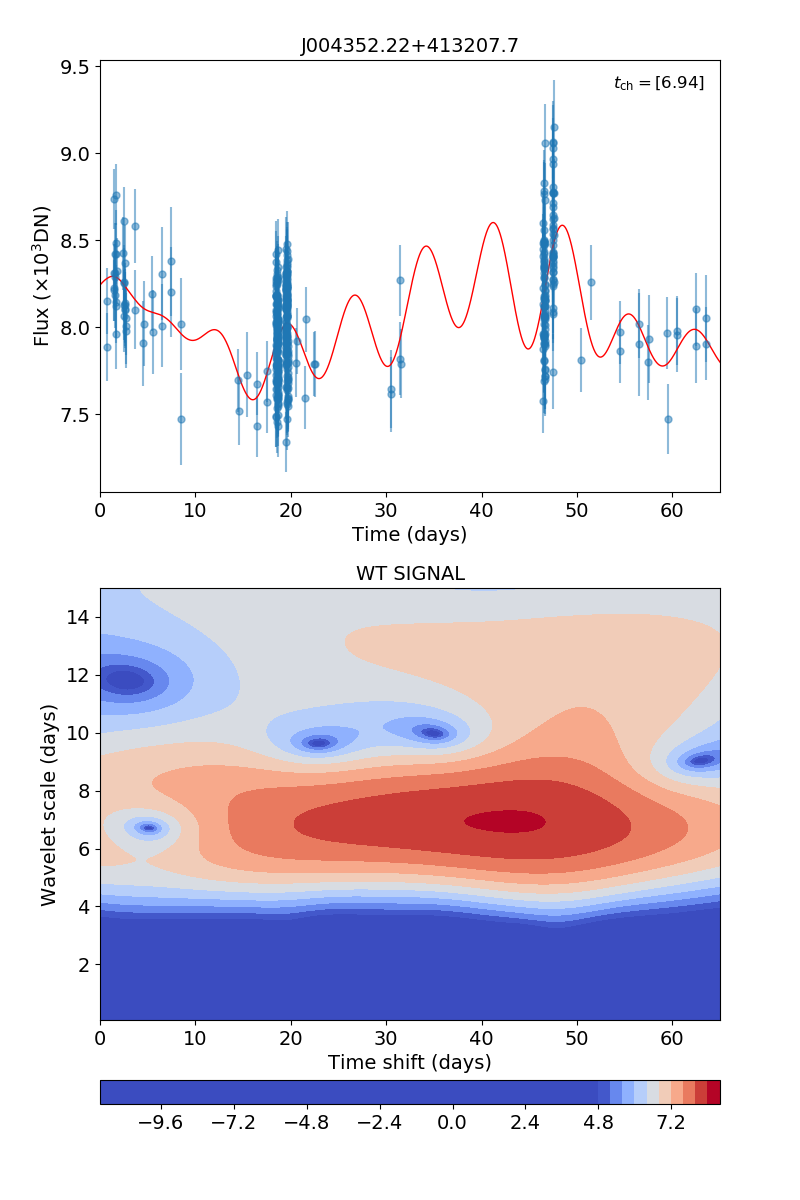}\hfill\includegraphics[width=60mm]{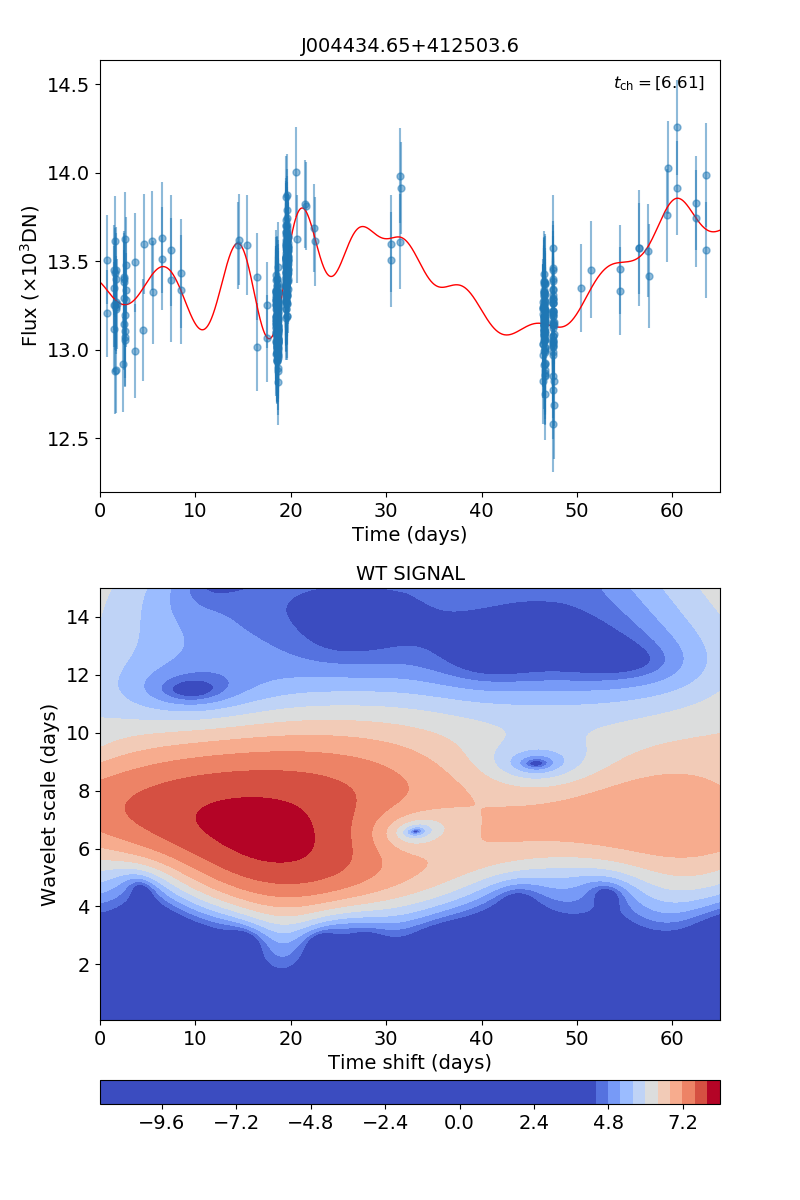}\includegraphics[width=60mm]{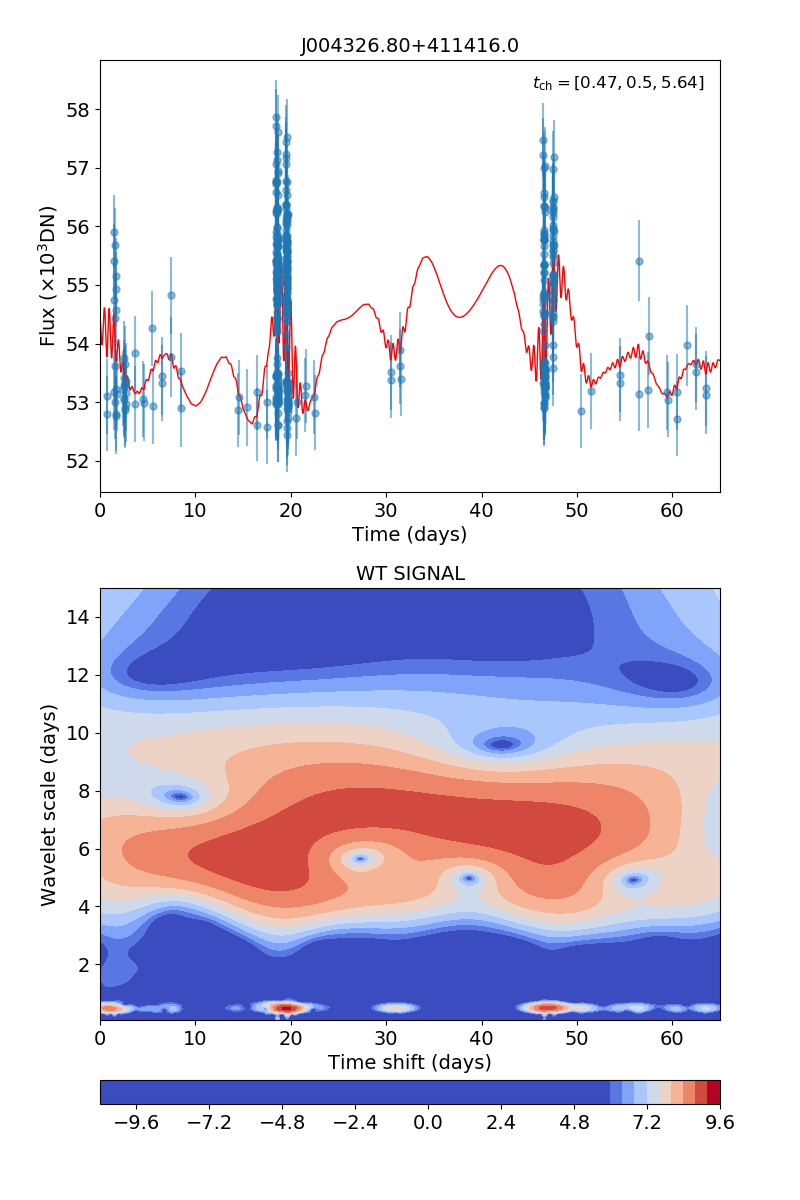}
\caption{Contd. The spectral types of the stars (going from left to right) are B0Ia, B1.5I:, B2.5Ia,  ({\it top} panels), and B8I, B5Ia+Neb, A:I ({\it bottom} panels).} 
\end{figure*}

\begin{figure*}
\centering
\ContinuedFloat
\includegraphics[width=60mm]{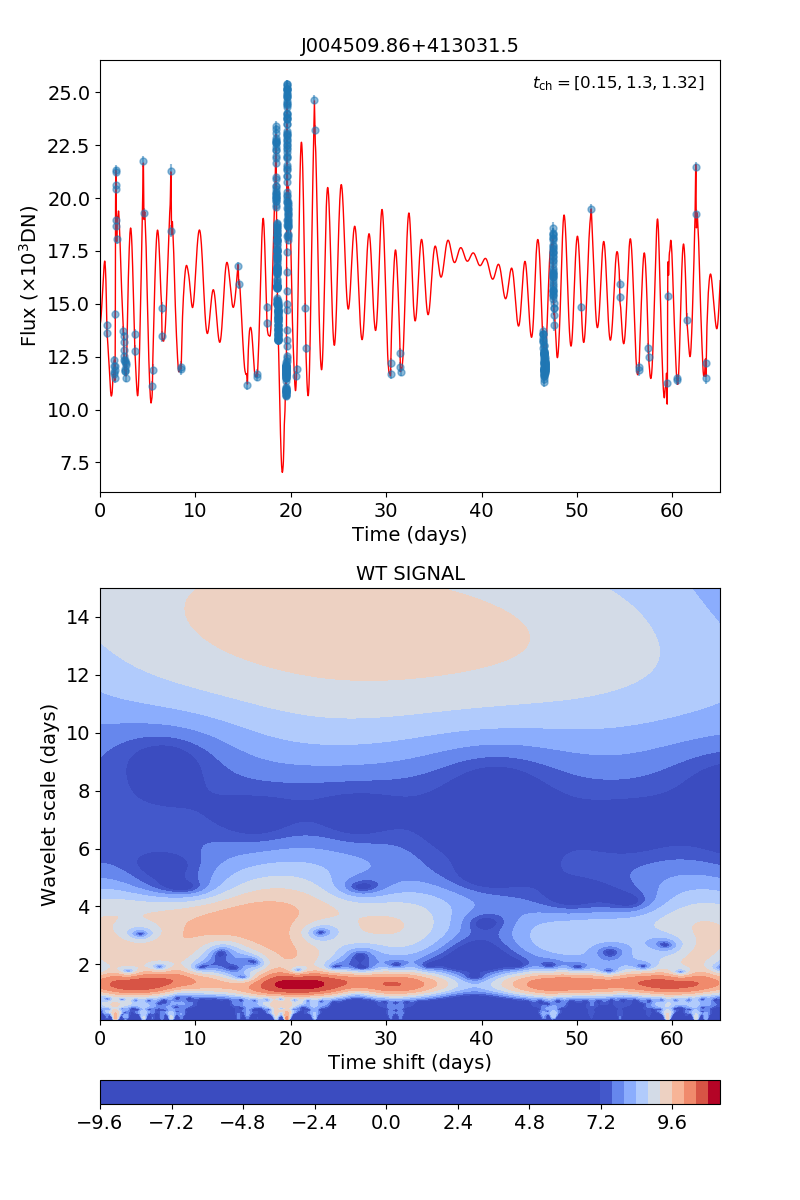}\hfill\includegraphics[width=60mm]{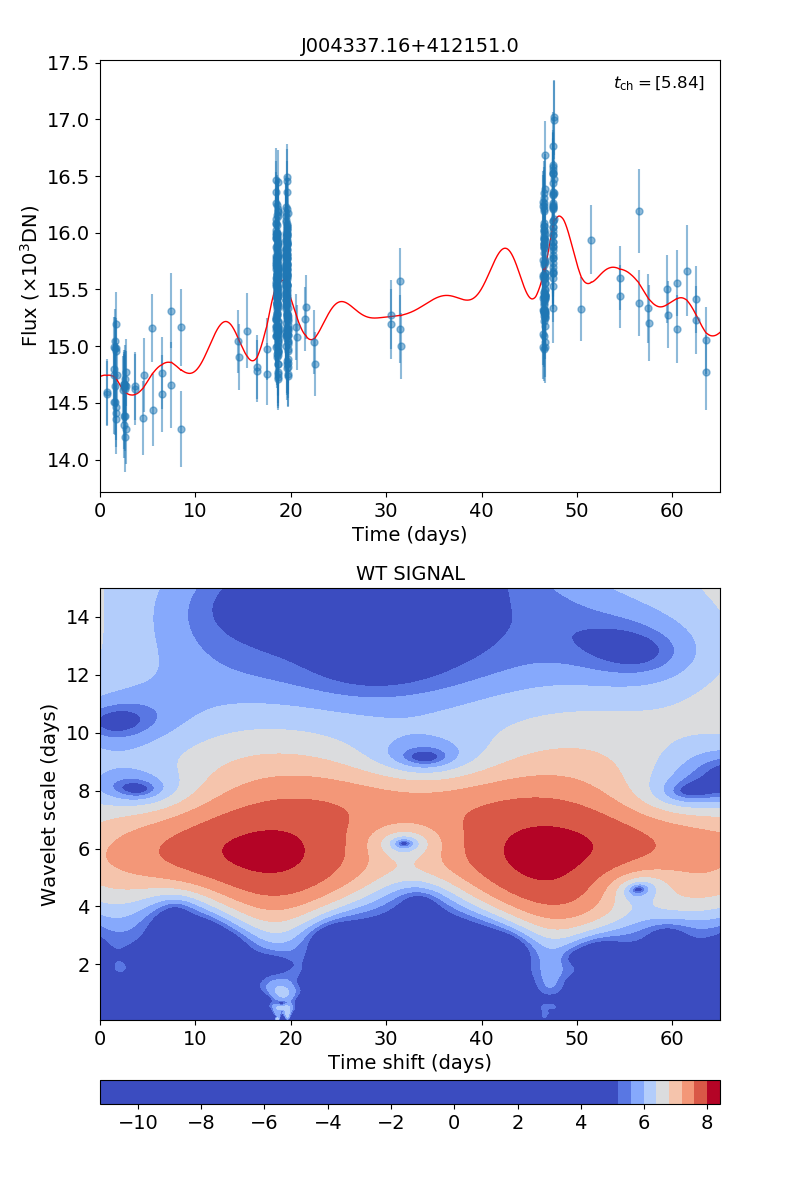}\hfill\includegraphics[width=60mm]{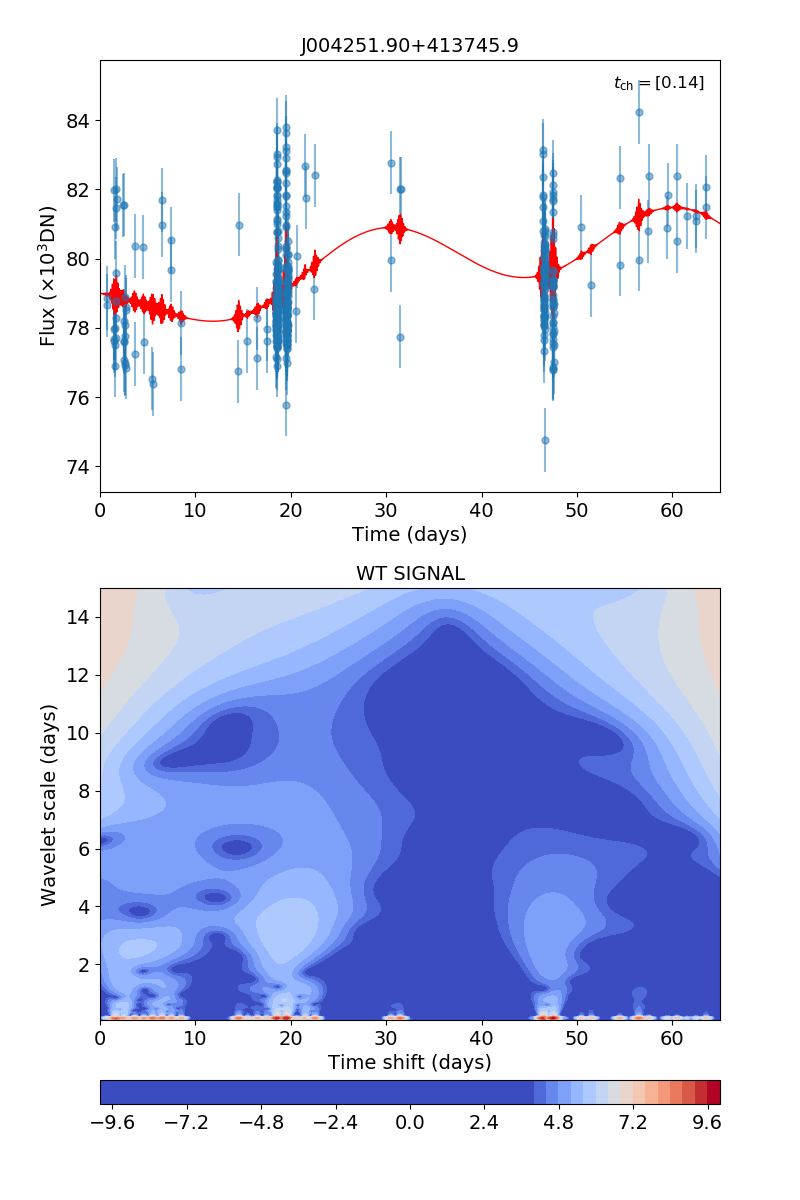}
\includegraphics[width=60mm]{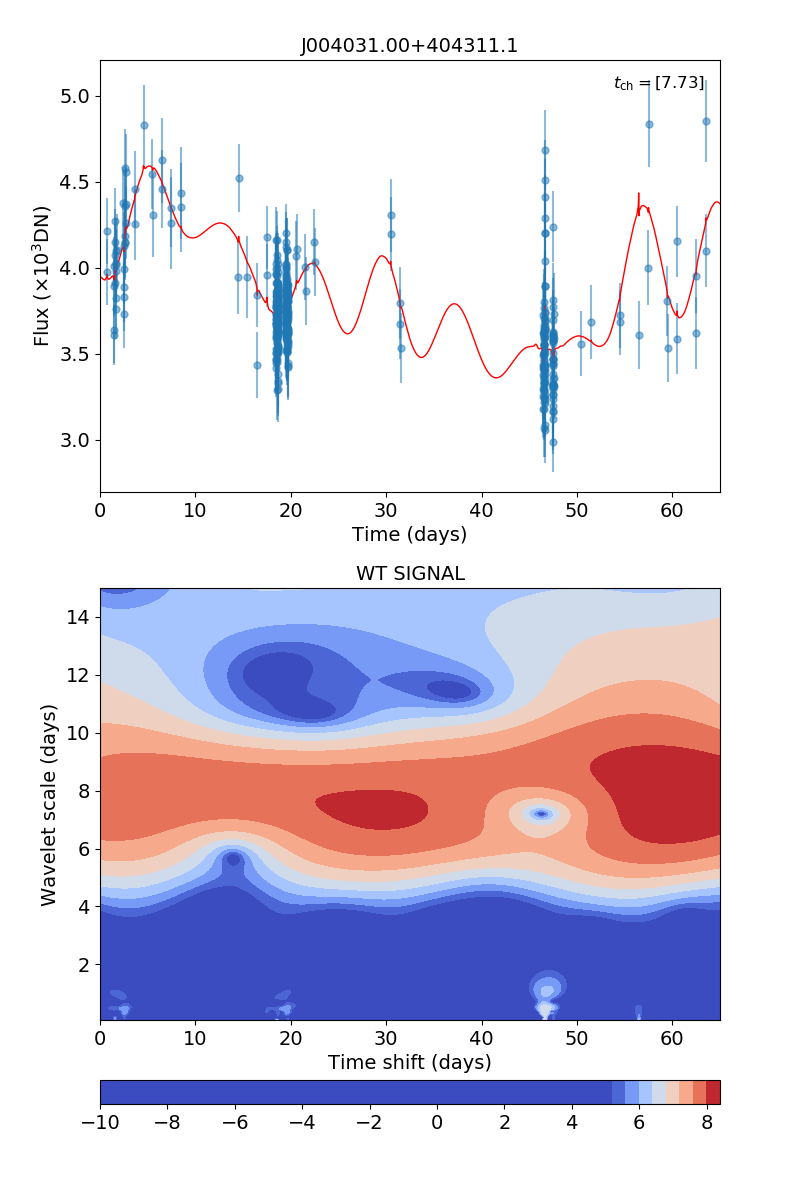}
\caption{Contd. The spectral types of the stars are YSG: (all three {\it top} panels) and M0I ({\it bottom}).} 
\end{figure*}

\section{Examples of Wavelet transform of simulated signals}\label{append:wt_sim}

\begin{figure*}
\centering
\includegraphics[width=60mm]{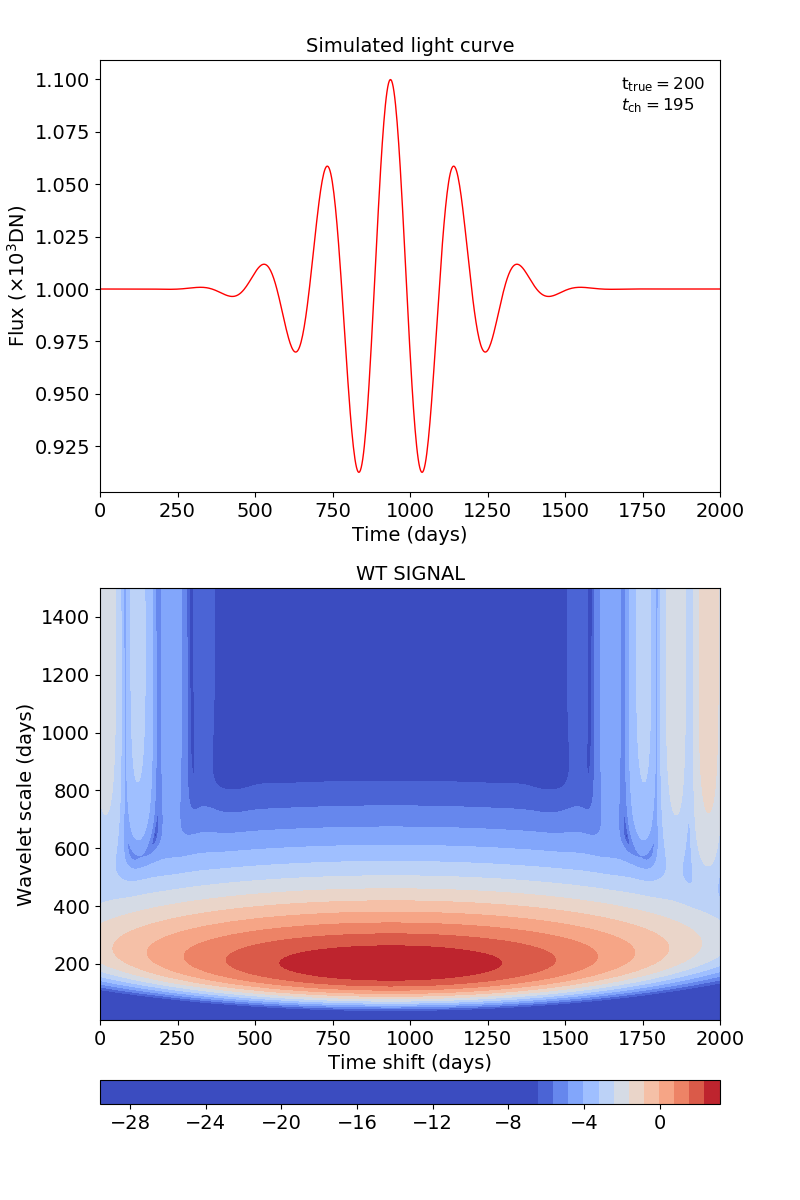}\hfill\includegraphics[width=60mm]{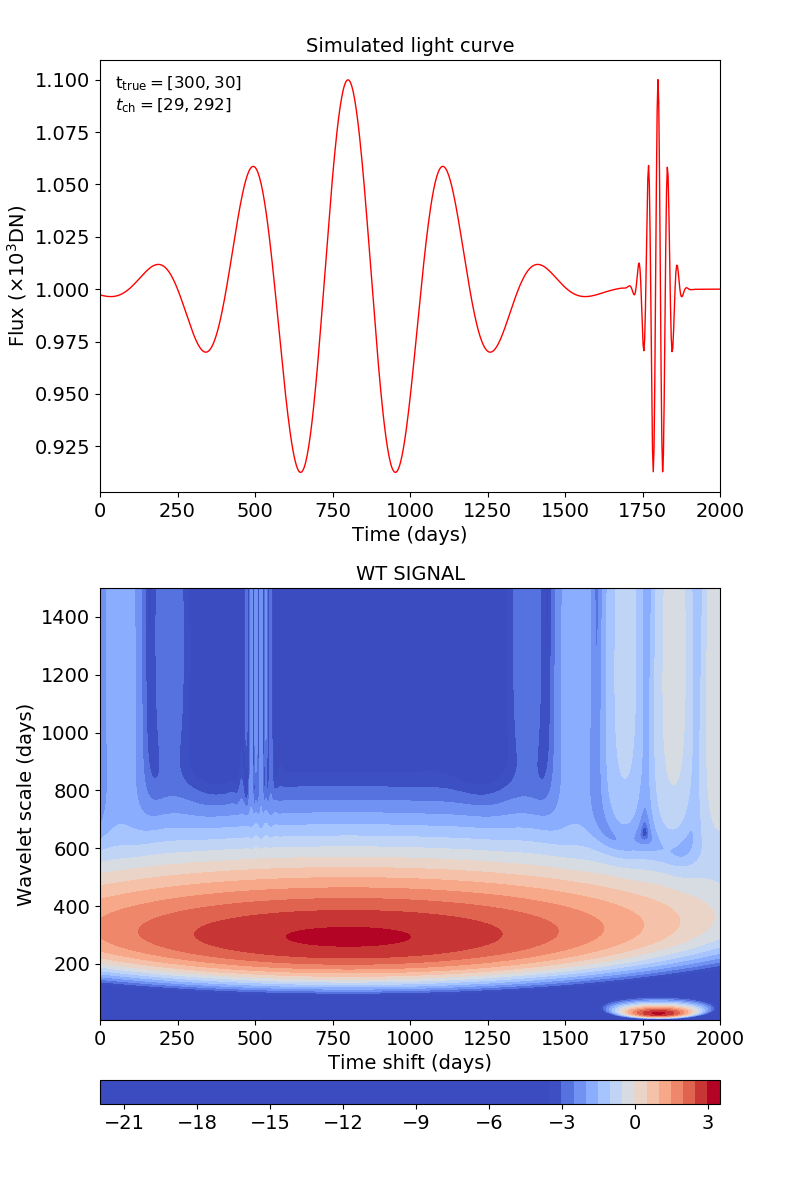}\hfill\includegraphics[width=60mm]{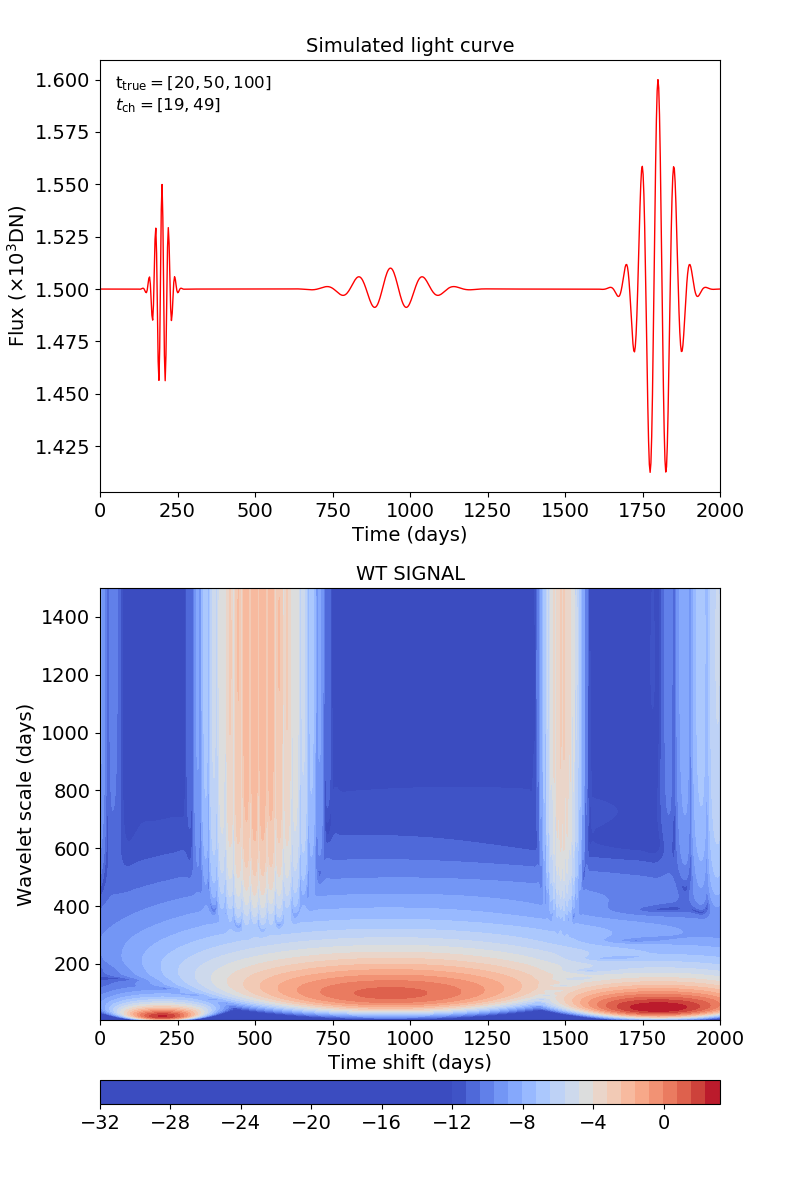}\\
\includegraphics[width=60mm]{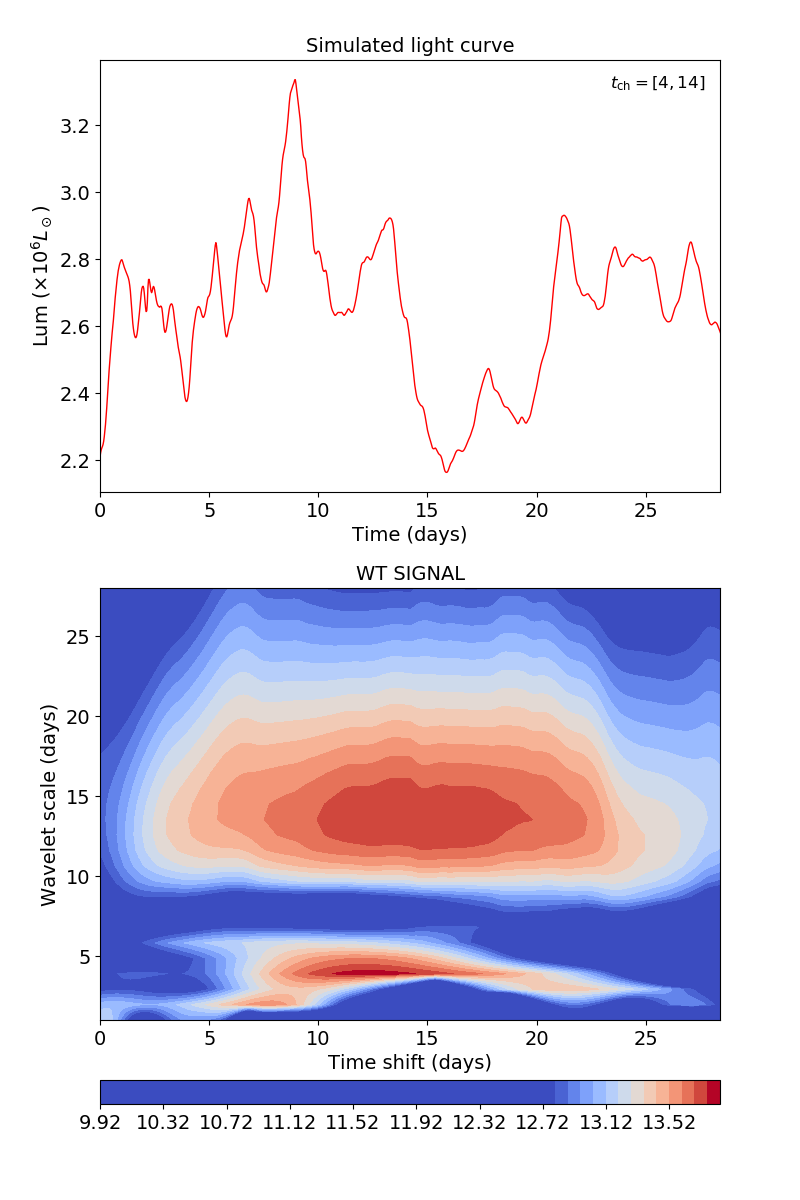}
\caption{Simulated light curves and their corresponding wavelet transform signal ({\it top} three panels). The input and recovered characteristic timescales, $t_{\rm true}$ and $t_{\rm ch}$, respectively, are also indicated in the legend. The bottom panel shows the theoretical lightcurve of the LBV model with $T_{\rm eff}=19000$~K and luminosity $\log(L/L_{\odot})=6.4$ from the 3D simulation of \citet{YanFei-2018}. The wavelet transform power is shown in logscale.} 
\label{fig:sims}
\end{figure*}

We perform the wavelet transform of some simulated signals with localized events to verify the recovery of the input characteristic timescales $t_{\rm true}$. To this end, we model the localized event with a wavelet and consider three light curves containing one, two and three wavelets, as shown in Fig.~\ref{fig:sims}. Each light curve extends over a baseline of 2000~days with a cadence of 2.5~days. The wavelet signal in the first light curve has an amplitude (in linear flux units) value of 100 and $t_{\rm true}$ of 200~days; for the second light curve, the two wavelets have the same amplitude of 100 but different $t_{\rm true}$ values of 300~days and 30~days; while the third light curve has three wavelets superposed with $t_{\rm true}$ values of 20~days, 100~days, and 50~days and corresponding amplitudes 50, 10, and 100. 

The respective wavelet transform power is shown below each simulated light curve, and the recovered timescale $t_{\rm ch}$ applying the same method as in Sect.~\ref{sec:time} is shown in the legend of each panel. As can be seen, we recover timescales similar to the input values for the three light curves. For the third light curve, the smallest amplitude $t_{\rm ch}$ is missed. This is due to the fact that, for the adopted background threshold, the island of power excess for this timescale has merged with the neighboring one. We experimented with different threshold values, but the threshold adopted in Sect.~\ref{sec:time} appears optimal and is hence used throughout our analysis. 

We find that the absolute magnitude of the wavelet transform coefficient, or the square root of the transform power, corresponding to the recovered timescale varies with the amplitude of the signal and inversely with the pixel size used in the Gaussian-Process reconstruction of the light curve. We determine the scaling factor by comparing the RMS amplitude of the simulated light curve, evaluated over an interval of width $2\times t_{\rm true}$, with the root of the transform power corresponding to $t_{\rm ch}$, and find it to be $1.7\times {\rm pixel}/2.5 ({\rm days})$. 

We also perform the wavelet transform of one of the theoretical LBV light curves from the 3D simulation of \citet{YanFei-2018} that shows stochastic varaibility. We find two timescales from our automated method of determining $t_{\rm ch}$, which are 4~days and 14~days with the variability amplitudes (relative to the average flux value of the light curve) being $\approx 5\%$ for both $t_{\rm ch}$ values.

\end{document}